\documentclass{article}
\usepackage[utf8]{inputenc}
\usepackage{amsmath}
\usepackage{amsthm}
\usepackage{graphicx}
\usepackage[mathletters]{ucs}
\usepackage{varioref}
\usepackage{units}
\usepackage{textcomp}
\usepackage{amsfonts}
\usepackage{bm}
\usepackage{amssymb}
\usepackage{mathalpha}
\usepackage{cancel}
\usepackage[unicode=true,
 bookmarks=true,bookmarksnumbered=false,bookmarksopen=false,
 breaklinks=false,pdfborder={0 0 1},backref=false,colorlinks=false]
 {hyperref}
\hypersetup{pdftitle={Time dispersion in bound states},
 pdfauthor={John Ashmead},
 pdfsubject={Estimate of time dispersion in bound states},
 pdfkeywords={time, quantum electrodynamics, quantum mechanics, relativity, experiments}}

\pdfoutput=1

\newcommand{\doi}[1]{DOI: \url{#1}}
\usepackage{latexsym}

\begin{document}
\title{Time dispersion in bound states}
\author{John Ashmead}
\maketitle
\begin{center}
Visiting Scholar, University of Pennsylvania, USA
\par\end{center}

\begin{center}
\textsl{jashmead@seas.upenn.edu}
\par\end{center}
\begin{abstract}
In quantum mechanics time is generally treated as a parameter rather
than an observable. For instance wave functions are treated as extending
in space, but not in time. But from relativity we expect time and
space should be treated on the same basis. What are the effects if
time is an observable? Are these effects observable with current technology?

In earlier work we showed we should see effects in various high energy
scattering processes. We here extend that work to include bound states.
The critical advantage of working with bound states is that the predictions
are significantly more definite, taking the predictions from testable
to falsifiable. 

We estimate the time dispersion for hydrogen as $.177$ attoseconds,
possibly below the current threshold for detection. But the time dispersion
should scale as the $3/2$ power of the principle quantum number $n$.
Rydberg atoms can have $n$ of order $100$, implying a boost by a
factor of $1000$. This takes the the time dispersion to $177$ attoseconds,
well within reach of current technology.

There are a wide variety of experimental targets: any time-dependent
processes should show effects. Falsification will be technically challenging
(due to the short time scales) but immediate and unambiguous. Confirmation
would have significant implications for attosecond physics, quantum
computing and communications, quantum gravity, and the measurement
problem. And would suggest practical uses in these areas as well as
circuit design, high speed biochemistry, cryptography, fusion research,
and any area involving change at attosecond time scales.
\end{abstract}

\newpage\tableofcontents{}

\newpage

\section{Introduction}

\label{sec:Introduction}
\begin{quote}
``...when we set aside non-locality and the measurement problem,
and ask how quantum theory’s formalism treats time, the answer is,
broadly speaking, quantum theory has nothing very special to say about
time---it just tags along with whatever is said by the cousin classical
theory, be it Newtonian or relativistic.'' -- Jeremy Butterfield
\cite{Butterfield:2014aa}
\end{quote}
\emph{Should quantum mechanics be applied along the time dimension
as it is in space? }The motivation for this question comes from special
and general relativity where time and space appear on an equal footing.
In the early days of quantum mechanics, the equivalence of time and
space was almost taken for granted: see the celebrated Bohr-Einstein
debates (especially the clock-in-a-box experiment \cite{Schilpp:1949oz})
or Heisenberg's development of the Heisenberg uncertainty principle
in time/energy \cite{Heisenberg:1930kb}. But the field has developed
in a different direction. As Hilgevoord \cite{Hilgevoord:1996bh,Hilgevoord:1998qu}
put it, in quantum mechanics ``time is a parameter not an operator''.

What might it mean to promote time back to being an operator? The
simplest approach is to extend the wave function to include time:
$\mathit{\psi}\left({\vec{x}}\right)\rightarrow\mathit{\psi}\left({t,\vec{x}}\right)$.
In the Feynman path integral approach this takes the form of including
paths that extend in time in addition to the usual three space dimensions.

We will refer to standard quantum mechanics as SQM. We will refer
to quantum mechanics with time taken as an operator as TQM. We are
using natural units throughout $\hbar=c=\epsilon_{0}=1$.

\paragraph{Is dispersion in time falsifiable with current technology?}

Our approach is to assume TQM is correct and look for testable, ideally
falsifiable, consequences. The requirements are:
\begin{enumerate}
\item Complete consistency with SQM in the domains where SQM is tested.
\item Complete covariance. 
\item Predictions falsifiable with current technology.
\end{enumerate}
The relative time scale is attoseconds. This scale is now achievable:
the shortest pulse times are now of order $43as$ (attoseconds) \cite{Kheifets:2020aa}.
Further, researchers (i.e. Ossiander et al \cite{Ossiander:2016fp})
have been able to detect even sub-attosecond effects. TQM is therefore
falsifiable with current technology, at least in principle. Since
we will need to estimate some quantities, we will need to find situations
where the time dispersion is expected to be $10as$ or greater, to
give a margin of one order of magnitude.

Since by hypothesis, all quantum effects in space -- interference,
coherence, tunneling, uncertainty, and most importantly entanglement
-- are also present in time TQM suggests a large number of experiments:
TQM is an experiment factory. If it is confirmed, then it will also
have a comparably large number of uses. Both aspects are discussed
below in subsections \ref{sec:Experimental-tests} and \ref{sec:Uses}.

\paragraph{Possible objections}
\begin{enumerate}
\item \emph{The extension of SQM to TQM is not well-defined.} We use the
Feynman path integral approach (FPI) \cite{Feynman:2010bt,Feynman:2015jk,Grosche:1998uj,Huang:1998bd,Kaku:1993xj,Kashiwa:1997xt,Kleinert:2009hw,Rivers:1987ma,Schulman:1981um,Swanson:1992ju,Zinn-Justin:2005nx}
to deal with this. Using the FPI, we extend the set of paths included
in the path integrals to go from paths in three dimensions to four.
The rest of the machinery is left essentially unchanged. As a result,
the predictions are unambiguous.
\item \emph{If dispersion in time is real, it should have been seen before.}
There are two explanations for this:
\begin{enumerate}
\item The estimates we get here are of order attoseconds, small enough to
easily explain why the effects of dispersion in time have not been
seen by accident. 
\item It can be very difficult to see something which you are not expecting
to see.\footnote{``I cannot think how I came to overlook it,'' said the Inspector,
with an expression of annoyance.

``It was invisible, buried in the mud. I only saw it because I was
looking for it.''

-- Mr. Sherlock Holmes in \emph{The Adventure of the Silver Blaze}
\cite{Doyle:1892dy}}
\end{enumerate}
\item \emph{Quantum electrodynamics (QED) is already covariant. Therefore
there is no need to extend quantum mechanics to include time as observable.}
A detailed examination, done in the earlier work, and summarized here,
shows that -- appearances not withstanding -- the paths in standard
Feynman diagrams are confined to on-shell trajectories. The integrals
are four-dimensional, but the fourth integral is a contour integral,
which samples the four dimension space only at the poles, not in general.
In TQM we include the off-shell paths; arguably more consistent with
the spirit of quantum mechanics. 
\item \emph{QED is fully confirmed. There is no room for dispersion in time.}
It \emph{is} confirmed to extraordinary precision. But the tests look
at longer times. The realm where TQM differs from SQM is at short
times with rapidly varying fields. We are only just starting to probe
the relevant domains.
\item \emph{TQM may ``work'' but will not be renormalizable. }With one
additional dimension to integrate over, the loop integrals may become
unrenormalizable. This question was dealt with in earlier work \cite{Ashmead:2023aa}.
In practice it turns out that the requirement of entanglement in time
entangles the initial wave functions with the loop integrals. If the
initial wave functions are normalized (as they must be), they act
as an implicit but organic regularizing factor. What appears to be
an objection turns out to be a significant advantage for TQM.
\end{enumerate}

\paragraph{Outline}
\begin{enumerate}
\item \emph{Restatement}: We summarize the previous work \cite{Ashmead:2019aa,Ashmead:2021aa,Ashmead:2023aa},
both to save the reader the trouble of flipping back and forth and
to present the earlier results in a way that serves as a springboard
for the analysis here.
\item \emph{Bound states}: We estimate the time dispersion in bound states.
This is the core of this work. We will show we expect the time dispersion
for hydrogen to be $.177as$ attoseconds (the typical time for a photon
to travel from nucleus to atomic electron) and that in general the
time dispersion should scale as the atomic radius $r^{3/4}$. Since
the atomic radius scales as the principal quantum number $n^{2}$,
and since Rydberg atoms \cite{Sibalic:2018aa} can have $n\sim100$
(and greater), the time dispersion should scale as $1000=100^{3/2}$
The time dispersion should therefore reach $177as$ or greater. This
is well past our target of $10as$.
\item \emph{Scattering}: We apply the results to the analysis of simple
scattering experiments.
\item \emph{Detection}: We look at the problem of the detection of a single
particle, the implications for measurement, possible experimental
tests, and possible for the time dispersion.
\item \emph{Discussion}: We finish by summarizing the argument here with
particular reference to falsifiability.
\end{enumerate}

\section{Restatement}

\label{sec:Restatement}
\begin{quote}
``Wheeler's often unconventional vision of nature was grounded in
reality through the principle of radical conservatism, which he acquired
from Niels Bohr: Be conservative by sticking to well-established physical
principles, but probe them by exposing their most radical conclusions.''
-- K. S. Thorne \cite{Thorne:2009wa}
\end{quote}

\subsection{Clock time and coordinate time}

\label{subsec:clock-time-and-coordinate-time}

\begin{figure}
\includegraphics[scale=0.67]{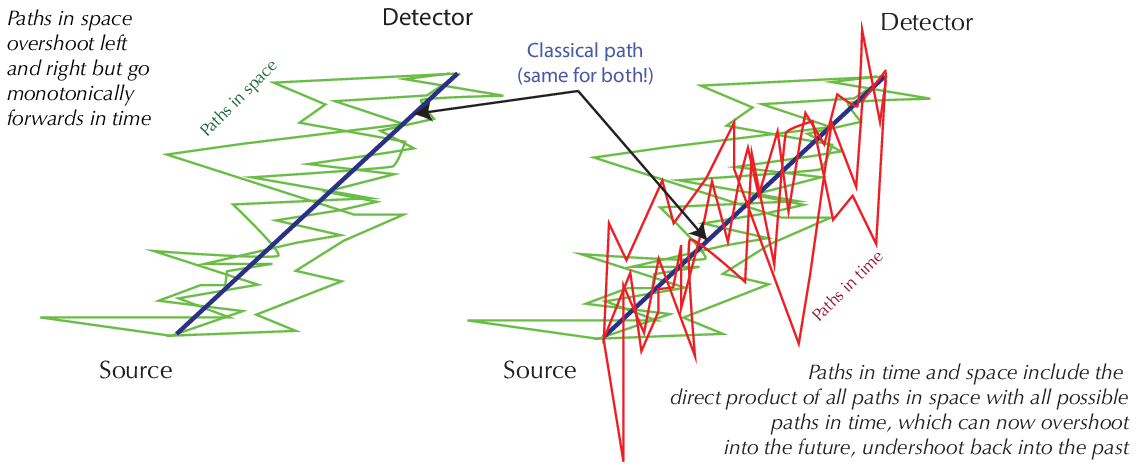}

\caption{\protect\label{fig:Paths-in-space;}Paths in space; paths in space
and time}
\end{figure}

We start with clock time as defined by laboratory clocks (\cite{Busch-2001,Muga:2002ft,Muga:2008vv}).
Clock time (also called laboratory time) has the properties normally
associated with time. It is a parameter, obeys Pauli's theorem \cite{Pashby:2014wu},
and so on. It is in charge of achieving conformance to our ordinary
intuitions about time, the respectable side of time as it were.

Now we introduce a second term, \emph{coordinate time}. This is time
as defined by covariance with respect to the three space dimensions.
Normally the paths in Feynman path integrals (FPI) are defined as
a series of points in a three dimensional grid. To extend FPIs to
four dimensions, we extend the usual three dimensional (3D) grid to
four dimensions. Now a path is defined as a series of points in four
rather than three dimensions. In figure 2.1 we illustrate this. The
four dimensional (4D) grid is a help to visualization and sometimes
to calculation as well (as \cite{Kashiwa:1997xt}). At an appropriate
point in the analysis we will take the limit as the grid spacing goes
to zero.

Effectively we are addressing the problem of time (\cite{Schulman:1997fd,Zeh:2001xb,Muga:2002ft,Muga:2008vv,Callender:2017rt})
by using a variation on divide and conquer: we are assigning the responsibilities
of time as parameter to clock time; of time as operator to coordinate
time.

\subsection{Single particle}

\label{subsec:single-particle}

Now we look at an FPI analysis of the quantum mechanics of a single
particle. One of the merits of the FPI approach is that it uses relatively
few ingredients. We discuss them in turn.

\paragraph{Paths}

The only change we are making to conventional FPI analyses is the
shift from a 3D to a 4D grid. To define the paths we break up the
clock time from $A\to B$ in $N$ time steps with each tick of size:

\begin{equation}
\varepsilon\equiv\frac{T}{N}\label{eq:slice-width-in-clock-time}
\end{equation}

At the end we will take the limit as $N$ goes to infinity or equivalently
$\epsilon\to0$.

Now we grid the space, first into a 3D grid. Let's say each grid cell
is a cube with dimensions $a^{3}$. 

At each clock tick, the path will be assigned a specific triad of
space coordinates. Each path is defined as a series of grid points,
one at each clock tick. To sum over the paths, we sum over the associated
measure:

\begin{equation}
\mathcal{D}\vec{x}\equiv\prod\limits_{n=0}^{N}{d{{\vec{x}}_{n}}}\label{eq:slice-metric}
\end{equation}
In some treatments the actual calculations are done using the limit
as $a\to0$, in other cases the grid simply serves as an aid to visualization.

To extend SQM to TQM we extend the 3D to the 4D grid. Each grid cell
is now a hypercube with hyper-volume $a^{4}$. The path measure is
now:

\begin{equation}
\mathcal{D}x\equiv\prod\limits_{n=0}^{N}{d{t_{n}}d{{\vec{x}}_{n}}}\label{eq:4D-slice-measure}
\end{equation}

And we extend the paths in parallel, the 3D to the 4D paths:

\begin{equation}
\pi\left({{\vec{x}}_{\tau}}\right)\to\pi\left({{t_{\tau}},{{\vec{x}}_{\tau}}}\right)\label{eq:slice-measure-1}
\end{equation}
We refer to the time dimension used in this way as coordinate time
$t$, with its properties defined with respect to space by covariance.

The resulting paths are in four dimensions. They curve around in time,
so can dart into the future or the past. Therefore the 4D grid has
to extend in past and in future far enough to include all computationally
significant paths. Note however that paths that stray too far from
the classical path -- in the free case a straight line from source
to detector -- are subject to cancellation by destructive interference.
So in practice the grid does not need to overshoot by that much. But
to that extent at least we are working in the block universe picture
\cite{Kirk:1983vn,Price:1996mr,Barbour:2000hl}: paths cannot dart
into the future unless the future exists.\footnote{Or back into the past unless the past in some sense is still present.}

\paragraph{Kernel}

Our primary object is to compute the kernel to go from $A\to B$.
This is given by the sum over all paths, weighted by the action, defined
as the integral of the Lagrangian along each path:

\begin{equation}
{K_{\tau}}\left({x'';x'}\right)=\int{\mathcal{D}{x_{\tau}}\exp\left({\imath\int\limits_{0}^{\tau}{d\tau'\mathcal{L}\left[{{x_{\tau}},{{\dot{x}}_{\tau}}}\right]}}\right)}\label{eq:4D-kernel-1}
\end{equation}

\paragraph{Lagrangian}

\label{par:Choice-of-Lagrangian-1}

We need a Lagrangian which is manifestly covariant, which correctly
models the behavior of a particle in an electro-magnetic field, and
which works equally well in 3D and 4D. We will use the following Lagrangian,
which we have from Goldstein \cite{Goldstein:1980ce} and also from
Feynman \cite{Feynman:1998wv}:

\begin{equation}
\mathcal{L}\left[{{x_{\tau}},{{\dot{x}}_{\tau}}}\right]=-\frac{1}{2}m{{\dot{x}}^{\mu}}{{\dot{x}}_{\mu}}-q{{\dot{x}}^{\mu}}{A_{\mu}}\left(x\right)-\frac{m}{2}\label{eq:4D-Lagrangian-1-1}
\end{equation}

This Lagrangian works for both 3D and 4D cases. In classical mechanics
the classical path is defined as the extremum of the variation of
the Lagrangian with respect to the relevant coordinates. With this
Lagrangian you get the same classical path whether you vary the Lagrangian
by the three space coordinates or by them plus time. The details are
covered in Goldstein. As a result we can use the same Lagrangian for
both SQM and TQM. This means that the only thing changed in going
from SQM to TQM is the paths. And since this is the defining feature
of the extension, this means that with this approach there are no
free parameters. And the extension to 4D is therefore completely defined
and therefore falsifiable.

\paragraph{Convergence}

\label{par:Convergence-of-slices}

Path integrals are normally computed by starting at a specific time,
then integrating slice-by-slice. The grid is helpful in visualizing
this. We break the paths into discrete steps numbered from $1$ to
$N$. We also replace the terms in the Lagrangian by their discretized
forms, i.e. for the free part:

\begin{equation}
{\mathcal{L}}^{DISCRETE}{=}{-}\frac{m}{2}\mathop{\sum}\limits_{{n}{=}{1}}^{{n}{=}{N}}{{\mbox{\ensuremath{\left({\mbox{\ensuremath{\frac{{t}_{n}{-}{t}_{{n}{-}{1}}}{\varepsilon}}}}\right)}}}^{2}{-}{\mbox{\ensuremath{\left({\mbox{\ensuremath{\frac{{\vec{x}}_{n}{-}{\vec{x}}_{{n}{-}{1}}}{\varepsilon}}}}\right)}}}^{2}}
\end{equation}

For this to make sense, the individual integrals have to converge.
The problem is that the $t$ and the $\vec{x}$ integrals enter with
opposite signs. In path integrals, convergence is normally forced
by adding small convergence factors, i.e. rewriting the mass as $m\to m+\imath\epsilon$.
But any such simple trick that makes one converge will cause the other
to blow up. And if we handle the time and space parts differently,
we violate covariance.

However if we are integrating against a initial Gaussian test function
(GTF), these tricks are not needed in the first place. The GTF itself
will keep each step convergent. GTFs are completely general: by using
Morlet wavelet analysis (MWA) we can decompose any normalizable wavelet
into sums over GTFs. So there is no loss of generality in computing
for a single GTF, then applying the results to an arbitrary (normalizable)
wave function. For further discussion see appendix \ref{sec:=000020Gaussian=000020test=000020functions}.

This give us convergence.

The limitation to normalizable initial wave functions is not an unacceptable
one. All physical wave functions are normalizable. And we are only
required to describe physically realizable experimental situations.
We may describe more, but we are not required to do so. See Bohr's
observation on this point \cite{Bohr:1935bt} in response to the EPR
paper \cite{Einstein:1935er}.

This fundamental requirement does mean that all TQM calculations --
to be well-defined -- must refer to a specific initial wave function.

\paragraph{Semi-classical approximation}

\label{par:restatement-Semi-classical-approximation}

Setting the variation of the Lagrangian to zero defines the classical
path. It also defines the solution of the integral by the stationary
phase method. So the first approximation to the kernel is given by
the action along the classical path:

\begin{equation}
{K}_{\mathit{\tau}}\left({x;{x}\prime}\right)\sim\exp\left({\imath{S^{CM}}}\right)
\end{equation}

Further, given $S^{CM}$ we can compute the kernel to quadratic order
in the coordinates directly from it, as the fluctuation factor \cite{Schulman:1981um,Kleinert:2009hw}: 

\begin{equation}
{K}_{\mathit{\tau}}\left({x;{x}\prime}\right){=}\sqrt{\frac{1}{\det\left({S^{CM}\left({x;{x}\prime}\right)}\right)}}\exp\left({\imath{S^{CM}}}\right)
\end{equation}

For example with the free Lagrangian:

\begin{equation}
{S}_{\mathit{\tau}}^{CM}\left({x;x'}\right){=}\frac{m}{{2}\tau}{\left({x-x'}\right)}^{2}{-}\imath\frac{m}{2}\tau
\end{equation}

We get the kernel:

\begin{equation}
{K}_{\mathit{\tau}}\left({x;x'}\right){=}\frac{1}{\sqrt{\det\left({\frac{{\partial}^{2}{S}^{CM}}{\partial{x}\partial{x}{'}}}\right)}}\exp\left({\imath{S}_{\mathit{\tau}}^{CM}}\right)
\end{equation}

It is a considerable strength of the FPI approach that there is such
a natural transition between quantum mechanics and classical mechanics.
The classical path is like the river going thru the center of a river
valley; like rivers the classical path finds the shortest path from
one place to another. And the slopes leading down to the river are
like the quantum fluctuations around the classical path. Depending
on the specifics of the Lagrangian the quantum fluctuations may be
tightly bound -- the river has a narrow gorge -- or spread out broadly
-- like an alluvial plane.

TQM is to SQM with respect to time as SQM is to classical mechanics
with respect to the three space dimensions. In TQM we add quantum
fluctuations in time to the quantum fluctuations in space that SQM
describes. Of course TQM is doing this in four-space and SQM is doing
this in three-space, while most rivers are fairly two-dimensional
in their behavior. But the point is clear.

\paragraph{Feynman-Stückelberg equation in time}

Usually we derive the FPI expression from the Schrödinger equation
\cite{Schulman:1981um,Feynman:1965bh,Kleinert:2009hw}. But here we
work in reverse, we go from the kernel to the Schrödinger equation.
To do this we take the time derivative of the kernel via a limiting
process:

\begin{equation}
\frac{\partial{\psi}}{\partial\tau}{=}\mathop{\lim}\limits_{\varepsilon\rightarrow{0}}\frac{{K}_{\tau{+}\varepsilon}{-}{K}_{\tau}}{\varepsilon}\psi
\end{equation}

and use this to compute the associated Schrödinger equation:

\begin{equation}
-2m\imath\frac{{\partial{\psi_{\tau}}}}{{\partial\tau}}=\left({\left({{p_{\mu}}-q{A_{\mu}}}\right)\left({{p^{\mu}}-q{A^{\mu}}}\right)-{m^{2}}}\right){\psi_{\tau}}\label{eq:single-FS/T-nonrel}
\end{equation}

This is a variation on the Feynman-Stückelberg equation \cite{Stueckelberg:1941aa,Stueckelberg:1941la,Feynman:1948,Feynman:1949sp,Feynman:1949uy,Feynman:1950rj}.
See also the Relativistic Dynamics (RD) program \cite{Feynman:1948,Fanchi:1993aa,Fanchi:1993ab,Land:1996aj,Horwitz:2015jk}.
Usually in the RD program $\tau$ is a free parameter; here it is
\emph{specifically} defined as the clock time. We will refer to this
equation with the specialization of $\tau$ to clock time as the Feynman/Stueckelberg
equation with time or the FS/T equation. Our use of clock time is
similar to Fanchi's historical time: \cite{Fanchi:1993ab}.

The FS/T functions as an interpolating equation between the Klein-Gordon
(KG) equation and the Schrödinger equation. The dependence on clock
time is typically at scales of nanoseconds (time taken by light to
cross a lab bench); the dependence on coordinate time is typically
at scales of attoseconds (time taken by light to cross an atom). At
the atomic level the FS/T looks like the KG equation; the left side
is approximately zero. But at the scale of nanoseconds the off-shell
fluctuations will tend to average out and the behavior of the wave
function will look more like that of the Schrödinger equation in the
text books. We discuss solutions for the FS/T in appendix \ref{Solution=000020of=000020free=000020equation=000020for=000020massive=000020particles}.

There is a good description of this sort of evolution in Tannor \cite{Tannor:2007qz}.
The specific scales -- attosecond for KG part, nanosecond for Schrödinger
equation -- are worked out in detail in earlier work.

\subsection{Quantum field theory }

\label{subsec:Spin-zero-fields}

At this point we have enough machinery to describe non-relativistic
single particles. But we need to probe at high energies to see short
time effects, and at high energies, particle creation is likely. Therefore
we have to extend this approach to quantum field theory, here limited
to QED \cite{Feynman:1998wv,Bjorken:1965el,Bjorken:1965mo,Sakurai:1967tl,Ramond:1990ab,Kaku:1993xj,Greiner:1994gf,Peskin:1995rv,Weinberg:1995hu,Weinberg:1995rl,Huang:1998bd,Greiner:2000vl,Itzykson:2005cv,McMahon:2008gt,Zee:2010oy,Klauber:2013tg,Lancaster:qy,Schwartz:2014tx,Schwichtenberg:2020vg}
and further limited to the spin-zero cases. (The effects we are interested
in are largely independent of spin and polarization.)

The two principle approaches to field theory are canonical quantization
(CQ) and the FPI. Unfortunately for our purposes the special role
of time is wired into the CQ approach fairly deeply, as in the definitions
of conjugate momenta. We therefore focus on the FPI approach.

\paragraph{Extending the single particle treatment to Feynman diagrams}

We can leverage the work done with the single particle case by working
with Feynman diagrams. These can be seen as graphs of individual particles
interacting at vertexes. If we extend the FS/T to describe relativistic
particles but keep the topology of the diagrams and the interactions
at the vertexes the same, we will have extended our single particle
treatment to field theory in a straightforward way. 

The simplest way to extend the FS/T to the relativistic case is to
replace the rest mass on the left side with the relativistic energy
$E$, which will reduce to the rest mass at non-relativistic energies.
A more sophisticated argument, based on a Machian approach, is discussed
in appendix \ref{Machian=000020approach=000020to=000020clock=000020time}.

The free FS/T becomes:
\begin{equation}
2\imath E\frac{\partial{\psi}_{\tau}}{\partial\tau}{=}{-}\left({p}^{\mathit{\mu}}{p}_{\mathit{\mu}}{-}{m}^{2}\right){\psi}_{\tau}
\end{equation}

where $m\to0$ for photons. We build up the solutions for this into
Feynman diagrams. With this approach, anything that can be described
with Feynman diagrams in SQM can be also be treated in TQM.

Note we are using an $S$ superscript, as ${\varphi}^{S}\left({\vec{x}}\right)$,
to mark objects as SQM. We use a $T$ superscript, as ${\varphi}^{T}\left({t}\right)$,
to mark objects as purely time or energy. We leave objects where time
and space, energy and momentum are entangled, as $\varphi\left({t,\vec{x}}\right),$
kept unmarked in keeping with our position that this is in fact the
normal case. To reduce clutter we sometimes leave these superscripts
off, when the distinction seems clear enough.

\paragraph{Precis of normal development of quantum electrodynamics}

\label{par:Normal-development-in-SQM}

We give an abbreviated treatment of SQM in a way that prepares for
the addition of coordinate time. The details are in the earlier work
\cite{Ashmead:2023aa}. In the literature, we have found particularly
helpful Klauber's very detailed approach \cite{Klauber:2013tg}, but
have also drawn on \cite{Feynman:1998wv,Bjorken:1965el,Bjorken:1965mo,Sakurai:1967tl,Ramond:1990ab,Kaku:1993xj,Greiner:1994gf,Peskin:1995rv,Weinberg:1995hu,Weinberg:1995rl,Huang:1998bd,Greiner:2000vl,Itzykson:2005cv,McMahon:2008gt,Zee:2010oy,Klauber:2013tg,Lancaster:qy,Schwartz:2014tx,Schwichtenberg:2020vg}. 

We start with the free solutions to the KG equation:

\begin{equation}
{\mathit{\psi}}_{\mathit{\tau}}^{\left({\vec{p}}\right)}\left({\vec{x}}\right){=}\frac{1}{\sqrt{2{VE}_{\vec{p}}}}\exp\mbox{\ensuremath{\left({{-}\imath{E}_{\mbox{\ensuremath{\vec{p}}}}\tau{+}\imath\mbox{\ensuremath{\vec{p}}}\cdot\mbox{\ensuremath{\vec{x}}}}\right)}}
\end{equation}

with:

\[
{E}_{\vec{p}}\equiv\sqrt{{m}^{2}{+}{\vec{p}}^{2}}
\]

In box normalization, the three momentum only takes on discrete values,
indexed by $i,j,k$. Box normalization is particularly helpful for
keeping track of the dimensions. The normalization factor $\frac{1}{\sqrt{2{E}_{\vec{p}}}}$
is customary in QED.

The field operators for SQM create and annihilate objects tracked
by these basis wave functions:

\begin{equation}
{\mathit{\psi}}_{\tau}^{S}\mbox{\ensuremath{\left({\mbox{\ensuremath{\vec{x}}}}\right)}}\equiv\mathop{\sum}\limits_{\vec{p}}{{a}_{\vec{p}}{\mathit{\psi}}_{\mathit{\tau}}^{\left({\vec{p}}\right)}\left({\vec{x}}\right){+}{a}_{\vec{p}}^{\dag}{\mathit{\psi}}_{\mathit{\tau}}^{\left({\vec{p}}\right){*}}\left({\vec{x}}\right)}\label{eq:sqn-field-operator}
\end{equation}

We use the occupation number representation for Fock space:

\begin{equation}
\left|{\left\{ {n_{\vec{p}}}\right\} }\right\rangle \label{eq:spinzero-fock-sqm}
\end{equation}
where $n$ is an integer from zero to infinity and the wave functions
are appropriately symmetrized. The creation and annihilation operators
are defined by their effects on Fock space:

\begin{equation}
{a_{\vec{p}}}\left|{n_{\vec{p}}}\right\rangle =\sqrt{{n_{\vec{p}}}}\left|{{n_{\vec{p}}}-1}\right\rangle ,a_{\vec{p}}^{\dag}\left|{n_{\vec{p}}}\right\rangle =\sqrt{{n_{\vec{p}}}+1}\left|{{n_{\vec{p}}}+1}\right\rangle 
\end{equation}
with the 3D commutators being:

\begin{equation}
\left[{{a_{\vec{p}}},a_{\vec{p}'}^{\dag}}\right]={\delta^{3}}\left({\vec{p}-\vec{p}'}\right)\label{eq:3D=000020commutators}
\end{equation}

We make no use of the usual interpretation of the $a$'s in terms
of harmonic oscillators.

The SQM Feynman propagator is defined as the time-ordered vacuum expectation
value of two of these field operators. This definition is key to evaluating
the $S$ matrix as a sum over Feynman diagrams: 

\begin{equation}
\imath\Delta_{{\tau_{x}}{\tau_{y}}}^{S}\left({\vec{x}-\vec{y}}\right)\equiv\left\langle {0\left|{T\left\{ {\phi_{{\tau_{x}}}^{S}\left({\vec{x}}\right),\phi_{{\tau_{y}}}^{S}\left({\vec{y}}\right)}\right\} }\right|0}\right\rangle 
\end{equation}

We expand the operators and use the commutators. To simplify the calculations,
we shift from discrete sums to continuous integrals by replacing $\Sigma\to\int,V\to{\left({2\pi}\right)^{3}}$. 

We use the vacuum product $\left\langle {0}\mathrel{\left|{\vphantom{00}}\right.\kern-\nulldelimiterspace}{0}\right\rangle =1$. 

Inside the ``vacuum sandwich'' most of the terms give zero and we
are left with contributions from terms that create a plane wave at
one point in time and propagate it to another time, giving sums over
factors of the form: 

\begin{equation}
\frac{\exp\mbox{ \ensuremath{\left({{-}\imath\omega_{\vec{p}}\Delta\tau}\right)}}}{{2}\omega_{\vec{p}}}\label{eq:sqm-wave-function}
\end{equation}

with:

\[
\omega_{\vec{p}}\equiv\sqrt{{m}^{2}{+}{\vec{p}}^{2}}
\]

A subtle but critical point about these is that they are on-shell,
$\omega_{\vec{p}}$ is the on-shell energy. In Klauber's treatment,
the sums are broken up into positive:

\begin{equation}
\imath\Delta_{xy}^{S+}\left({\vec{x}-\vec{y}}\right)=\frac{1}{{{\left({2\pi}\right)}^{3}}}\int{d\vec{p}\frac{{e^{-\imath{\omega_{\vec{p}}}\tau_{xy}+\imath\vec{p}\cdot\left({\vec{x}-\vec{y}}\right)}}}{{2{\omega_{\vec{p}}}}}}\label{eq:3D=000020Feynman=000020prop=000020as=000020pure=000020number}
\end{equation}
and negative frequency components:

\begin{equation}
\imath\Delta_{xy}^{S-}\left({\vec{x}-\vec{y}}\right)=\frac{1}{{{\left({2\pi}\right)}^{3}}}\oint{d\vec{p}\frac{{e^{\imath{\omega_{\vec{p}}}\tau_{xy}-\imath\vec{p}\cdot\left({\vec{x}-\vec{y}}\right)}}}{{2{\omega_{\vec{p}}}}}}\label{eq:3D=000020Feynman=000020negative=000020prop}
\end{equation}

Note these are on-shell 3D propagators. We will refer to this as the
``unpacked'' form. We do not see much of this form in the literature
however. 

\begin{figure}
\includegraphics[scale=0.67]{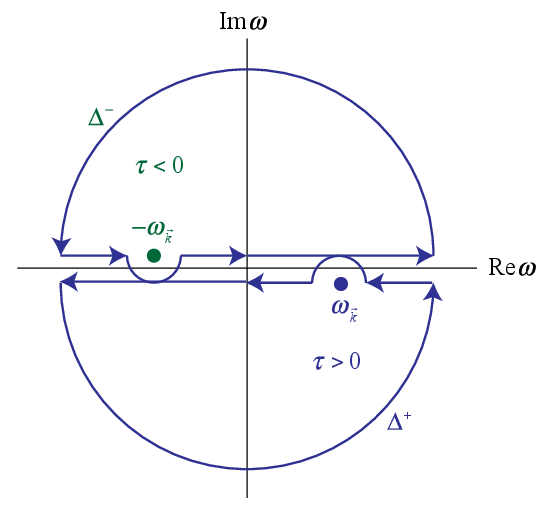}

\caption{\protect\label{fig:Feynman-boundary-conditions}Feynman boundary conditions
in clock time}
\end{figure}

The next step is to rewrite the  $\frac{1}{2{\omega}_{\vec{p}}}$
as a contour integral over $\omega$, with poles at $\pm\omega_{\vec{p}}$.
The integral over $\vec{p}$ becomes an integral over the four vector
$\omega,\vec{p}$, and the denominator is a relativistic invariant:

\begin{equation}
\imath\Delta_{\omega}^{S}\left({\vec{p}}\right)=\frac{\imath}{{{\omega^{2}}-{{\vec{p}}^{2}}-{m^{2}}+\imath\varepsilon}}\label{eq:spinzero-sqm-prop-p}
\end{equation}

This is the normal form for a propagator in field theory. To promote
to photons, fermions, and so on, we add polarization vectors and spinors,
but that is all just wrapping. Since the integral over the clock frequency
$\omega$ is a contour integral it is really only seeing the value
of $\omega$ at the on-shell poles. Particles described in this way
are called ``virtual'' particles. In a deep way, TQM is about promoting
these virtual particles to real ones, making the fourth integral,
now over the coordinate frequency $w$, a real not a contour integral.

\paragraph{Quantum electrodynamics with time dispersion}

To extend the SQM approach to time we make a series of straight-forward
changes. We require that free solutions obey the free FS/T (rather
than the KG equation), we promote the box normalization from cube
to hyper cube $V\to TV$, and we replace the on-shell energy in the
denominator with the coordinate energy $\frac{1}{\sqrt{2{E}_{\vec{p}}}}\rightarrow\frac{1}{\sqrt{2E}}.$

The free solutions are now:

\begin{equation}
{\mathit{\psi}}_{\mathit{\tau}}^{p}\mbox{\ensuremath{\left({x}\right)}}{=}\frac{1}{\sqrt{2TVE}}\exp\mbox{\ensuremath{\left({{-}\imath{\mathit{\varpi}}_{p}\mathit{\tau}{-}\imath{E}{t}{+}\imath\mbox{\ensuremath{\vec{p}}}\cdot\mbox{\ensuremath{\vec{x}}}}\right)}}\label{eq:wf-tqm}
\end{equation}

We define the clock frequency $\varpi_{p}$ as:

\begin{equation}
{\varpi}_{p}\equiv{-}\frac{{E}^{2}{-}{\vec{p}}^{2}{-}{m}^{2}}{2E}
\end{equation}

which gives the free solution of the FS/T in momentum space:

\begin{equation}
{2}{E}\imath\frac{\partial}{\partial\mathit{\tau}}{\mathit{\psi}}_{\mathit{\tau}}\left({E,\vec{p}}\right){=}\left({{-}{E}^{2}{+}{\vec{p}}^{2}{+}{m}^{2}}\right){\mathit{\psi}}_{\mathit{\tau}}\left({E,\vec{p}}\right)
\end{equation}

We define the field operator as sum over wave functions:

\begin{equation}
{\phi_{\tau}}\left({t,\vec{x}}\right)=\sum\limits_{E,\vec{p}}{\frac{1}{{\sqrt{TV}}}\frac{1}{{\sqrt{2{E}}}}\left({{a_{E,\vec{p}}}{\operatorname{e}^{-\imath{\varpi_{p}}\tau-\imath Et+\imath\vec{p}\cdot\vec{x}}}+a_{E,\vec{p}}^{\dag}{\operatorname{e}^{\imath{\varpi_{k}}\tau+\imath Et-\imath\vec{p}\cdot\vec{x}}}}\right)}\label{eq:tqm-field-operator}
\end{equation}

If we are using box normalization then the four momentum $E,\vec{p}$
is indexed by $h,i,j,k$ with $h$ the index in time. The field operator
is:

\begin{equation}
\mathit{\psi}\mbox{\ensuremath{\left({t,\mbox{\ensuremath{\vec{x}}}}\right)}}\equiv\mathop{\sum}\limits_{p}{{a}_{p}{\mathit{\psi}}_{\mathit{\tau}}^{p}\left({x}\right){+}{a}_{p}^{\dag}{\mathit{\psi}}_{\mathit{\tau}}^{p*}\left({x}\right)}\label{eq:tqm-operator}
\end{equation}

It is given in the occupation representation by:

\begin{equation}
\left|{\left\{ {n_{p}}\right\} }\right\rangle \label{eq:spinzero-fock-tqm}
\end{equation}
where $n$ is an integer from zero to infinity and the wave functions
are appropriately symmetrized. 

The creation and annihilation operators are defined by their effects
in Fock space:

\begin{equation}
{a_{p}}\left|{n_{p}}\right\rangle =\sqrt{{n_{p}}}\left|{{n_{p}}-1}\right\rangle ,a_{p}^{\dag}\left|{n_{\vec{p}}}\right\rangle =\sqrt{{n_{p}}+1}\left|{{n_{p}}+1}\right\rangle 
\end{equation}
with 4D commutators:

\begin{equation}
\left[{{a_{p}},a_{p'}^{\dag}}\right]={\delta^{4}}\left({p-p'}\right)\label{eq:4D=000020commutator}
\end{equation}
All other commutators are zero. Again, we make no use of the usual
interpretation of the $a$'s in terms of harmonic oscillators.

The TQM propagator is defined as:

\begin{equation}
\imath{\Delta_{xy}}\left({x-y}\right)=\left\langle 0\left|T\left\{ {{\phi_{x}}\left(x\right),{\phi_{y}}\left(y\right)}\right\} \right|0\right\rangle 
\end{equation}

As before, inside the ``vacuum sandwich'' most of the terms give
zero and we are left with contributions from terms that create a plane
wave at one point in clock time and propagate it to another point
in clock time, giving sums over factors of the form: 

\begin{equation}
\frac{\exp\left({{-}\imath{\varpi}_{p}\Delta\mathit{\tau}}\right)}{2E}\label{eq:unpacked-propagator-tqm}
\end{equation}

which we write in turn as the integral:

\begin{equation}
\imath\Delta_{xy}^{+}\left({x-y}\right)=\frac{1}{{{\left({2\pi}\right)}^{4}}}\int{{d^{4}}p\frac{{e^{-\imath{\varpi_{p}}\tau_{xy}-\imath p\left({x-y}\right)}}}{2E}}
\end{equation}

As with SQM, for $\tau_{x}<\tau_{y}$ we can get the results for the
propagator by interchanging $x\leftrightarrow y$:

\begin{equation}
\imath\Delta_{xy}^{-}\left({x-y}\right)=-\frac{1}{{{\left({2\pi}\right)}^{4}}}\int{{d^{4}}p\frac{{e^{-\imath{\varpi_{p}}\tau_{xy}-\imath p\left({x-y}\right)}}}{2E}}
\end{equation}

We can write the propagator in terms of $\omega$: 

\begin{equation}
\exp\left({-\imath{\varpi_{p}}\tau}\right)\theta\left(\tau\right)-\exp\left({-\imath{\varpi_{p}}\tau}\right)\theta\left({-\tau}\right)=\frac{\imath}{{2\pi}}\int\limits_{-\infty}^{\infty}{d\omega\exp\left({-\imath\omega\tau}\right)}{\Delta_{\omega}}\left(p\right)\label{eq:key-prop}
\end{equation}
with:

\begin{equation}
\imath{\Delta_{\omega}}\left(p\right)=\frac{1}{2E}\left({\frac{\imath}{{\omega-{\varpi_{p}}+\imath\varepsilon}}+\frac{\imath}{{\omega-{\varpi_{p}}-\imath\varepsilon}}}\right)\label{eq:spinzero-tqm-packed}
\end{equation}

A contour integral over the clock frequency will recover the previous
dependence on clock time in equation \ref{eq:unpacked-propagator-tqm}.

\paragraph{Interactions}

For QED the interactions have the form:

\begin{equation}
q\bar{\psi}\left({p'}\right)A^{\nu}\left(k\right){\gamma_{\nu}}\psi\left({p}\right)\label{eq:tqm-vertex}
\end{equation}

These are formally identical in TQM and SQM as they have no explicit
dependence on clock time.

\paragraph{Comparison of SQM and TQM}

\subparagraph{Initial and final wave functions}

In TQM the initial wave functions have, by construction, dispersion
in time. In SQM they do not. The situation at the detector for TQM
is not without its subtleties: we discussed it at length in the earlier
work,  and have a short summary in the discussion of detection in
subsection \ref{subsec:Time-of-arrival}.

\subparagraph{Propagators}

The propagator in TQM is basically the propagator in SQM but with
``temporal fuzz'', uncertainty in time. In the non-relativistic
case or in scattering problems with wave packets well-localized in
momentum space the differences are not that great in practice: the
coordinate time functions largely as an extra spatial dimension. In
the bound case, as we will see, the situation is more complex.

\subparagraph{Energy conservation at vertexes}

At each vertex, the integral over the paths in the $x,y,z$ directions
create a $\delta^{3}$ function in three momentum. Taking the limit
as the clock time goes to $\pm\infty$ creates the fourth $\delta$
function, in energy. As a result, SQM does not compel conservation
of energy except in the long time limit, i.e. scattering. There is
nothing to keep one from looking at short times, but at short times
we do not have manifest energy conservation. Taking the long (clock
time) limit is the almost invariable practice, but it means that the
short time coverage of SQM is not all it might be.

This is critical here because it is precisely at short times that
the strongest effects of $t$ as operator will appear.

In TQM we have four coordinate space integrals at each vertex, giving
$\delta$ functions in all four components of the four momentum. We
can take the limit as clock time goes to infinity; this will give
conservation of clock frequency as well. But we do not need to do
this. And in fact not letting the clock time go to infinity lets us
more easily model the behavior of the wave function on its way to
the detector.

\subparagraph{Ultraviolet (UV) divergences in loop diagrams}

\label{par:Ultraviolet-(UV)-divergences}

While the loop diagrams in TQM do have one more variable to integrate
over, they are still convergent. The loop integral is entangled with
the wave function of the initial wave function(s). The time/energy
part of those initial wave functions (which are by assumption normalized)
is thereby mixed into the loop integral, acting as a de facto regularizing
factor. See analysis in \cite{Ashmead:2023aa}. Note the result of
the loop calculation is not something you then ``renormalize away'';
it is a number, dependent on the original wave functions, which may
be measurable.

\paragraph{Recipe for Feynman diagrams with time dispersion}

With these changes, the usual recipe for Feynman diagrams is unchanged.
All SQM diagrams are TQM diagrams and vice versa. The symmetries are
the same. The loop integrals are the same, but with one more variable
of integration. You might expect this would make them more infinite;
in fact the entanglement of the loop integrals with the initial wave
functions makes them self-regularizing. 

\subsection{Meaning of clock time}

\label{subsec:Meaning-of-clock}

\paragraph{Choice of specific laboratory frame }

If we develop TQM in different laboratories, the laboratories will
in general be using slightly different reference frames. 

This is unlikely to have much practical effect on experiments aimed
at falsifying TQM. There is not much difference between even laboratories
in space and on planet. The special relativistic corrections are from
the orbital speed, of order $v=4km/s$. Given the speed of light is
$300,000km/sec$, the time dilation factor $\mathit{\gamma}\equiv\frac{1}{\sqrt{{1}{-}\frac{{v}^{2}}{{c}^{2}}}}\sim1+{10}^{{-}{10}}\approx1$.
The general relativistic corrections are of the same order of magnitude.

Even if this is not likely to be of practical concern, we would still
like to phrase TQM in a manifestly covariant way. We can do this by
first writing:

\begin{equation}
{2}{E}^{PARTICLE}\imath\frac{\partial}{\partial\mathit{\tau}}\rightarrow{2}{E}^{PARTICLE}{\mathcal{E}}^{LAB}
\end{equation}

and then noting that in the rest frame of the laboratory this is the
product of the momentum four vectors for the laboratory and the particle:

\begin{equation}
{2}{E}^{PARTICLE}{\mathcal{E}}^{LAB}\rightarrow{2}{p}^{PARTICLE}\cdot{\mathcal{P}}^{LAB}
\end{equation}

Making this replacement we have a manifestly covariant form for the
FS/T on the left and on the right:

\begin{equation}
{2}{p}{\mathcal{P}}^{LAB}{\mathit{\psi}}^{PARTICLE}{=}\left({{p}^{2}{-}{m}^{2}}\right){\mathit{\psi}}^{PARTICLE}\label{eq:invariant-fst}
\end{equation}

This is a variation on the Machian argument made in appendix \ref{Machian=000020approach=000020to=000020clock=000020time}.

\paragraph{Two different perspectives on a single time}

We were able to develop the FPI approach to TQM by splitting time
up into two types of time. This is a variation on the traditional
``divide and conquer'' strategy. But (as all good strategists know)
``divide and conquer'' needs to be followed by ``unite and triumph''.
How to combine coordinate time and clock time? 

Consider the laboratory. It has a wave function, $\Psi$. Define its
clock time as the expectation of $\Psi$'s coordinate time operator:

\begin{equation}
\mathit{\tau}\equiv\left\langle {\Psi\left|{t}\right|\left.{\Psi}\right\rangle }\right.\label{eq:clocktime-as-expectation}
\end{equation}

Now we have only one kind of time. To be sure, we are using a different
(and less detailed) set of approximations for the wave function of
the laboratory than for the particle wave functions. We discuss this
question further in the subsection on measurement \ref{subsec:Measurement}.

\section{Bound states}

\label{sec:Bound-states}

``The problem of creating something which is new, but which is consistent
with everything which has been seen before, is one of extreme difficulty.''
-- Richard P. Feynman\emph{, The Feynman Lectures in Physics} (v
II p 20-10 of \cite{Feynman:1965ah})

\subsection{Overview}

\label{subsec:Overview}

This is the core section of this work. We will compute to first approximation
the ground state wave function of a hydrogen atom, using the normal
SQM space part as the starting point, then asking what the corresponding
time part should look like.

The calculations are non-trivial. To keep focus on the core problem
we will make the following simplifications:
\begin{enumerate}
\item Consider only the non-relativistic (NR) case.
\item Consider only the hydrogen atom and only its ground state.
\item Look only at the effect of the proton on the electron, but not the
reverse action of the electron on the proton. We will not go to the
center-of-mass frame.
\item And we will ignore all complications of spin and polarization: we
will replace the photon, electron, and proton with three spin zero
particles $A,B,C$ corresponding to each. We will refer to the $A,B,C$
particles as photon, electron, and proton however. We will tag various
quantities with superscripts $A,B,C$ when this is useful to distinguish
them, for instance we have the respective dispersions in time ${\mathit{\sigma}}_{t}^{\left({A}\right){2}},{\mathit{\sigma}}_{t}^{\left({B}\right){2}},{\mathit{\sigma}}_{t}^{\left({C}\right){2}}$.
\end{enumerate}
While we are focusing on the simplest possible case, it should be
clear enough from the development here how to extend this treatment
to cover the general case.

This examination is an essential step in the development of TQM. In
TQM we have to specify an initial wave function. The only requirement
is that the initial wave function be normalizable. But if we have
all this flexibility about the input, we lose specificity in the outputs.
We still have testability, but we lose falsifiability.

But if we start with a specific bound state, then only certain time
dispersions will fit. Technical complications will keep us from achieving
as much precision as we might like, but we will have no trouble getting
to an order-of-magnitude (OOM) estimate, all we need for falsifiability.

In the NR case we had a straight-forward line of attack: promote paths
from 3D to 4D, leave everything else the same, turn the crank. In
QED, we also had a straight-forward line of attack: promote the NR
propagator to a relativistic propagator by promoting the rest mass
to the relativistic energy, leave everything else the same, turn the
crank. 

Matters are not quite so simple here. We are looking for a fixed point
of the equations; factors that are insignificant for single point
interactions may become decisive when the interactions are repeated
for infinite times.

We will employ a pincer attack. We will move ``up'' from the FS/T
equation and ``down'' from the associated exchange diagram in QED
to trap our bound state in the middle.

Our ultimate goal is to compute first the effective potential in time
and then the corresponding dispersion in time.

We will employ a series of approximations. We will start by estimating
the size of the bound state wave function using the ``entropic estimate'',
use this to calculate the time part of the potential energy, then
use that to revise our estimate of the time part. We will finish by
showing the result takes us to falsifiability.

We begin by looking at two \emph{a priori} estimates of the time part.
Neither is satisfactory (they are merely \emph{a priori} estimates)
but they will help to frame the subsequent calculations.

\subsection{Initial estimates of time dispersion}

\label{subsec:Initial-estimate-of-WF}

\paragraph{Estimate from symmetry between time and space}

\label{par:Estimate-from-symmetry}

The simplest estimate of the dispersion in time is to take advantage
of our assumed symmetry between time and space to use essentially
the same measure for each. The characteristic radius of a hydrogen
atom is $a_{0}$, the Bohr radius, $52.9pm$ (picoseconds). The equivalent
in time is the time for a photon to travel that distance, about $.177as.$
We can think of this as the Bohr time, the basic unit of time in this
analysis. In terms of the fine structure constant $\alpha$ and the
mass of the electron it is $\frac{1}{\alpha m_{e}}$.

From a falsifiability point of view this is a bit on the small side.
As noted earlier, the shortest time measurement we are aware of \cite{Ossiander:2016fp}
is 1/25th of the atomic unit of time $(24.2as)$ or about $.968as$,
six times larger than the Bohr time.

Cesium, with the largest radii of all the elements whose radii have
been measured, has a radius about $265pm$ or five times that of the
hydrogen atom. Treating time and space as full symmetrical takes us
up to about $.887as$, so (barely) within reach.

We will refer to this as the estimate from symmetry or the naive estimate:
assume time and space are equal, see what results. The final result
will be similar to this but will supply a mechanism.

\paragraph{Estimate from uncertainty in energy}

\label{par:Estimate-from-uncertainty}

We showed in earlier work that if we can estimate the uncertainty
in energy for a wave function we can estimate its uncertainty in time.
We estimate the wave function by using the Lagrange multiplier method
to pick out the wave function with maximum entropy subject to known
constraints. The constraints are total energy and the uncertainty
in energy. The first is known; the second can be estimated from the
known uncertainty in momentum. Details in appendix \ref{sec:Entropic-estimate}.

For the ground state we get (eqn \ref{eq:ground-state-energy-uncertainty}):

\begin{equation}
\Delta{E}={\mathit{\alpha}}^{2}{m}_{e}={27}{.}{2}{eV}
\end{equation}

From the posited Heisenberg uncertainty principle in time/energy $\Delta{E}\Delta{t}\sim\frac{1}{2}$
and $\hbar{=}658.2{eVas}$ we have: 

\begin{equation}
\ensuremath{\Delta{t}=\frac{658.2eVas}{{2}\cdot{27}{.}{2}eV}={12}{.}{1}{as}}\label{eq:2nd-guess-time-uncertainty}
\end{equation}

This is within reach of current technology, although not by a large
margin. 

We will refer to this as the entropic estimate. It will turn out to
be too optimistic for the hydrogen atom. But then the entropic estimate
was developed for scattering problems which have a different character.

\paragraph{Estimate of the wave functions}

In either case once we have an estimate of the uncertainty in time/energy
we can estimate the associated wave functions as, in energy:

\begin{equation}
{\psi^{T}}_{0}\mbox{ {\ensuremath{\left({E}\right)}}}{=}\sqrt[4]{\frac{1}{{\pi\sigma}_{E}^{2}}}{e}^{{-}\frac{{\mbox{ {\ensuremath{\left({{E}{-}\mbox{ {\ensuremath{\bar{E}}}}}\right)}}}}^{2}}{2{\sigma}_{E}^{2}}}{,}\hspace{0.33em}{\sigma}_{E}^{2}=2{\left({\Delta{E}}\right)^{2}}\label{eq:generic-wf-energy}
\end{equation}

And in time:

\begin{equation}
{\psi^{T}}_{0}\mbox{ \ensuremath{\left({t}\right)}}\equiv\sqrt[4]{\frac{1}{{\pi\sigma}_{t}^{2}}}{e}^{-\frac{{\mbox{ \ensuremath{\left({{t}{-}{t}_{0}}\right)}}}^{2}}{2{\sigma}_{t}^{2}}}{,}{\sigma}_{t}^{2}=2{\left({\Delta{t}}\right)}^{2}\label{eq:generic-wf-time}
\end{equation}

And we can therefore construct a zeroth-order estimate of the atomic
wave functions as:

\begin{equation}
{\psi}_{0}\mbox{ \ensuremath{\left({E,\mbox{ \ensuremath{\vec{p}}}}\right)}}\approx{\varphi}_{0}^{T}\mbox{ \ensuremath{\left({E}\right)}}{\psi}_{0}^{S}\mbox{ \ensuremath{\left({\mbox{ \ensuremath{\vec{p}}}}\right)}}
\end{equation}

Note with this parameterization, the Heisenberg uncertainty principle
is wired into the definition of the wave functions in time and energy
by the correspondence:

\[
{\mathit{\sigma}}_{E}^{2}{=}\frac{1}{{\mathit{\sigma}}_{t}^{2}}
\]

The main problem with these estimates is not the specific sizes, but
rather than they are \emph{a priori} estimates. To provide a suitable
target for experimental effort, we need to derive the wave function
in time; we need to supply dynamics.

\subsection{Two particle equation}

\label{subsec:Center-of-mass-frame}

\begin{figure}
\includegraphics[scale=0.5]{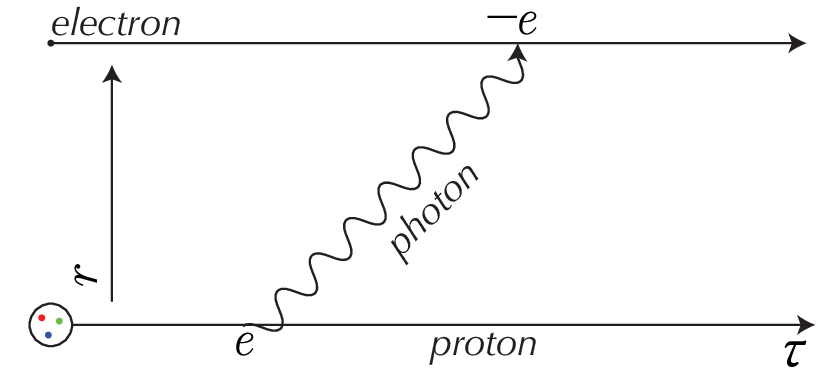}

\caption{\protect\label{fig:Effective-potential}Central potential in Schrödinger
equation}
\end{figure}

\paragraph{Standard quantum mechanics}

The usual starting point for an analysis of the atomic wave function
is Schrödinger's equation:

\begin{equation}
{\imath\frac{\partial}{\partial\tau}{\psi}_{\mathit{\tau}}^{\mathrm{S}}{=}\left({\frac{{\vec{p}}^{2}}{2m}{-}\frac{{e^{2}}}{4\pi r}}\right){\psi}_{\mathit{\tau}}^{\mathrm{S}}}\label{eq:seqn-sqm-1}
\end{equation}

The relevant diagram in QED is the exchange diagram:

\begin{equation}
\begin{array}{c}
{\imath\frac{\partial}{\partial\tau}{\bar{\varphi}}_{\tau}^{B}\mbox{\ensuremath{\left({{\mbox{\ensuremath{\vec{x}}}}_{1}}\right)}}{=}\frac{{\vec{p}}_{A}^{2}}{2{m}_{A}}{\bar{\varphi}}_{\tau}^{B}\mbox{\ensuremath{\left({{\mbox{\ensuremath{\vec{x}}}}_{1}}\right)}}{-}{e}^{2}\mathop{\int}\limits_{{-}\infty}^{\mathit{\tau}}{{d}\mathit{\tau}{'}\int{{d}{\vec{x}}_{2}{G}_{\mathit{\tau}\mathit{\tau}{'}}^{\mathit{S}}\left({{\vec{x}}_{1}{-}{\vec{x}}_{2}}\right){\bar{\varphi}}_{\mathit{\tau}}^{C}\left({{\vec{x}}_{2}}\right)}}}\\
{\imath\frac{\partial}{\partial\tau}{\bar{\varphi}}_{\tau}^{C}\mbox{\ensuremath{\left({{\mbox{\ensuremath{\vec{x}}}}_{2}}\right)}}{=}\frac{{\vec{p}}_{B}^{2}}{2{m}_{B}}{\bar{\varphi}}_{\tau}^{C}\mbox{\ensuremath{\left({{\mbox{\ensuremath{\vec{x}}}}_{2}}\right)}}{-}{e}^{2}\mathop{\int}\limits_{{-}\infty}^{\mathit{\tau}}{{d}\mathit{\tau}{'}\int{{d\vec{x}}_{1}{G}_{\mathit{\tau}\mathit{\tau}{'}}^{\mathit{S}}\left({{\vec{x}}_{2}{-}{\vec{x}}_{1}}\right){\bar{\varphi}}_{\mathit{\tau}}^{B}\left({{\vec{x}}_{1}}\right)}}}
\end{array}\label{eq:sqm-two-particle}
\end{equation}

We note several obvious differences between the Schrödinger equation
picture and the exchange picture. In the Schrödinger equation we have:
\begin{enumerate}
\item \emph{A single particle equation. }We can get to this from the exchange
picture by shifting to the center of mass (COM) frame. Here we will
assume that we can ignore the back-reaction of the electron on the
proton, so we will drop the second of the twinned equations. To extend
the approach here to atoms beyond hydrogen we will need to shift to
the COM frame. In practice we should be able to handle this by replacing
the electron mass with the reduced electron mass.
\item \emph{A potential rather than a series of photons. }The sum over the
photons can be taken as an effective potential.
\item \emph{An instantaneous potential, not a retarded. }We can get from
the retarded potential to the instantaneous Coulomb potential by a
gauge transform (see \cite{Jackson:2002ab}). Here we prefer to stay
with the retarded potential.
\item \emph{The use of the probability density of the other particle as
the source rather than the wave function.} This is completely natural
when starting with a solar system like model for the atom (as Bohr
and to a lesser extend Schrödinger did). However it makes less sense
in the QED picture.
\end{enumerate}

\paragraph{Quantum mechanics with dispersion in time}

The exchange approach translates directly to TQM:

\begin{equation}
\begin{array}{c}
{\imath\frac{\partial}{\partial\tau}{\varphi}_{\tau}^{B}\mbox{\ensuremath{\left({{x}_{1}}\right)}}{=}{-}\frac{{p}_{A}^{2}{-}{m}_{A}^{2}}{2{m}_{A}}{\varphi}_{\mathit{\tau}}^{B}\mbox{\ensuremath{\left({{x}_{1}}\right)}}{-}{e}^{2}\mathop{\int}\limits_{{-}\infty}^{\infty}{{d}\mathit{\tau}{'}\int{{dx}_{2}{G}_{\mathit{\tau}\mathit{\tau}{'}}^{\mathit{}}\left({{x}_{1}{-}{x}_{2}}\right){\varphi}_{\mathit{\tau}}^{C}\left({{x}_{2}}\right)}}}\\
{\imath\frac{\partial}{\partial\tau}{\varphi}_{\tau}^{C}\mbox{\ensuremath{\left({{x}_{2}}\right)}}{=}{-}\frac{{p}_{B}^{2}{-}{m}_{B}^{2}}{2{m}_{B}}{\varphi}_{\mathit{\tau}}^{C}\mbox{\ensuremath{\left({{x}_{2}}\right)}}{-}{e}^{2}\mathop{\int}\limits_{{-}\infty}^{\infty}{{d}\mathit{\tau}{'}\int{{dx}_{1}{G}_{\mathit{\tau}\mathit{\tau}{'}}^{\mathit{}}\left({{x}_{2}{-}{x}_{1}}\right){\varphi}_{\mathit{\tau}}^{B}\left({{x}_{1}}\right)}}}
\end{array}\label{eq:tqm-two-particle}
\end{equation}

To go from SQM to TQM we need to:
\begin{enumerate}
\item \emph{Include coordinate time. }The defining characteristic of TQM.
\item \emph{Use the free FS/T rather than the free Schrödinger equation.}
\item \emph{Take a different approach to the use of that Hamiltonian.} In
SQM the Hamiltonian defines the energy of the wave function. In TQM
the time and space parts balance each other out, with the total Hamiltonian
averaging out to approximately zero over sufficient clock time.
\item \emph{Set the limits in the integral on clock time to $\pm\infty$}.
Bear in mind that paths can go backwards: to ensure we capture all
paths we have to look at the possibility that we are starting in the
future. In practice, this is expected to play a minor role.
\end{enumerate}

\subsection{Interaction terms}

\label{subsec:Exchange-diagram}

\begin{figure}
\includegraphics[scale=0.5]{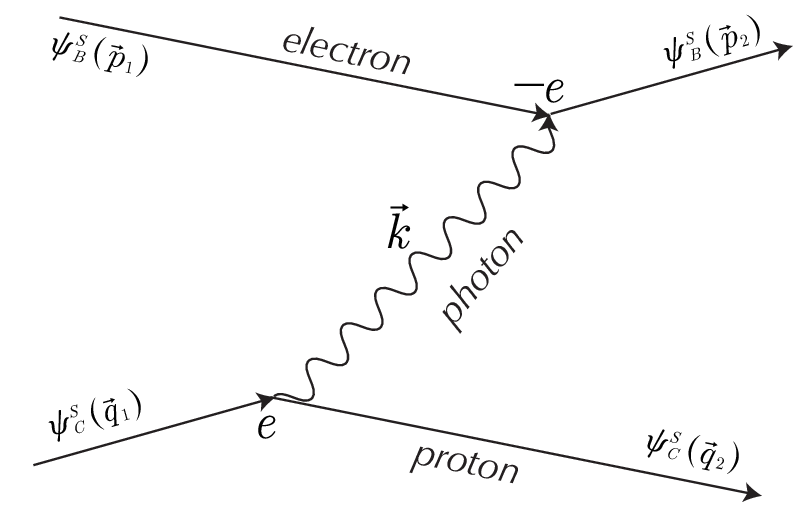}

\caption{\protect\label{fig:The-sumo-wrestler}Exchange diagram in QED}
\end{figure}

\paragraph{Emission term}

The initial interaction term -- a $C$ emits an $A$ -- is:

\begin{equation}
{e}{\mathit{\Psi}}_{C}{}^{\dag}{\varphi^{\dag}}{\mathit{\Psi}}_{C}\label{eq:cac}
\end{equation}

Per the restatement, each such term is composed of several elements:

\emph{1. The coupling constant $e.$} This is the same in SQM and
TQM.

\emph{2. Products of three (SQM) or four (TQM) plane waves, one such
product for each particle.} In the construction of Feynman diagrams,
the integrals over these plane waves give corresponding delta functions
in momenta. In SQM the plane waves are in three-space so only give
conservation of three momenta at each vertex. The normal fourth delta
function, in energy, comes from taking the limit as clock time from
$-\infty\to\infty$. This means we either have to sacrifice conservation
of energy at vertexes or restrict (for correctness) our investigations
to problems where taking the limits to $\pm\infty$ is acceptable.
In practice, the problem is ignored, without apparent harmful effects
to this point.

In TQM the plane waves are four dimensional, so that all four components
are conserved at each vertex:

\begin{equation}
{\mathit{\delta}}^{4}\left({{q}_{2}{+}{k}{-}{q}_{1}}\right)\label{eq:cac-delta-function}
\end{equation}

There is no need to take the limit as clock time goes to $\pm\infty$;
In fact TQM is most interesting when we look at short clock times;
when we do \emph{not} take that limit.

\emph{3. Spin and polarization operators.} As noted, they are not
essential for our purposes here.

\emph{4. Products of three raising and lowering operators.} The only
relevant term is the one that gives emission of a photon:

\begin{equation}
{c}_{2}{}^{\dag}{a}_{k}^{\dag}{c}_{1}
\end{equation}

Here we annihilate a $c$, create an $a$, and create a $c$. 

\paragraph{Absorption term}

The absorption at the electron side is the reciprocal of the source.
The interaction term looks like:

\begin{equation}
{-}{e}{\mathit{\psi}}_{B}{}^{\dag}{\varphi}{\mathit{\psi}}_{B}\label{eq:bab-interaction}
\end{equation}

The associated plane waves contribute, when integrated, an appropriate
delta function in momentum:

\begin{equation}
{\mathit{\delta}}^{4}\left({{p}_{2}{-}{k}{-}{p}_{1}}\right)\label{eq:bab-delta-function}
\end{equation}

At the core of the interaction are the three Fock space operators.
Here the only relevant one is the one that represents the absorption
of a photon:

\begin{equation}
{b_{2}}^{\dag}{a}_{k}{b_{1}}\label{eq:bab-operator}
\end{equation}

We annihilate a $b$ and an $a$ and create a $b$. 

\paragraph{Normalization}

If we look at the normalization for the $B$ and $C$ factors we see
-- in the NR limit -- a pair of square roots of the mass. These
are part of the normalization of the proton and electron wave function.
They are constant so are not a concern here:

\begin{equation}
\frac{1}{\sqrt{2{m}_{B,C}}}\frac{1}{\sqrt{2{m}_{B,C}}}\rightarrow{1}\label{eq:mass-normalization-is-eaten}
\end{equation}

There is no non-relativistic limit for photons. However for the photon,
our $A$ particle, we normalize to the frequency. The emission and
absorption will each contribute a power of one over the square root
of the energy. This is part of the conventional definition of the
QED propagator, so we will absorb these into that:

\begin{equation}
\frac{1}{\sqrt{2{w}_{A}}}\frac{1}{\sqrt{2{w}_{A}}}{G}_{\mathit{\tau}}^{NR}\rightarrow{G}_{\mathit{\tau}}^{QED}
\end{equation}

We have thereby rationalized the normalization factors between the
NR and the QED pictures.

\paragraph{Use of probability density at the source}

In the usual formulation of the Schrödinger equation the source of
the photon is not the initial wave function of the proton but its
probability density.

When we integrate over all possible initial $c$'s and final $c$'s,
we get a function of the form:

\begin{equation}
\int{{dq}_{2}{dq}_{1}{\mathit{\psi}}_{C}^{*}\left({{q}_{2}}\right){\mathit{\psi}}_{C}\left({{q}_{1}}\right)}\label{eq:c2c-probability-density}
\end{equation}

We will assume that over a bit of time, this averages out to the probability
density for $C$:

\begin{equation}
\int{{dq}{\mathit{\psi}}_{C}^{*}\left({q}\right){\mathit{\psi}}_{C}\left({{q}}\right)}{=}\int{{dq}{\mathit{\rho}}_{C}\left({{q}_{}}\right)}
\end{equation}

Effectively we are assuming that the exchanged photon has very little
momentum compared to that of the nucleus, i.e. that $\mathit{\delta}\left({{q}_{2}+{k}-{q}_{1}}\right)\approx\mathit{\delta}\left({{q}_{2}-{q}_{1}}\right)$.
Effectively the delta function ``eats'' the second q integral, reducing
the integral to an integral over the probability density. We will
see below that there is good reason to suppose ${k}\ll{q}_{1}{,}{q}_{2}$
since $k$ is of order KeV while the $q$'s are of order MeV.

The expression for the retarded potential is then:

\begin{equation}
{V}_{\mathit{\tau}}^{S}\mbox{ \ensuremath{\left({\mbox{{ \ensuremath{\vec{x}}}}}\right)}}{=}{-}{e}^{2}\mathop{\int}\limits_{{-}\infty}^{\infty}{{d}\tau{'}{d}\vec{x}{'}{G}_{\tau\tau'}^{S}\mbox{ \ensuremath{\left({\vec{x};\vec{x}'}\right)}}{\rho}_{\mathit{\tau}}^{\mathrm{S}}\mbox{ \ensuremath{\left({\mbox{ \ensuremath{\vec{x}}}'}\right)}}}\label{eq:set-clock-time-to-zero}
\end{equation}

The kernel $G^{S}$ here is the kernel for the photon carrying the
force. The TQM expression is completely parallel, except for the inclusion
of the coordinate time integral on the right:

\begin{equation}
{V}_{\tau}\mbox{ \ensuremath{\left({t,\mbox{ \ensuremath{\vec{x}}}}\right)}}{=}{-}{e}^{2}\mathop{\int}\limits_{{-}\infty}^{\infty}{{d}\tau{'}{dtd}\vec{x}{'}{G}_{\tau\tau{'}}\mbox{ \ensuremath{\left({t,\mbox{ \ensuremath{\vec{x}}};t',\mbox{ \ensuremath{\vec{x}}}'}\right)}}{\rho}_{\tau}\mbox{ \ensuremath{\left({t',\mbox{ \ensuremath{\vec{x}}}'}\right)}}}
\end{equation}

Again we have rationalized the difference between the NR and the QED
pictures, at the cost of the assumption that recoil from the emitted
photons is not drastically distorting the bound state wave functions.

\subsection{Initial photon wave function}

\label{subsec:Initial-photon-wave}

As noted, our ultimate goal is to compute first the effective potential
in time and then the corresponding dispersion in time.

In TQM we cannot in general compute the potential independently of
the initial wave function.

In a general exchange diagram, the scattered particles could go everywhere.
But if we are looking at a specific bound state, the particles have
to stay within the bound state. This implies significant limits on
the exchanged photons. If the proton gives off a photon of four momentum
$k$, then we know the proton went from four momentum $q\to q'$ with
$q'+k=q$.

We focus on the emitted photon, which has to have a distribution of
the form:

\begin{equation}
{\varphi}_{A}\left({k}\right)\sim\int{{dq}{'}{dq}\left\langle {{\mathit{\psi}}_{C}\left({q'}\right)\left|{\mathit{\delta}\left({{q}{'}{+}{k}{-}{q}}\right)}\right|{\mathit{\psi}}_{C}\left({q}\right)}\right\rangle }\label{eq:initial-photon-wf}
\end{equation}

The initial and final proton wave functions are the same, so if the
photon is emitted from a $q'q$ pair in one case it must also be emitted
from a reciprocal $q'q$ pair in another case. We have a kind of detailed
balance principle, imposed by the requirement that we are looking
at a well-defined bound state.

Now to get an exact result we need to have a complete wave function
for the proton. This is a complicated bag of three quarks whose full
description is far beyond the scope of this paper. But for our purposes
we can estimate it as a GTF with average radius in space/momentum
given by the radius of the proton. And the radius in time/energy given
by the entropic estimate. We take the radius of the proton as $\text{.841fm}$
(femtometers) \cite{NIST-2025-proton-radius} or $28.0ys$ (yoctoseconds).
Taking the radius as giving the uncertainty in position, we use the
uncertainty principle to estimate the uncertainty in momentum as $117.5MeV$. 

This gives us an estimate for the dispersion in momentum:

\begin{equation}
{\mathit{\sigma}}_{\vec{q}}^{\left({C}\right){2}}{=}\frac{1}{{\mathit{\sigma}}_{\vec{r}}^{\left({C}\right){2}}}
\end{equation}

In energy:

\begin{equation}
{\mathit{\sigma}}_{E}^{\left({C}\right){2}}{=}{\mathit{\sigma}}_{\vec{q}}^{\left({C}\right){2}}
\end{equation}

And therefore in time:

\begin{equation}
{\sigma}_{t}^{\mbox{ \ensuremath{\left({C}\right)}}{2}}{=}\frac{1}{{\sigma}_{E}^{\mbox{ \ensuremath{\left({C}\right)}}{2}}}{=}{\sigma}_{\vec{r}}^{\mbox{ \ensuremath{\left({C}\right)}}{2}}\approx{28}{ys}
\end{equation}

The integral (eqn \ref{eq:initial-photon-wf}) factors into energy
and space parts. 

\paragraph{Energy integral}

We can do the integral in energy directly. We are taking $\tau=0$
for simplicity:

\begin{equation}
{\varphi}_{0}^{\left({A}\right)}\mbox{ \ensuremath{\left({w}\right)}}{=}\sqrt{\frac{1}{2{\mathit{\sigma}}_{E}^{\hspace{0.33em}\left({C}\right){2}}}}\int{{d}{E}{'}{d}{E}\exp\left({{-}\frac{{\left({{E}{'}{-}{E}_{0}}\right)}^{2}}{2{\mathit{\sigma}}_{E}^{\hspace{0.33em}\left({C}\right){2}}}}\right)\exp\left({{-}\frac{{\left({{E}{-}{E}_{0}}\right)}^{2}}{2{\mathit{\sigma}}_{E}^{\hspace{0.33em}\left({C}\right){2}}}}\right)\delta\mbox{ \ensuremath{\left({{E}{'}{+}{w}{-}{E}}\right)}}}\label{eq:=000020photon-initial-wf-energy}
\end{equation}

which gives:

\begin{equation}
{\varphi}_{0}^{\left({A}\right)}\mbox{ \ensuremath{\left({w}\right)}}{=}\exp\left({{-}\frac{{w}^{2}}{4{\mathit{\sigma}}_{E}^{\hspace{0.33em}\left({C}\right){2}}}}\right)\label{eq:photon-initial-wf-2}
\end{equation}

The key point here is the large size of the dispersion in photon frequency
-- with correspondingly reduced dispersion in time. The wave function
is not exactly a plane wave but a very broad GTF, a kind of gently
curved plane wave like the surface of a lens on a major telescope.
As a result, including it in the integral ensures the energy integral
is finite but does not give much useful information about the time
dispersion (except that it is not a mathematical point, rather more
of a small but not infinitesimal dot).

\paragraph{Momentum integral}

The equivalent calculation for the three momentum part is significantly
more complex because the magnitude of the photon momentum includes
an allowance for the angular separation of $\vec{q},\vec{q}'$:

\begin{equation}
\left|{\vec{k}}\right|{=}\sqrt{{\left|{\vec{q}}\right|}^{2}{+}{\left|{\vec{q}'}\right|}^{2}{-}{2}\left|{\vec{q}}\right|\left|{\vec{q}'}\right|\cos\left({\mathit{\theta}}\right)}\label{eq:angular-factor}
\end{equation}

However the result is essentially the same. The uncertainty in $\left|{\vec{k}}\right|$
is of order $100MeV$ and the corresponding uncertainty in position
of the same order of magnitude as the original size of the proton.
Again useful for insuring convergence of integrals, but far below
any plausible threefold of detectibility.

\paragraph{Summary}

The key points are:
\begin{enumerate}
\item The initial photon wave function is finite, which will cause the integral
of the propagator applied to the wave function to be finite.
\item The photon wave function in time/space starts off about $10^{5}$
smaller than the atomic wave function, so will not supply a useful
source of dispersion in time. The dispersion in time at the position
of the atomic electron will come almost entirely from the TQM propagator,
which allows off-shell components. The further from the nucleus, the
greater the dispersion.
\end{enumerate}

\paragraph{Suppression of off-shell frequencies}

\label{par:Suppression-of-off-shell}

However there is an interesting implication of this. We have been
assuming the wave functions are a direct product of the time/energy
part and the space/momentum part. But now consider the time to get
from the nucleus to the atomic electron. This is of order $a_{0}=.177as$.
Since $\hbar=658.2eV$ the corresponding frequency in energy units
is $3872eV$ or about $4KeV$. The FS/T imposes a penalty given by
$\varpi_{k}\tau$ on frequencies that stray too far from their associated
$\kappa$. The denominator in the $\varpi$ of order $w$ (MeV) times
$\tau^{-1}$ or 4KeV. The numerator is of two quantities both presumably
of order MeV, so we have $\frac{\left(MeV\right)^{2}}{MeV\ 4KeV}$.
As a result, frequencies more than 4KeV away from the associated $\kappa$
will be vigorously suppressed by destructive interference, forcing
the frequencies to stay within a few $KeV$ of the momentum part,
even on the short trip out to the atomic electron. There will be aggressive
entanglement and focusing of the frequency part as given by the expression: 

\begin{equation}
\varpi\tau\sim\frac{{\left({\mathit{\kappa}{+}\mathit{\delta}{w}}\right)}^{2}{-}{\mathit{\kappa}}^{2}}{{2}\mathit{\kappa}\hspace{0.33em}{\mathit{\tau}}^{{-}{1}}}{\sim}\frac{\mathit{\delta}{w}}{{\mathit{\tau}}^{{-}{1}}}
\end{equation}

So the frequency is unlikely to matter much if it is more than a few
$KeV$ from its companion $\kappa$, a small fraction of the nominal
size of the frequency. Off-shell photons will be strongly filtered
on their way to their electron. 

\paragraph{Two simplifications}

So we get two simplifications from this analysis. The initial wave
function for the photon may be treated as a GTF, ensuring that integrals
of otherwise divergent propagators converge. And we can expand $w$
in powers of $\delta w$ around $\kappa$.

\subsection{Three views of the photon propagator}

\label{subsec:Three-views-of}

In our investigations of the propagator for the photon we have found
ourselves in the position of the six blind men and the elephant in
the well-known story. Each blind man feels a different part -- side,
trunk, tusk, leg, ear, tail -- of the elephant; each gets a different
impression -- wall, snake, spear, tree, fan, rope. In the various
versions of this story (which has been around for over 2000 years)
the blind men are typically criticized for their limited perspectives.
But we prefer to think of these perspectives not as false but rather
as incomplete. All are useful and taken together can give a reasonable
picture of the elephant.

We will look at three candidate perspectives here. Each approach will
have difficulties which keep us from using it naively; but each will
offer insights which will help us construct a reasonable solution
in the following subsection:
\begin{enumerate}
\item Working in strict parallel to SQM,
\item Taking seriously the notion of coordinate time as a 4th spatial dimension,
\item And, jumping to the far side of the problem, infer from the necessary
properties of the solution in time what an effective potential might
look like.
\end{enumerate}
We will keep three blind men in reserve.

\subsubsection{Photon propagator}

\label{subsec:Photon-propagator}

\begin{figure}
\includegraphics[scale=0.67]{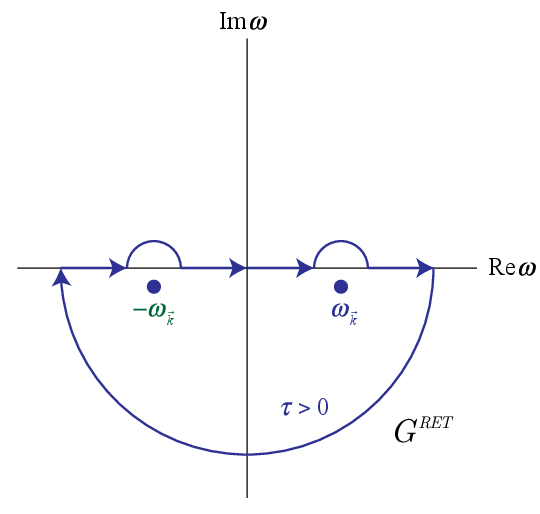}

\caption{\protect\label{fig:Feynman-boundary-photon}Photon boundary conditions
in clock time (figure derived from Jackson's figure 12.7 \cite{Jackson:1999ng})}
\end{figure}

The goal here is to get the photon propagator for TQM using an argument
parallel to the argument used in SQM.

\paragraph{SQM photon propagator}

We are using the derivation in Jackson \cite{Jackson:1999ng} as our
SQM reference.\footnote{For ease of comparison in this subsection we are using Jackson's conventions
for the Fourier transform rather than our usual. The only place this
makes a difference here is where Jackson has ${f}\left({t}\right){=}\frac{1}{{2}\mathit{\pi}}\int{{dw}\exp\left({{-}\imath{wt}}\right)}\hat{f}\left({w}\right)$
we have ${f}\left({t}\right){=}\frac{1}{\sqrt{{2}\mathit{\pi}}}\int{{dw}\exp\left({{-}\imath{wt}}\right)}{\hat{f}}\left({w}\right)$.
We document our conventions in appendix \ref{sec:=000020Gaussian=000020test=000020functions}.}

The starting point is the Klein-Gordon equation:

\begin{equation}
\Box\mathit{\psi^{S}}\left(\tau,\vec{x}\right){=}{0}\label{eq:klein-gordon}
\end{equation}

Where the d'Alembertian is defined as:

\[
\Box\equiv\frac{{\partial}^{2}}{\partial{\mathit{\tau}}^{2}}{-}{\nabla}^{2}
\]

And where the associated Green's function is defined as:

\begin{equation}
\Box{G}_{\mathit{\tau}\mathit{\tau}{'}}{}^{S}\mbox{ \ensuremath{\left({\mbox{ \ensuremath{\vec{x}}};\mbox{ \ensuremath{\vec{x}}}'}\right)}}{=}\delta\mbox{ \ensuremath{\left({\tau{-}\tau{'}}\right)}}{\delta}^{3}\mbox{ \ensuremath{\left({\mbox{ \ensuremath{\vec{x}}}{-}\mbox{ \ensuremath{\vec{x}}}{'}}\right)}}
\end{equation}

Since the Green's function is only a function of the difference in
initial and final positions we can write: 

\begin{equation}
{G}^{S}\mbox{ \ensuremath{\left({\tau{,}\mbox{ \ensuremath{\vec{x}}}{;}\tau{',}\mbox{ \ensuremath{\vec{x}}}{'}}\right)}}\rightarrow{G}_{\mathit{\tau}{-}\mathit{\tau}{'}}^{S}\left({\vec{x}{-}\vec{x}{'}}\right)
\end{equation}

\begin{equation}
{G}_{\mathit{\omega}\mathit{\omega}{'}}^{S}\left({\vec{k};\vec{k}'}\right)\rightarrow{G}_{\mathit{\omega}}^{S}\left({\vec{k}}\right)\mathit{\delta}\left({\mathit{\omega}{-}\mathit{\omega}{'}}\right){\mathit{\delta}}^{3}\left({\vec{k}{-}\vec{k}{'}}\right)
\end{equation}

We shift to the momentum space form. We get:

\begin{equation}
{G}^{S}\mbox{ \ensuremath{\left({\mathit{\omega}{,}\mbox{ \ensuremath{\vec{k}}}}\right)}}{=}{-}\frac{1}{{\mathit{\omega}}^{2}{-}{\kappa}^{2}}
\end{equation}

With:

\[
\kappa\equiv\left|\vec{k}\right|
\]

The inverse Fourier transform from $\omega\to\tau$ is divergent.
We resolve this problem by treating it as a contour integral. We have
poles at $\omega=\pm\kappa$ (see figure \ref{fig:Feynman-boundary-photon}).
For the retarded Green's function we need clock time positive so we
close the contour below:

\begin{equation}
\begin{array}{lll}
\oint{{d}\omega\frac{\exp\mbox{ \ensuremath{\left({{-}\imath\omega\tau}\right)}}}{{\omega}^{2}{-}{\kappa}^{2}}} & = & {2}\pi\imath{Res}\mbox{ \ensuremath{\left({\mbox{ \ensuremath{\frac{\exp\mbox{ \ensuremath{\left({{-}\imath\omega\tau}\right)}}}{{\omega}^{2}{-}{\kappa}^{2}}}}}\right)}}\\
 & = & {-}\frac{{2}\mathit{\pi}}{\mathit{\kappa}}\sin\left({\mathit{\kappa}\mathit{\tau}}\right)
\end{array}
\end{equation}

The result is:

\begin{equation}
{G}_{\tau}^{S}\mbox{ \ensuremath{\left({\mbox{ \ensuremath{\vec{r}}}}\right)}}{=}{-}\frac{\mathit{\theta}\left({\mathit{\tau}}\right)}{\left({2}\mathit{\pi}\right)^{3}}\int{\frac{d\vec{k}}{\kappa}\exp\mbox{ \ensuremath{\left({\imath\mbox{ \ensuremath{\vec{k}}}\cdot\mbox{ \ensuremath{\vec{r}}}}\right)}}\sin\mbox{ \ensuremath{\left({\kappa\tau}\right)}}}
\end{equation}

We do the angular integral, getting: 

\begin{equation}
{G}_{\tau}^{S}\mbox{ \ensuremath{\left({\mbox{ \ensuremath{\vec{r}}}}\right)}}{=}{-}\frac{\mathit{\theta}\left({\mathit{\tau}}\right)}{2\pi^{2}r}\mathop{\int}\limits_{0}^{\infty}{{d}\kappa}sin\left({\kappa{r}}\right)sin\left({\kappa\tau}\right)
\end{equation}

Note this last integral is not completely well-defined itself. Since
the integrand is even in $\kappa$, we expand the range to $\mp\infty$
and then expand the sines as exponentials to get:

\begin{equation}
{G}_{\mathit{\tau}}^{S}\left({\vec{r}}\right){=}\frac{\mathit{\theta}\left({\mathit{\tau}}\right)}{8\pi^{2}r}\mathop{\int}\limits_{{-}\infty}^{\infty}{{d}\mathit{\kappa}}\left({\exp\left({\imath\left({\mathit{\tau}{-}{r}}\right)\mathit{\kappa}}\right){-}\exp\left({\imath\left({\mathit{\tau}{+}{r}}\right)\mathit{\kappa}}\right)}\right)
\end{equation}
This gives a pair of delta functions:

\begin{equation}
{G}_{\tau}^{S}\mbox{ \ensuremath{\left({\mbox{ \ensuremath{\vec{r}}}}\right)}}{=}\frac{\theta\mbox{ \ensuremath{\left({\tau}\right)}}}{{4}\pi{r}}\mbox{ \ensuremath{\left({\delta\left({{r}{-}\tau}\right){+}\delta\left({{r}{+}\tau}\right)}\right)}}
\end{equation}

From the left side we get a retarded potential; from the right an
advanced. We drop the advanced since because of the theta function
it can never hit:

\begin{equation}
{G}_{\mathit{\tau}}^{S}\left({\vec{r}}\right){=}\frac{\mathit{\theta}\left({\mathit{\tau}}\right)\pi}{{4\pi r}}\mathit{\delta}\left({\mathit{\tau}{-}{r}}\right)\label{eq:sqm-propagator-in-space}
\end{equation}

This is the familiar form for a retarded potential:\footnote{Using the definition of the fine structure constant in natural units:
$\alpha=\frac{{e}^{2}}{{4}\pi}.$}

\begin{equation}
{-}\frac{{e}^{2}}{{4}\mathit{\pi}{r}}\rightarrow{-}\frac{\mathit{\alpha}}{r}\mathop{\int}\limits_{{-}\infty}^{\mathit{\tau}}{{d}\mathit{\tau}{'}\mathit{\delta}\left({\mathit{\tau}{-}\mathit{\tau}{'}{-}{r}}\right)}\label{eq:sqn-potential-in-space}
\end{equation}

We have returned to the Schrödinger equation, albeit with a retarded
potential rather than an instantaneous one.

\paragraph{TQM photon propagator}

We now apply essentially the same approach to the TQM photon propagator.
We start with the free FS/T equation. The Green's function is defined
by:

\begin{equation}
\left({2w\imath\frac{\partial}{\partial\mathit{\tau}}+{k}_{\mathit{\mu}}{k}^{\mathit{\mu}}}\right){G}_{\mathit{\tau}\tau'}\left({x;x'}\right){=}\mathit{\mathit{\delta}\left(\tau-\tau'\right)\delta}^{4}\left({x;x'}\right)\label{eq:photon-greens}
\end{equation}

In momentum space:\footnote{We return to our conventions for the Fourier transform.}

\begin{equation}
{-}\mbox{ \ensuremath{\left({{2}\omega{w}{+}{w}^{2}{-}{\mbox{ \ensuremath{\vec{k}}}}^{2}}\right)}}{G}_{\omega}{\left(w,\vec{k}\right)=}\frac{1}{{\sqrt{{2}\pi}}^{5}}
\end{equation}

Or:

\begin{equation}
{G}_{\omega}{\left(w,\vec{k}\right)=}{-}\frac{1}{{\sqrt{{2}\pi}}^{5}}\frac{1}{\mbox{ \ensuremath{\left({{2}\omega{w}{+}{w}^{2}{-}{\mbox{ \ensuremath{\vec{k}}}}^{2}}\right)}}}
\end{equation}

The next problem is to take the five Fourier transforms to get back
to coordinate space. To get to the clock time we use, as earlier,
contour integration. We have only one root for $\omega$, which simplifies
the contour integral. We use retarded boundary conditions:

\begin{equation}
{G}_{\tau}\mbox{ \ensuremath{\left({k;k'}\right)}}{=}\frac{{e}^{{-}{i}\tau{\varpi}_{k}}}{2w}{\mathit{\delta}}^{4}\left({{k}{-}{k}{'}}\right)\Theta\mbox{ \ensuremath{\left({\tau}\right)}}\label{eq:tqm-green-momentum-space}
\end{equation}
We next take the Fourier transform from coordinate frequency $w$
to coordinate time $t$:

\begin{equation}
{\mathcal{F}}^{{-}{1}}\left[{{G}_{\mathit{\tau}}\left({{w}{,}\mathit{\kappa}}\right){,}{w}{,}{t}}\right]\equiv\frac{1}{\sqrt{2\pi}}\int{{dw}\exp\left({{-}\imath{wt}}\right)}\frac{\exp\left({\imath\frac{w}{2}\mathit{\tau}{-}\imath\frac{{\mathit{\kappa}}^{2}}{2w}\mathit{\tau}}\right)}{2w}
\end{equation}

We have:

\begin{equation}
{G}_{\tau}\mbox{ \ensuremath{\left({{t}{,}\kappa}\right)}}{=}{-}\frac{{i}\sqrt{{2}\pi}\mbox{ \ensuremath{\left({\mbox{ \ensuremath{\sqrt{{(}\tau\mbox{ \ensuremath{{-}}}{2}{t}{)}^{2}}}}{+}{2}{t}{-}\tau}\right)}}{J}_{0}\mbox{ \ensuremath{\left({\kappa\mbox{ \ensuremath{\sqrt{\tau}}}\mbox{ \ensuremath{\sqrt[4]{{(}\tau\mbox{ \ensuremath{{-}}}{2}{t}{)}^{2}}}}}\right)}}}{{8}{t}{-}{4}\tau}
\end{equation}

Or:

\[
{G}_{\tau}\mbox{ \ensuremath{\left({{t}{,}\kappa}\right)}}{=}{a}\left({\mathit{\tau}{,}{t}_{\mathit{\tau}}}\right){J}_{0}\mbox{ \ensuremath{\left({\kappa{\mathrm{b}}\left({\mathit{\tau}{,}{t}_{\hspace{0.33em}\mathit{\tau}}}\right)}\right)}}
\]

With $a,b$ defined as:

\begin{equation}
{a}_{\mathit{\tau}}\left({t_{\tau}}\right){=}{-}\imath\sqrt{\frac{\pi}{2}}\mathit{\theta}\left(\tau+{2t_{\tau}}\right)
\end{equation}

\begin{equation}
{b}_{\mathit{\tau}}\left({t_{\tau}}\right){=}\mathit{\tau}\sqrt{\left|{1}{+}\frac{{2}t_{\tau}}{\mathit{\tau}}\right|}
\end{equation}

And with the relative time $t_{\tau}$ defined by:

\[
{{t}{=}\mathit{\tau}{+}t_{\tau}}
\]

The relative time represents the dispersion in time at each instant
in clock time. The SQM limit is $t_{\tau}=0$:

\begin{equation}
{a}_{\tau}\mbox{ \ensuremath{\left({0}\right)}}{=}{-\imath}\sqrt{\frac{\pi}{2}}\theta\mbox{ \ensuremath{\left({\tau}\right)}}
\end{equation}

\begin{equation}
{b}_{\tau}\mbox{ \ensuremath{\left({0}\right)}}{=}\tau
\end{equation}

This has the expected theta function in clock time. And the argument
of the Bessel function is the same as the argument of the sines above.
Now we look at the Fourier transform from momentum to space:

\begin{equation}
{\mathcal{F}}^{{-}{1}}\left[{{G}_{\mathit{\tau}}\left({t,\vec{k}}\right){,}\vec{k}{,}\vec{r}}\right]{=}\mathop{\int}\limits_{0}^{\infty}{{\mathit{\kappa}}^{2}{d}\mathit{\kappa}}\mathop{\int}\limits_{{-}{1}}^{1}{{d}\cos\mathit{\theta}}\mathop{\int}\limits_{0}^{{2}\mathit{\pi}}{{d}\mathit{\phi}}\exp\left({\imath\hspace{0.33em}\mathit{\kappa}{r}\cos\mathit{\theta}}\right){G}_{\mathit{\tau}}\mbox{ \ensuremath{\left({{t}{,}\kappa}\right)}}
\end{equation}

The angular part is unchanged from above. This gives the expected
$1/r$ dependence: 

\begin{equation}
{\mathcal{F}}^{{-}{1}}\mbox{ \ensuremath{\left[{{G}_{\tau}\left({k}\right){,}{k}{,}{r}}\right]}}{=}-{2}\mathit{\pi}\imath\frac{1}{r}\mathop{\int}\limits_{0}^{\infty}{\kappa{d}\kappa}\sin\left({\mathit{\kappa}{r}}\right){G}_{\tau}\mbox{ \ensuremath{\left({{t}{,}\kappa}\right)}}\label{eq:sqm-based-tqm-space-form}
\end{equation}

However the Bessel function goes as $1/\sqrt{\kappa}$ as $\kappa\to\infty$,
not enough to cancel the $\kappa$ in the numerator. Where the SQM
equivalent was arguably convergent, the TQM form is not convergent. 

\paragraph{Assessment}

The lack of convergence of the propagator is not that troubling; if
we take the inverse Fourier transform of the propagator as applied
to one of our GTFs, the integral will converge. This is the same approach
we used to make the initial FPI integrals converge.

But the bizarre shape of the argument of the $\theta$ function is
troubling. At clock time zero this is saying that the dispersion in
time is required to be strictly positive. This is contrary to the
nature of paths in coordinate time, which go as freely into the past
as into the future. 

The implication is that our parallel development of TQM has stayed
too close to its original: we need to have ``fuzzy'' starting conditions
rather than ``crisp'' ones. In terms of Wheeler's ``radical conservatism''
we have been too conservative; we have not been radical enough. There
are various ways of handling this; we have found none entirely convincing.
We will therefore put this approach aside for the moment.

\subsubsection{Pseudo-Euclidean approach}

\label{subsec:Pseudo-Euclidean-approach}

Consider the two aspects of time: clock time and coordinate time.
We can see clock time as ``outer time'': it is used to describe
averages over the paths and in general the macroscopic variables used
at the laboratory level. We can see coordinate time as ``inner time'':
it deals with the individual paths. To better align our variables
with our intuitions we shift variables, writing the microscopic variables
as differences between them and the corresponding macroscopic variables.
We also take advantage of the (partial) resemblance of coordinate
time to a fourth spatial coordinate.

We start by spelling out the full expression for the Green's function:

\begin{equation}
{G}_{\mathit{\tau}}\left({t,\vec{r}}\right){=}\frac{1}{{\sqrt{{2}\mathit{\pi}}}^{5}}\int{{dwd\vec{k{d}\mathit{\omega}}\frac{\exp\left({{-}\imath\mathit{\omega}\mathit{\tau}}\right)\exp\left({{-}\imath{wt}}\right)\exp\left({\imath\vec{k}\cdot\vec{r}}\right)}{2\omega w+w^{2}-\vec{k^{2}}}}}
\end{equation}

We hypothesize that:

\begin{equation}
\begin{array}{c}
{\mathit{\omega}\approx{\mathit{\omega}}_{\vec{k}}\equiv\kappa}\\
{{t}\approx\tau}
\end{array}
\end{equation}

We therefore define:

\begin{equation}
\begin{array}{rll}
{\bm{\omega}} & \equiv & w+\omega\\
\mathbf{{w}} & \equiv & -w
\end{array}
\end{equation}

With these changes the Green's function becomes:

\begin{equation}
{G}_{\tau}\mbox{ \ensuremath{\left({t,\mbox{ \ensuremath{\vec{r}}}}\right)}}{=}\frac{1}{{\sqrt{{2}\pi}}^{5}}\int{d\mathbf{w}d\vec{k}{d}\bm{\omega}\frac{\exp\mbox{ \ensuremath{\left({{-}\imath\bm{\omega}\tau}\right)}}\exp\mbox{ \ensuremath{\left({\imath{\mathbf{k}}\cdot{\mathbf{d}}}\right)}}}{{\bm{{\omega}}}^{\mathbf{2}}\mathbf{{-}}{\mathbf{\bm{\kappa}}}^{\mathbf{2}}}}
\end{equation}

with three ancillary definitions and an equality:

\begin{equation}
\begin{array}{rll}
\bm{\kappa} & \equiv & \sqrt{{w^{2}+\vec{k}^{2}}}\\
{\mathbf{k}} & \equiv & \left({\mathbf{w},\vec{k}}\right)\\
\mathbf{d} & \equiv & \left({t_{\tau}{,}\vec{r}}\right)\\
{\mathbf{k}}\cdot{\mathbf{d}} & = & {\mathbf{w}}t_{\tau}{+}\vec{k}\cdot\vec{r}
\end{array}
\end{equation}

$\mathbf{k},\mathbf{d}$ are Euclidean not Lorentz vectors. 

We are here using $t_{\tau}$ as the 4th component of a Euclidean
four-vector. We do the integral over $\bm{\omega}$ as a contour integral
reducing (with retarded boundary conditions) the Green's function
to:

\begin{equation}
\oint{{d}\mathbf{\bm{{\omega}}}\frac{\exp\mbox{ \ensuremath{\left({{-}\imath\mathbf{\bm{{\omega}}}\tau}\right)}}}{{\bm{{\omega}}}^{\mathbf{2}}\mathbf{{-}}{\mathbf{\bm{\kappa}}}^{\mathbf{2}}}}{=}{-}\frac{{2}\pi}{{\mathbf{\bm{\kappa}}}}exp\mbox{ \ensuremath{\left({\mathbf{\bm{{\kappa}}}\tau}\right)}}
\end{equation}

We have an integral in four Euclidean dimensions rather than three:

\begin{equation}
{G}_{\tau}\mbox{ \ensuremath{\left({t,\mbox{ \ensuremath{\vec{r}}}}\right)}}{=}\frac{-1}{{\sqrt{{2}\pi}}^{3}}\int{{\mathbf{\bm{{\kappa}}}}^{3}{d^{3}}\Omega d\bm{k}\frac{\exp\left({{-}\imath{\bm{{\kappa}}}\mathit{\tau}}+\imath{\mathbf{k}}\cdot{\mathbf{d}}\right)}{2\mathbf{\bm{{\kappa}}}}}
\end{equation}

We take the angle between $\bm{\kappa}$ and $\bm{d}$ as $\theta$.
The integral over this is:

\begin{equation}
\mathop{\int}\limits_{{-}{1}}^{1}{{d}\cos\left({\mathit{\theta}}\right)}\exp\left({{-}\imath{\mathbf{k}}\cdot{\mathbf{d}}}\right){=}\frac{\exp\left({\imath{\mathbf{kd}}}\right){-}\exp\left({\imath{\mathbf{kd}}}\right)}{\imath{\mathbf{kd}}}{=}{2}\frac{\sin\left({\mathbf{kd}}\right)}{\mathbf{kd}}
\end{equation}

The remaining two angular integrals contribute a factor of $2\pi^{2}$,
leaving us with:

\begin{equation}
{G}_{\tau}\mbox{ \ensuremath{\left({t,\mbox{ \ensuremath{\vec{r}}}}\right)}}{=}{-}\sqrt{\frac{\mathit{\pi}}{2}}\frac{1}{\mathbf{d}}\mathop{\int}\limits_{0}^{\infty}{{\mathbf{k}}{d}{\mathbf{k}}\sin\left({\mathbf{kd}}\right)}
\end{equation}

The additional angular factor in front, $\sqrt{\frac{\mathit{\pi}}{2}}$,
essentially results from the combination of one more delta function
and one more angle to integrate over. And of course the remaining
integral is not convergent.

If we could ignore the resulting angular factor of $\sqrt{\frac{\pi}{2}}$,
we could replace the $1/r$ in SQM by $1/d$ in TQM:

\begin{equation}
\frac{1}{r}\rightarrow\frac{1}{\mathbf{{d}}}{=}\frac{1}{\sqrt{t_{\tau}^{2}{+}{\vec{r}}^{2}}}
\end{equation}

We refer to this set of variables as pseudo-Euclidean: the underlying
variables are Lorentzian, but the distance is formed using the Euclidean
metric. We can make the potential appear still more symmetric in time
by writing in terms of the start and end points in time as well as
in space:

\begin{equation}
\frac{1}{r}\equiv\frac{1}{\sqrt{{\left({{\vec{r}}_{B}{-}{\vec{r}}_{C}}\right)}^{2}}}\rightarrow\frac{1}{\mathbf{d}}{=}\frac{1}{\sqrt{{\left({t_{B}{-}t_{C}}\right)}^{2}{+}{\left({{\vec{r}}_{B}{-}{\vec{r}}_{C}}\right)}^{2}}}
\end{equation}

In an obvious notation we define $t_{BC}\equiv t_{B}-t_{C}$. We expand
in powers of $t_{BC}$:

\begin{equation}
\frac{1}{\sqrt{{t}_{BC}^{2}{+}{r}_{BC}^{2}}}\approx\frac{1}{{r}_{BC}}{-}\frac{1}{2}\frac{{t}_{BC}^{2}}{{r}_{BC}^{3}}
\end{equation}

\subparagraph{Assessment}

The additional angular factor is difficult to explain away. It suggests
we do not have four full dimensions to work with. Instead this suggests
the off-shell components are sufficiently suppressed that we are dealing
with three-and-a-fraction dimensions rather than four. This is consistent
with the earlier analysis that only a small fraction of the possible
photons actually contribute to the potential.

Further the $t_{\tau}^{2}$ correction may also have the wrong sign.
To lowest order in $t_{\tau}$ it should look like a harmonic oscillator
potential term, since to lowest order most restoring forces look like
a harmonic oscillator's potential term. This implies it should same
sign as the $E^{2}$ term. But if we look at the entire FS/T we get:

\begin{equation}
{-}\frac{{E}^{2}}{2m}{+}\frac{{\vec{p}}^{2}}{2m}{-}\frac{\mathit{\alpha}}{r}{+}\frac{1}{2}\frac{{t}_{\mathit{\tau}}^{2}}{{r}^{3}}
\end{equation}

where the $E^{2}$ and $t_{\tau}^{2}$ terms come in with opposite
signs.

\subsubsection{Global harmonic oscillator}

\label{subsec:GHO}

Let us turn this last point around. Suppose we approximate the behavior
of the time side of the FS/T as a harmonic oscillator. What properties
should it have? This will help give us a target. Let us ignore for
the moment any dependence on the radius; treat the time part as some
kind of an average harmonic oscillator.

We focus on the time/energy part of the FS/T. We hypothesize we are
looking at a harmonic oscillator averaged over time:\footnote{We are using $\mathcal{{E}}$ to refer to the eigenvalue of the harmonic
oscillator Hamiltonian, since we have already used up $E$ to mean
the complementary operator $\imath\frac{\partial}{\partial t}$ to
$t$.}

\begin{equation}
{\mathcal{E}}^{\mathrm{T}}{\psi}^{\mathrm{T}}{=}\frac{{E}^{2}}{2m}{\psi}^{\mathrm{T}}{+}\frac{m{\Omega}^{2}}{2}{t}^{2}{\psi}^{\mathrm{T}}
\end{equation}

We have the average energy in time equal but opposite to energy in
space:

\begin{equation}
\mathcal{{E}}^{S}+\mathcal{{E}}^{T}=0
\end{equation}

The energy in space is:

\begin{equation}
\mathcal{{E}^{S}}{=}{-}\frac{m{\alpha}^{2}}{2}
\end{equation}
This gives the energy in time as the negative of this. The corresponding
energy of an oscillator is:

\begin{equation}
{\mathcal{E}}_{0}^{\mathrm{T}}=\frac{1}{2}\Omega
\end{equation}

This gives $\Omega=\alpha^{2}m$. The ground state solution is:

\begin{equation}
{\mathit{\psi^{T}}}_{0}\left({t}\right){=}{\left({\frac{{m}\Omega}{\mathit{\pi}}}\right)}^{1/4}{e}^{{-}\frac{{m}\Omega{t}^{2}}{2}}
\end{equation}

With dispersion:

\begin{equation}
{\mathit{\sigma}}_{t}^{\left(B\right)2}{=}\frac{1}{{m}\Omega}
\end{equation}

and uncertainty:

\begin{equation}
{\left({\Delta{t}}\right)}^{2}=\frac{1}{{2}{m}\Omega}
\end{equation}

Spelled out:

\begin{equation}
{\mathit{\sigma_{t}}^{\left(B\right)2}}{=}\frac{1}{{\left({{m}\mathit{\alpha}}\right)}^{2}}{=}{a}_{0}^{2}\label{eq:gho-estimate-for-sigma-b}
\end{equation}

which is the original naive estimate. So we have recovered this by
a second method.

We will refer to this as the ``global harmonic oscillator'' or GHO
estimate.

\subparagraph{Assessment}

It is interesting that knowledge of the energy in the time direction
also gives one a zeroth estimate of the dispersion in the time direction.
The problem here is we still have no ``how''.

\subsection{Effective potential}

\label{subsec:Effective-potential}

\begin{figure}
\includegraphics[scale=0.75]{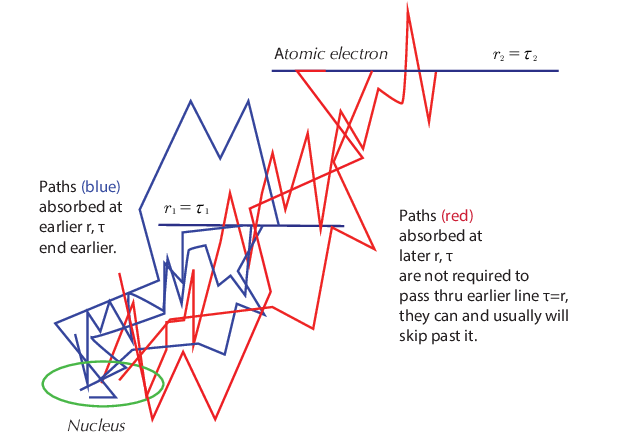}

\caption{\protect\label{fig:intermediate-paths}Absorption of paths at different
times and radii.}
\end{figure}

We return to the analysis of the propagator. We will:
\begin{enumerate}
\item Compute the photon propagator,
\item Apply the photon propagator to the photon wave function at the nucleus
to get the photon wave function at the point of interaction with the
atomic electron,
\item Map the interaction amplitude (QED approach) into an equivalent harmonic-oscillator-like
potential for use in the FS/T,
\item And thereby infer the frequency of the harmonic oscillator potential.
\end{enumerate}
This will give us an effective potential as experienced by the atomic
electron.

\paragraph{Compute the photon propagator}

As noted (eqn \ref{eq:sqm-based-tqm-space-form}), the momentum space
form is better defined than the coordinate space form; we will start
with the momentum space form. We first split the propagator into energy
and momentum parts. In this way we can write the energy part as a
pre-factor to known SQM results:

\begin{equation}
{G}_{\tau}\mbox{ \ensuremath{\left({w,\mbox{ \ensuremath{\vec{k}}}}\right)}}{\approx}{G}_{\tau}^{T}\mbox{ \ensuremath{\left({w,\mbox{ \ensuremath{\vec{k}}}}\right)}}{G}_{\tau}^{S}\mbox{ \ensuremath{\left({\mbox{ \ensuremath{\vec{k}}}}\right)}}
\end{equation}

The overall propagator -- time plus space -- goes as $\exp\left({{-}\imath{\varpi}_{k}\mathit{\tau}}\right)$.
The requirement in TQM is that the clock frequency be approximately
zero, especially over longer times. Since the space part has an $\exp\left({-}\imath\mathit{\kappa}\mathit{\tau}\right)$
attached we need to add a counter-balancing $\exp\left(\imath\mathit{\kappa}\mathit{\tau}\right)$
to the time part:

\begin{equation}
{G}_{\mathit{\tau}}^{T}\left({{w}{,}\mathit{\kappa}}\right){G}_{\mathit{\tau}}^{S}\left({\mathit{\kappa}}\right){=}\left({\frac{{2}\mathit{\kappa}}{2w}\exp\left({{-}\imath{\varpi}_{k}\mathit{\tau}+}\imath\mathit{\kappa}\mathit{\tau}\right)}\right)\left({\frac{\exp\left({{-}\imath\mathit{\kappa}\mathit{\tau}}\right)}{{2}\mathit{\kappa}}}\right)
\end{equation}

We note that the SQM propagator can be written as (eqn \ref{eq:sqm-propagator-in-space}):

\begin{equation}
\frac{\mathit{\delta}\left({\mathit{\tau}{-}{r}}\right)}{r}\label{eq:space-propagator}
\end{equation}

In general we get the TQM version of anything by starting with the
SQM form and adding dispersion in time. So we expect that the full
propagator in TQM will be approximately this but with dispersion in
time. The natural extension of a delta function is a narrow exponential. 

Further the SQM propagator suggests that once we know the radius we
know the clock time and vice versa. Picture the radius from nucleus
to atomic electron as broken up into steps of size $\Delta r$. We
will lay out one clock time grid line at each step. If the absorption
is to take place at the \emph{n-th} radial step, we will compute the
TQM propagator with a value of the clock time (relative to a starting
clock time of zero) equal to the radius at that step. So the propagator
to get to radius $r$ will have clock time $\tau=r$:

\begin{equation}
{G}_{\mathit{\tau}}\left({t,r}\right){=}{\left.{{G}_{\mathit{\tau}}\left({t,r}\right)}\right|}_{\mathit{\tau}{=}{r}}
\end{equation}

If we are later computing the Green's function at the \emph{m-th}
radial step we will treat the \emph{n-th} Green's function as fully
general:

\begin{equation}
{G}_{m}\left({{t}_{m},{r}_{m};{t}_{0},{r}_{0}}\right){=}\int{{dt}_{n}{dr}_{n}}{G}_{\mathit{mn}}\left({{t}_{m},{r}_{m};{t}_{n},{r}_{n}}\right){G}_{n}\left({{t}_{n},{r}_{n};{t}_{0},{r}_{0}}\right)
\end{equation}

so the intermediate Green's function is \emph{not} confined to $\tau_{n}=r_{n}$.
The paths to the \emph{m-th} step will include ``off-shell'' ($\tau\ne r$)
intermediate Green's functions.\footnote{The interaction of a path with the atomic electron acts as a kind
of mini-measurement: the absorbed paths are pinned down at the point
of interaction and transfer their momentum to the atomic electron,
but the other paths, those not so absorbed, swirl on past it, to be
absorbed in their turn.} This lets us simplify the calculations while still preserving their
essential character: paths arriving at clock time $\tau=r$ will include
paths arriving from both past and future as well as paths not necessarily
traveling at a constant speed in time.\footnote{In TQM the speed of light $c$ necessarily goes from being fixed and
absolute to being itself subject to quantum fluctuations.}

Per earlier arguments we expect that frequencies more than a few KeV
from their associated $\kappa$ will be suppressed by destructive
self-interference. Therefore it is reasonable to write the frequency
in terms of the difference between it and the associated momentum
$\delta w\equiv w-\kappa.$

Factors outside the exponential will not be nearly as significant
as factors within so it is acceptable to approximate:

\begin{equation}
\frac{2w}{{2}\kappa}{=}\frac{{2}\mbox{ \ensuremath{\left({\kappa{+}\delta{w}}\right)}}}{{2}\kappa}\approx{1}
\end{equation}

Within the exponential we keep only terms through the quadratic:

\begin{equation}
{G}_{\mathit{\tau}}^{T}\left({{w}{,}\mathit{\kappa}}\right)\approx\exp\left(\imath\kappa\tau+{\imath\delta w\tau{-}\imath\frac{{\left({\mathit{\delta}{w}}\right)}^{2}}{{2}\mathit{\kappa}}\mathit{\tau}}\right)
\end{equation}

The quadratic term is the decisive one: it will determine the dispersion
at the atomic electron. The factor $\frac{\kappa}{\tau}$ appears
formally as an effective dispersion, albeit an imaginary one.

We have one more simplification to make before we do the actual calculation.
We will need to integrate over $\kappa$ to get the full wave function.
This integration entangles the frequency part with the momentum part.
We can disentangle the frequency and momentum parts by replacing $\kappa$
by its average: $\kappa\to\bar{\kappa}$. We would like to estimate
the average value of $\kappa$ of the photons arriving at position
$r$ at time $\tau$. From dimensional arguments we expect $\bar{\kappa}\sim\frac{1}{r}$.We
will therefore take as a trial hypothesis that:

\begin{equation}
\bar{\kappa}=\frac{\mu}{r}
\end{equation}

with $\mu$ dimensionless, expected of order one, and to be determined.
Additional discussion in appendix \ref{sec:bar-kappa}.

With the energy and momentum parts disentangled, and the energy part
in quadratic form, we now compute the propagator in time. First we
fold back in the delta function in frequency:

\begin{equation}
{G}_{\tau}^{T}\mbox{ \ensuremath{\left({w,\mbox{ \ensuremath{\bar{\kappa}}};w',\mbox{ \ensuremath{\bar{\kappa}}}}\right)}}={G}_{\tau}^{T}\mbox{ \ensuremath{\left({{w}{,}\bar{\kappa}}\right)}}\delta\mbox{ \ensuremath{\left({{w}-{w}{'}}\right)}\ensuremath{}}
\end{equation}

We write out the formal inverse Fourier transform from $w,w'\to t,t'$
coordinates:

\begin{equation}
{{G}_{\tau}^{T}\mbox{ \ensuremath{\left({t,t';\mbox{ \ensuremath{\bar{\kappa}}}}\right)}}=\frac{1}{\sqrt{{2}\pi}}\int{{dw}\exp\left({-\imath{wt}}\right)}\frac{1}{\sqrt{{2}\pi}}\int{{dw}{'}\exp\left({\imath{w}{'}{t}{'}}\right)}{G}_{\tau}^{T}\mbox{ \ensuremath{\left({w,\mbox{ \ensuremath{\bar{\kappa}}};w',\mbox{ \ensuremath{\bar{\kappa}}}}\right)}}}
\end{equation}

We change coordinates from $w\to\delta w$ and from $t\to t_{\tau}$,
bearing in mind that as we are starting at clock time $\tau=0$ we
have $t'=t'_{\tau}$. To simplify we define the difference in relative
time between the nucleus and the atomic electron:

\begin{equation}
\Delta{t}_{\mathit{\tau}}\equiv{t}_{\mathit{\tau}}-{t}_{\mathit{\tau}}{'}
\end{equation}

We get:
\begin{equation}
{G}_{\tau}^{T}\mbox{ \ensuremath{\left({\Delta{t}_{\mathit{\tau}}{;}\mbox{ \ensuremath{\bar{\kappa}}}}\right)}}=\frac{1}{\sqrt{{2}\pi}}\exp\mbox{ \ensuremath{\left({-\imath\mbox{ \ensuremath{\bar{\kappa}}}\Delta{t}_{\mathit{\tau}}}\right)}}\frac{1}{\sqrt{{2}\pi}}\int{{d}\mathit{\delta}{w}\exp\mbox{ \ensuremath{\left({-\imath\mathit{\delta}{w}\Delta{t}_{\mathit{\tau}}-\imath\mbox{ \ensuremath{\frac{{\mbox{ \ensuremath{\left({\delta{w}}\right)}}}^{2}}{2\bar{\kappa}}}}\tau}\right)}}}
\end{equation}

This gives the Green's function as the product of three factors:

\begin{equation}
{G}_{\tau}^{T}\mbox{ \ensuremath{\left({\Delta{t}_{\mathit{\tau}}{;}\mbox{ \ensuremath{\bar{\kappa}}}}\right)}}=\exp\mbox{ \ensuremath{\left({-\imath\mbox{ \ensuremath{\bar{\kappa}}}\Delta{t}_{\mathit{\tau}}}\right)}}\sqrt{\frac{\imath\bar{\kappa}}{{2}\pi\mathit{\tau}}}\exp\mbox{ \ensuremath{\left({\imath\mbox{ \ensuremath{\frac{{\left({\Delta{t}_{\mathit{\tau}}}\right)}^{2}}{{2}\mathit{\tau}}}}\bar{\mathit{\kappa}}}\right)}}
\end{equation}

The first factor says that if we start with a relative time at the
nucleus that relative time will be carried forward with the frequency
$\bar{\kappa}$. This will merely carry the time dispersion at the
nucleus to the position of the atomic electron. Since that time dispersion
is of order $10^{-5}$ of the estimated time dispersion of atomic
electron, we ignore it.\footnote{Compare the remaining two terms to the kernel for a particle of mass
$m$ traveling for a time $\tau$ (eqn \ref{eq:kernel-in-time}):

\begin{equation}
{K}_{\tau}\mbox{ \ensuremath{\left({{t}'';{t}\prime}\right)}}{=}\sqrt{{-}\imath\frac{m}{{2}\pi\tau}}{e}^{{-}\imath{m}\frac{{\mbox{ \ensuremath{\left({{t}''{-}{t}'}\right)}}}^{2}}{{2}\tau}}
\end{equation}

The parallel is close: $\bar{\kappa}$ functions as an effective mass
for the photon, the difference in coordinates becomes a difference
in relative time, the clock time is unchanged. The sign of the quadratic
part is opposite.}

We rewrite the photon's Green's function as:

\begin{equation}
{G}_{\Delta\tau}^{T}\mbox{ \ensuremath{\left({{\mathrm{\Delta t}}_{\tau}{,}\bar{\kappa}}\right)}}{=}\sqrt{\frac{\imath}{{\bar{\sigma}}_{\mathrm{t}}^{\left(A\right)2}}}{e}^{{-}\imath\bar{\kappa}{\mathrm{\Delta t}}_{\tau}{+}\imath\frac{\left({\mathrm{\Delta t}}_{\tau}\right)^{2}}{2{\bar{\sigma}}_{\mathrm{t}}^{\left(A\right)2}}}\label{eq:greens-function-in-time}
\end{equation}

with:

\begin{equation}
{\bar{\sigma}}_{\mathrm{t}}^{\mbox{\ensuremath{\left({A}\right)}}{2}}\equiv\frac{\Delta\tau}{\bar{\kappa}}=\frac{{r}^{2}}{\mathit{\mu}}\label{eq:photon-dispersion}
\end{equation}

where the $A$ is to remind us that this dispersion in time is the
dispersion associated with particle $A$, our spinless photon. Since
the function depends only on the difference in clock time we can write
it in terms of $\Delta\tau$, although it will usually be simpler
to write it as just $\tau$.

Working together, our three blind men have achieved a simple and reasonable
picture of the 4D elephant.

\paragraph{Effective potential in time and space}

We next apply the photon propagator to the initial wave photon function.
For the initial photon wave function, it will be convenient to use
the traditional approach to the nucleus: a delta function at the origin.
This keeps a strict parallel between time and the traditional approach
in space. (And eliminates the need for an initial photon wave function): 

\begin{equation}
{\mathit{\rho}}_{0}\left({{t}_{0},\vec{r}}\right)=\mathit{\delta}\left({0,\vec{0}}\right)
\end{equation}

letting us replace the difference between the relative time at electron
and nucleus with just the relative time at the electron:

\begin{equation}
\left(\Delta{t}_{\tau}^ {}\right)^{2}\rightarrow{t}_{\tau}^{2}
\end{equation}

For both space parts and time parts we apply the associated Green's
function. The Green's function in space is given by eqn \ref{eq:space-propagator}
above. The Green's function in time is given by eqn \ref{eq:greens-function-in-time}.
To compute the potential at clock time $\tau$, time $t$ (or relative
time $t_{\tau}$), and radius $r$ we integrate over the possible
source clock times, coordinate times, and source space positions:

\begin{equation}
{V}\left({t,\vec{r}}\right)=-\mathit{\alpha}\mathop{\int}\limits_{-\infty}^{\infty}{{d}\mathit{\tau}}\int{dt'd\vec{r}'}{G}_{\mathit{\tau}}^{T}\left({{t}-\mathit{\tau}{,}\bar{\kappa}}\right)\frac{\mathit{\delta}\left({\mathit{\tau}-\left|{\vec{r}-\vec{r}{'}}\right|}\right)}{r'}\mathit{\delta}\left({t'}\right){\mathit{\delta}}^{3}\left({\vec{r}'}\right)
\end{equation}

The five delta functions nicely cancel the five integrals leaving
us with:

\begin{equation}
{V}\left({{t}_{\mathit{\tau}},r}\right)=-\frac{\mathit{\alpha}}{r}{G}_{\mathit{\tau}}^{T}\mbox{ \ensuremath{\left({{\mathrm{t}}_{\tau},\bar{\kappa}}\right)}}
\end{equation}

From the QED point of view, we have an interaction matrix element
for atomic electron and photon wave functions:

\begin{equation}
-\mathit{\alpha}\left\langle {{\psi}_{r}^{\mathrm{B}}\left|{\frac{1}{r}}\right|{\psi}_{r}^{\mathrm{B}}}\right\rangle \left\langle {{\psi}_{\mathrm{t}}^{\mathrm{B}}\left|{{G}_{\mathit{\tau}}^{T}\mbox{ \ensuremath{\left({{\mathrm{t}}_{\tau},\bar{\kappa}}\right)}}}\right|{\psi}_{\mathrm{t}}^{\mathrm{B}}}\right\rangle 
\end{equation}

where the Green's function in time acts as a photon wave function:

\begin{equation}
{\varphi}_{\mathit{\tau}}^{A}\mbox{ \ensuremath{\left({{\mathrm{t}}_{\tau},\bar{\kappa}}\right)}}={G}_{\mathit{\tau}}^{T}\left({{t}_{\mathit{\tau}},\bar{\kappa}}\right)
\end{equation}

while from the FS/T point of view we have a complex potential:

\begin{eqnarray}
{V}\left({{t}_{\mathit{\tau}},r}\right) & = & -\frac{\mathit{\alpha}}{r}{\varphi}_{\mathit{\tau}}^{A}\mbox{ \ensuremath{\left({{\mathrm{t}}_{\tau},\bar{\kappa}}\right)}}
\end{eqnarray}

We have one interaction but two different languages for describing
it. The effect of the space part is wired into the shape of the usual
$\psi_{nlm}$ wave functions, dealt with implicitly in the standard
treatments. But meshing the time part with the rest of the FS/T will
require some more work.

\paragraph{Translate the interaction amplitude from QED terms to FS/T terms}

We have an expression for the interaction which is quadratic in relative
time. If this were a calculation in QED, we would integrate against
the initial wave function. But in FS/T it is more natural to represent
it as an operator acting on a general initial wave function. We start
with the FS/T kernel:

\begin{equation}
{K}\sim\exp\mbox{ \ensuremath{\left({\imath\hspace{0.33em}\mbox{ \ensuremath{\int{{d}\tau\int{{d}^{4}{x}{\mathcal{L}}\left[{{x}_{\mathit{\mu}},{\mbox{ \ensuremath{\dot{x}}}}_{\mathit{\mu}}}\right]}}}}}\right)}}
\end{equation}

The Lagrangian for a harmonic oscillator in $x$ is given by:

\begin{equation}
{\mathcal{L}}^{X}\left[{x}\right]{=}\frac{1}{2}{m}{\dot{x}}^{2}{-}\frac{m{\Omega}_{x}^{2}}{2}{x}^{2}
\end{equation}

giving the Euler-Lagrange equation:

\begin{equation}
\frac{\partial}{\partial\mathit{\tau}}\frac{\mathit{\delta}{\mathcal{L}}}{\mathit{\delta}\dot{q}}{-}\frac{\mathit{\delta}{\mathcal{L}}}{\mathit{\delta}{q}}{=}{0}\rightarrow{m}\ddot{x}{+}{m}{\Omega}_{x}^{2}{x}{=}{0}
\end{equation}

The kernel for an harmonic oscillator is well-known \cite{Schulman:1981um,Kleinert:2009hw}.
Taking its limit as $\tau\to\epsilon$ we get:
\begin{equation}
\sqrt{\frac{-\imath{m}}{{2}\mathit{\pi}\varepsilon}}\exp\left({\imath\frac{{\dot{x}}^{2}}{2m}\epsilon-\imath\frac{m{\Omega}^{2}}{2}{x}^{2}}\epsilon\right)
\end{equation}

The ``in time'' version enters with opposite sign:

\begin{equation}
{\mathcal{L}}^{T}\mbox{ \ensuremath{\left[{t}\right]}}{=}{-}\frac{1}{2}{m}{\dot{t}}^{2}{+}\frac{m{\Omega}_{t}^{2}}{2}{t}^{2}
\end{equation}

The short time kernel for this is:

\begin{equation}
\sqrt{\frac{\imath{m}}{{2}\mathit{\pi}\varepsilon}}\exp\left({-\imath\frac{{\dot{t}}^{2}}{2m}\epsilon+\imath\frac{m{\Omega}^{2}}{2}{t}^{2}}\epsilon\right)
\end{equation}

\paragraph{Infer the frequency of the harmonic oscillator}

\label{par:Infer-the-frequency}

We return to the effective potential, writing it as: 

\begin{equation}
{V}\mbox{ \ensuremath{\left({{t}_{\tau},r}\right)}}=-\frac{\alpha}{r}\sqrt{\frac{\imath}{{\bar{\sigma}}_{\mathrm{t}}^{\mbox{ \ensuremath{\left({A}\right)}}{2}}}}\exp\mbox{ \ensuremath{\left({\imath\mbox{ \ensuremath{\frac{{\mbox{ \ensuremath{\left({\Delta{\mathrm{t}}_{\tau}}\right)}}}^{2}}{2{\bar{\sigma}}_{\mathrm{t}}^{\mbox{ \ensuremath{\left({A}\right)}}{2}}}}}}\right)}}
\end{equation}

We compare the kernel for the photon to the short time kernel for
the harmonic oscillator in time. We get:

\begin{equation}
\frac{m{\Omega}_{\tau}^{2}}{2}\Longleftrightarrow\frac{\alpha}{2r{\bar{\mathit{\sigma}}}_{t}^{\left({A}\right){2}}}\label{eq:harmonic-oscillator-comparison}
\end{equation}

So we have $\Omega^{2}\sim r^{-3/2}$.

But what we are interested in is the dispersion of the relative time
coordinate of the atomic electron ${\mathit{\sigma}}_{t}^{\left({B}\right){2}}$.
Per the discussion of the GHO, we have:

\begin{equation}
{\mathit{\sigma}}_{t}^{\left({B}\right){2}}=\frac{1}{m{\Omega}_{\mathit{\tau}}}\label{eq:electron-dispersion}
\end{equation}

So the effective dispersion in time will go as $r^{3/4}$.

Stringing the factors together we get:

\begin{equation}
{\mathit{\sigma}}_{t}^{\left({B}\right){2}}=\frac{1}{\sqrt{\mathit{\mu}}}\sqrt{{a}_{0}{r}^{3}}
\end{equation}

With the average value of $r\approx a_{0}$, we match the results
for the GHO (eqn \ref{eq:gho-estimate-for-sigma-b}) if we take $\mu=1$.
So we take the dispersion in time of the atomic electron as:

\begin{equation}
{\mathit{\sigma}}_{t}^{\left({B}\right){2}}=\sqrt{{a}_{0}{r}^{3}}\label{eq:sigma-bold}
\end{equation}

Since this is the critical result of this analysis we will use a bold
$\bm{\sigma}$ for it:

\begin{equation}
{\mathbf{\bm{\sigma}}}_{r}^{2}=\sqrt{{a}_{0}{r}^{3}}\label{eq:bold-sigma}
\end{equation}

The square of the uncertainty in time is one-half this. Taking the
typical $r$ as the Bohr radius $.177as$ we get the uncertainty in
time as $.125as$ or about an eighth of an attosecond.

It has the right dimensions of course. Of the three powers of $r$,
one comes from the effects of the Coulomb potential $-\alpha\frac{1}{r}$,
two from the dispersion of the photon wave function in time. The $a_{0}$
is needed to the make the overall dimensions come out; it is the simplest
possible way to do this.

\subsection{Ground state}

Given an estimate for the effective potential as a function of $r$
it is straightforward to estimate the wave function at each value
of $r$.

\begin{equation}
{\psi}^{T}\mbox{ \ensuremath{\left({t,r}\right)}}={\mbox{ \ensuremath{\left({\mbox{{\large \ensuremath{\frac{1}{\pi{\mathbf{\bm{\sigma}}}_{r}^{2}}}}}}\right)}}}^{1/4}{e}^{-\frac{{t}_{r}^{2}}{2{\mathbf{\bm{\sigma}}}_{r}^{2}}}
\end{equation}

This is the ``local harmonic oscillator'' or LHO estimate. It represents
an imperfectly disentangled time part: it depends on the radius after
all. 

The easiest way to get the values of specific properties is to take
the value of the local wave functions at the most typical radius $r=a_{0}$:

\begin{equation}
{\psi}^{T}\mbox{ \ensuremath{\left({t,{a}_{0}}\right)}}={\mbox{ \ensuremath{\left({\mbox{{\large \ensuremath{\frac{1}{\pi a_{0}^{2}}}}}}\right)}}}^{1/4}{e}^{-\frac{{t}_{r}^{2}}{2{a}_{0}^{2}}}
\end{equation}

This matches the GHO earlier. 

This gives us some confidence that -- whether or not TQM is a correct
description of reality -- at least the treatment here is a reasonable
first order approximation of TQM.

The expected dispersion for hydrogen atoms appears to be below the
current threshold for detection. However the $r^{3/4}$ dependence
coupled with the $r\sim n^{2}$ dependence on the principle quantum
number $n$ gives ($\bm{\sigma}\sim n^{3/2}$). In the context of
Rydberg atoms, where $n$ can be 100 or greater, there are some striking
possibilities for experimental test and falsification. Discussed further
below.

There are a number of significant features of the LHO and of this
analysis:
\begin{enumerate}
\item The estimated dispersion goes to zero as the radius goes to zero.
This is expected: the closer to the nucleus the more tightly the electron
should be bound in time.
\item And contrariwise, the further from the nucleus, the more loosely bound
in time the electron is. Also expected.
\item The potential is attractive. If it were not, that would almost be
a refutation of TQM, as there could then be no stable dispersion in
time.
\item The FS/T and the QED approaches are in harmony. Given that the mechanism
in the FS/T is an upside down harmonic oscillator while the mechanism
in QED is a gradual increase in the off-shell component of the photon
wave function, this was by no means guaranteed.
\item The predicted effects are small enough that they are unlikely to have
been seen by chance.
\end{enumerate}
If any of these five points -- especially the last three -- were
not the case, we would have an argument against TQM. Their derivation
therefore strengthens the case for TQM. Only experiment can decide
whether TQM is correct or not of course, but such confirmatory arguments
strength the case for investing the effort needed to design and perform
the necessary experiments.

\subsection{Discussion}

\label{subsec:bound-discussion}

\paragraph{Extensions}

We have been very aggressive in our use of simplifications: $\tau\approx r,\kappa\approx\bar{\kappa,}n\to1$,
and so on. There are many ways to tighten these calculations up:
\begin{enumerate}
\item Make an explicit calculation of the effective potential.
\item Calculate the first and higher order perturbation corrections.
\item Switch to the center-of-mass (COM) frame so as to include the back
reaction of the electron on the nucleus.
\item Include the effects of the atomic number $Z$. For instance we might
expect this would reduce the $a_{0}\to\frac{a_{0}}{Z}$ in equation
\ref{eq:bold-sigma}.
\item Include spin, polarization.
\item Include relativistic elements.
\item Include the effects of the finite radius of the nucleus and of the
Lamb shift.
\end{enumerate}

\paragraph{TQM}

In this work, to maximize falsifiability, we have started with SQM
and then pushed out in the time direction. But a correct treatment
should start with time and space as equal partners. Several aspects
require particular attention.

\subparagraph{Virtual photons are real in TQM}

One of the curious aspects of standard treatments of QED is that mass
corrections and other UV loop problems are normally postponed till
later in the course. But the self-energy corrections due to, say,
the electron's photons are likely to play a larger role in the total
picture than the exchange photons. The self-energy photons do not
need to travel far to be absorbed, their source and destination are
the same place. But the exchange photons have to cross a full Bohr
radius to be absorbed. Logically the self-energy photons will be more
important to the electron's energy budget; logically they should be
treated first. The reasons for the usual order are 1) the UV divergences
were discovered later historically and 2) are harder to manage technically.
But the point remains.

To continue this line of thinking, once a photon has been emitted,
there is no way to tell if it is a self-energy or an exchange photon.
Photons -- like the anonymous heroes in some westerns -- have no
past and no future. The photon is emitted to the photon cloud and
absorbed from it. The distinction between self-energy and exchange
is meaningless.

Further there is no reason to have only one photon in the cloud at
a time. Photons are bosons and therefore naturally gregarious (unlike
those Western heroes). They are most appropriately treated as coherent
states and from a statistical thermodynamics viewpoint.

This implies that in TQM an atom has an associated photon cloud which
is an essential part of its description. The electron and nucleus
interact with the photon cloud but not directly with each other.

This also provides a natural way to track decoherence. If the atom
is surrounded by a sea of photons -- and even atoms in deep space
are surrounded by the cosmic microwave background -- that can be
treated as a additional contribution to the atom's own personal cloud.

\subparagraph{Variational approaches should look for maximum entropy rather than
minimum energy}

One perhaps slightly unnerving aspect of this treatment is how we
are treating the time part of the FS/T as an upside down harmonic
oscillator. What does this mean for the variational method of determining
the ground state? Are we to look for the highest energy on the time
side when we look for the lowest energy on the space side? To some
extent, yes.

But instead of looking for the lowest/highest energy we should look
for the maximum entropy when both time and space are included. This
meshes both approaches. 

When we are looking at only the space side, looking for minimum energy
and looking for maximum entropy are likely to give similar results.
If the electron and nucleus have the minimum energy then the rest
of the energy is available for the photon cloud. For the same available
energy, there will be in general be more photon states than fermion
states. (Consider for one thing very low energy photons.) So the maximum
total entropy will come from putting as much of the available energy
as possible in the photon cloud, taking it away from the fermions.

\subparagraph{Existing calculational approaches may need to be modified}

We expect existing approaches may require modification to apply here.
For instance, in TQM the usual loop diagrams are regularized organically,
by the associated GTF in time. In SQM renormalization is usually done
in powers of the coupling constant. But in TQM we will need to track
the effects of the regularizing GTF as well. To put this in standard
terms: the energy dispersion in TQM acts in parallel to the $\Lambda$
cutoff in SQM. But now the cutoff is a physical parameter -- it may
be large but it is not infinite. So we will need to track \emph{both}
the powers of the coupling constant and of the actual cutoff .

\section{Scattering}

\label{sec:Interactions}
\begin{quotation}
``The world changed from having the determinism of a clock to having
the contingency of a pinball machine.'' -- Heinz R. Pagels \cite{Pagels:1982aa}
\end{quotation}

\subsection{Interaction terms}

\label{subsec:Interaction-terms}

We will look here at only the simplest possible scattering problems,
in semi-classical approximation. By semi-classical we mean an approximation
that treats the electrons and nucleons as individual particles, but
the photons in terms of Fock space or coherent states (\cite{Sakurai:1967tl}).

We will use the direct product/disentangled approximation for the
4D wave function.

\paragraph{Incoming wave function}

In TQM the wave functions can be any normalizable wave functions.
But we will assume they are:
\begin{enumerate}
\item Direct products of a time and a space part. The space part will be
modeled by a plane wave defined by $\vec{k}$; the time part will
be modeled by a GTF,
\item Centered in frequency on the $\vec{k}$,
\item And centered in time on the clock time.
\end{enumerate}
As a result they will look something like:

\begin{equation}
{\varphi}_{\mathit{\tau}}\left({t}\right)\exp\left({\imath\vec{k}\cdot\vec{r}}\right)
\end{equation}

with the GTF:

\begin{equation}
{\varphi}_{\mathit{\tau}}\left({t}\right)\sim\exp\left({{-}\imath\omega_{\vec{k}}t-\frac{{\left({t_{\tau}}\right)}^{2}}{2{\mathit{\sigma}}_{t}^{2}}}\right)
\end{equation}

\paragraph{Interaction}

When the incoming GTF encounters the atom it interacts with the atomic
wave function in accordance with the interaction term:

\begin{equation}
{-}{e}\bar{\psi}\mbox{ \ensuremath{\left({p'}\right)}}{A}_{\tau}^{\nu}\mbox{ \ensuremath{\left({k}\right)}}{\gamma}_{\nu}\psi\mbox{ \ensuremath{\left({p}\right)}}
\end{equation}
The interaction terms in both SQM and TQM are formally identical.
The difference is the definition of the four vectors in the two cases:

\begin{equation}
\begin{array}{c}
{{p}^{S}{=}\left({{\mathit{\omega}}_{\vec{p}},\vec{p}}\right)}\\
{{p}^{T}{=}\left({w,\vec{p}}\right){,}\hspace{0.33em}{w}\approx{\mathit{\omega}}_{\vec{p}}}
\end{array}
\end{equation}

The polarization and spinors in TQM are essentially the same as those
in SQM, with the obvious change that any appearances of $E_{\vec{p}}$
in an SQM Dirac spinor will be replaced by the more general $E$ in
the TQM spinor.

We will assume we can take as given the SQM part of the interaction,
letting us focus on the time part. As in the discussion of interactions
earlier, the interaction will be associated with a delta function
in three momentum in SQM and in four momentum in TQM. 

We will look at the cases:
\begin{enumerate}
\item \emph{Absorption of a particle.} This can serve as a model for detection.
\item \emph{Emission of a particle.} This can serve as a model for creating
particles for subsequent use in various experiments. This lets us
calibrate the starting wave functions in TQM experiments.
\item \emph{Resonant scattering of a particle.} Where one particle is absorbed
and another emitted in such quick succession that characteristics
of the initial particle are carried through to the emitted particle.
\end{enumerate}
There are many other cases of course, but this set lets us cover the
basic principles.

A striking aspect of the analyses here is the absence of any quantum
jumps (\cite{Bohm:2010ab,Brun:1996bv,Cerf:1997ux,Minev:2019aa,Minev:2019ua,Schulman:2001hl,Zeh:1993wy,Zurek:2012aa,Zurek:2014aa}).
By running the FS/T step by step we see the wave function evolve smoothly
but rapidly during the interaction, then slowly but smoothly as it
decoheres to its final quasi-stationary state. We are replacing the
mysterious quantum jumps with comprehensible and continuous evolution
in clock time.

\subsection{Absorption of a photon}

\label{subsec:Absorption}

\begin{figure}
\includegraphics[scale=0.6]{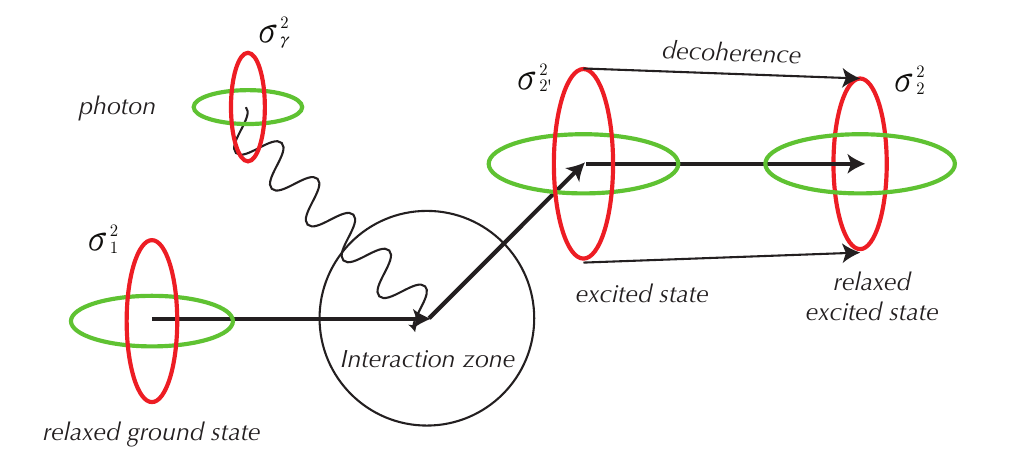}

\caption{\protect\label{fig:Absorption-of-a}Absorption of a photon}
\end{figure}

Let's suppose that the atom starts in its ground state $n=1$. It
has had time to relax so that the dispersion of the ground state energy
is given by $\sigma_{1}$.\footnote{In this section we are using the subscript on $\sigma$ to identify
the state.} We will take the incoming photon as having a dispersion in energy
of $\sigma_{\gamma}$. The analysis of the space part has given us
the amplitude to transition to, say, atomic state 2. 

To the current level of approximation the time and space parts factor.
The SQM interaction is given by:

\begin{equation}
\left\langle {\psi_{2}^{S}\left|{{e}^{{-}\imath\vec{k}\cdot\vec{x}}{\varepsilon}^{\left({\mathit{\alpha}}\right)}\cdot\vec{p}}\right|\psi_{1}^{S}}\right\rangle 
\end{equation}
and the TQM part by:

\begin{equation}
\left\langle {{\mathit{\psi}}_{2}^{T}\left({{E}_{2}}\right)\left|{{\varphi}^{T}\left({w}\right)}\right|{\mathit{\psi}}_{1}^{T}\left({{E}_{1}}\right)}\right\rangle 
\end{equation}

For the time part we integrate over the coordinate time giving as
usual a delta function in the coordinate energy:

\[
\mathit{\delta}\left({{E}_{2}{-}{E}_{1}{-}{w}}\right)
\]

To get the outgoing $\psi_{2}^{T}$ we integrate over the incoming
$\psi_{1}^{T}$ and photon wave function. Because of the delta function
this will convolute the two Gaussians. Because we taking a convolution
in energy we will get a Gaussian whose dispersion in energy is given
by:

\begin{equation}
\sigma_{2'}^{2}=\sigma_{1}^{2}+\sigma_{\gamma}^{2}
\end{equation}

where $\sigma_{2'}^{2}$ is the resulting dispersion in energy. If
we take $\sigma_{2}^{2}$ as the stable value of the dispersion of
state 2, we will expect that $\sigma_{2'}^{2}$ is not in general
equal to $\sigma_{2}^{2}$. Over time, however, the dispersion should
relax to the stable value:

\begin{equation}
{\sigma}_{2'}^{2}\rightarrow{\sigma}_{2}^{2}
\end{equation}

We will refer to the stable dispersion as the ``relaxed'' dispersion.
The mechanism for this we assume is some form of decoherence with
the environment of the atom (external decoherence) . While we can
model the transition to 2' explicitly, we would have to have a model
of the decoherence before we can be similarly explicit about the relaxation.
If we are dealing with an atom more complex than hydrogen, we would
also have decoherence (internal decoherence \cite{Braun:2006oq})
coming from interactions with the other atomic electrons. We expect
this will be too complex to compute in the general case.

\subsection{Emission of a photon}

\label{subsec:Emission}

\begin{figure}
\includegraphics[scale=0.6]{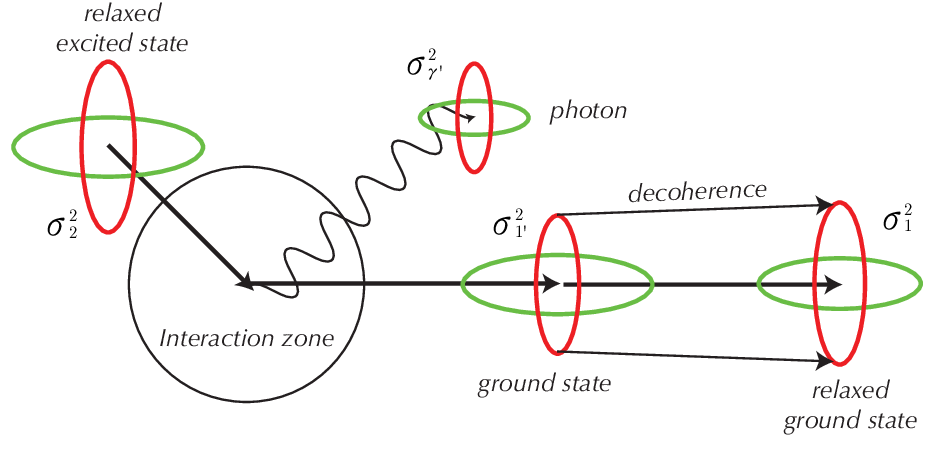}

\caption{\protect\label{fig:Emission-of-a}Emission of a photon}
\end{figure}

The analysis of emission is the inverse of the analysis for absorption.
We start with a state 2 with time dispersion $\sigma_{2}^{2}$. Using
the amplitudes already calculated for the SQM side, we have an amplitude
to transition to state 1, with relaxed dispersion $\sigma_{1'}^{2}$.
The outgoing photon will have dispersion $\sigma_{\gamma}$. 

Then we will have the dispersion in the ground state again in general,
but not the relaxed dispersion:

\begin{equation}
{\mathit{\sigma}}_{1'}^{2}{=}{\mathit{\sigma}}_{2}^{2}{-}{\mathit{\sigma}}_{\mathit{\gamma}}^{2}
\end{equation}

Note the minus sign. The ground state is getting the ``leftover''
dispersion from the photon.

Over time we expect the time dispersion of state 1 will relax to the
relaxed value:

\begin{equation}
{\sigma}_{1'}^{2}\rightarrow{\sigma}_{1}^{2}
\end{equation}

But this will happen after the photon has left. 

We can measure the properties of the photon. We will not in general
know the time of emission so time-of-arrival may not help with direct
measurement of the time dispersion. However we can look at diffraction
patterns created by the photon to measure its dispersion in energy,
which should be enough to get the dispersion in time. A statistical
samples worth of measurements of $\sigma_{\gamma}$ can let us calibrate
specific transitions as sources of photons for subsequent use.

\subsection{Resonant scattering of a photon}

\label{subsec:Scattering-of-photon}

\begin{figure}
\includegraphics[scale=0.55]{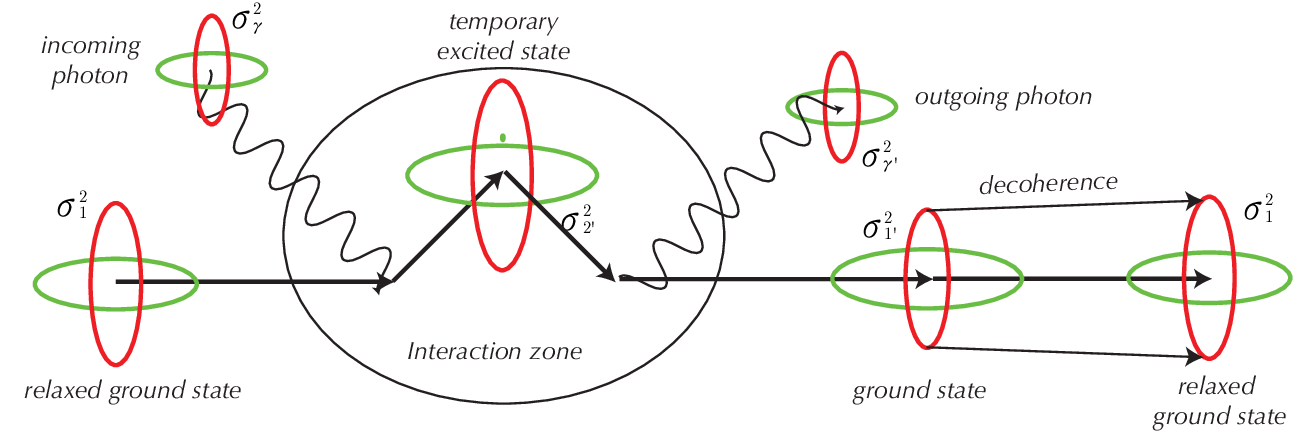}

\caption{\protect\label{fig:Resonant-scattering-of}Resonant scattering of
a photon}
\end{figure}

To complete this analysis we look at the case of a photon temporarily
absorbed by an atom but then promptly re-emitted. This is the sort
of thing that happens in refraction, for instance. 

It will be the sum of the previous two, but with little time to relax
in between. The dispersions will go as:

\begin{equation}
\begin{array}{c}
{{\mathit{\sigma}}_{2'}^{2}{=}\hspace{0.33em}{\mathit{\sigma}}_{1}^{2}{+}{\mathit{\sigma}}_{\mathit{\gamma}}^{2}}\\
{{\mathit{\sigma}}_{1'}^{2}{=}{\mathit{\sigma}}_{2'}^{2}{-}{\mathit{\sigma}}_{\mathit{\gamma}{'}}^{2}}
\end{array}
\end{equation}

If there is enough time between absorption and emission the intermediate
dispersion will evolve towards the relaxed dispersion for state 2:

\begin{equation}
{\sigma}_{2'}^{2}\rightarrow{\sigma}_{2}^{2}
\end{equation}

At that point we are past resonant scattering and into ordinary absorption-followed-by-emission.

In any case, examination of the time/energy dispersion of the emitted
photon may now tell us about the time/energy dispersion of the initial
photon. And since we will in general be able to control the time of
the initial absorption we can look more directly at the dispersion
of that photon.

\emph{Therefore we can get a controlled source of particles for use
as a starting point for TQM experiments.}

\section{Detection}

\label{sec:Detection}
\begin{quote}
“It is quite wrong to try founding a theory on observable magnitudes
alone. In reality the very opposite happens. It is the theory which
decides what we can observe.”
\begin{quotation}
-- Albert Einstein (as cited on p154 of Kay \cite{Kay:2024aa})
\end{quotation}
\end{quote}

\subsection{Time-of-arrival measurements}

\label{subsec:Time-of-arrival}

\begin{figure}
\includegraphics[scale=0.6]{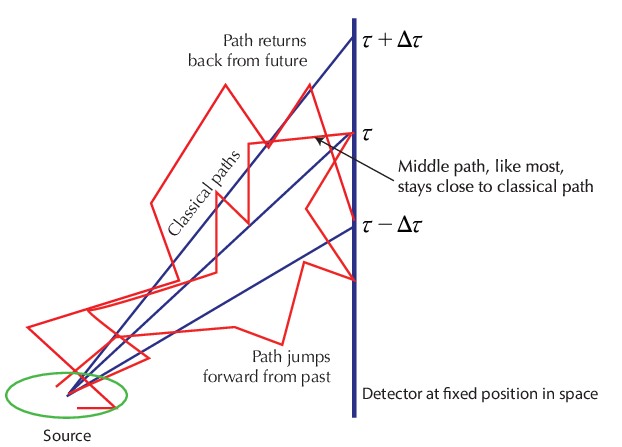}

\caption{\protect\label{fig:Detection-of-a}Detection of a particle}
\end{figure}

\begin{quotation}
The archetypal measurement for TQM is of time-of-arrival (TOA). If
TQM is correct, then we should see greater dispersion in the time
of arrival; if not, less. In TQM paths explore future and past just
as they do left/right, up/down, forward/back. More dispersion in time-of-arrival
is unavoidable. We explored the details of this in the paper \cite{Ashmead:2021aa}.

A critical element of the analysis is that the quantum paths can not
entirely be disentangled from either detector or source. They have
to overlap with the source to be emitted and with the detector to
absorbed, so they are necessarily entangled at each end. As a practical
matter, we normally treat the the source, detector, and paths as separate
elements, but their deeply entangled status has to be kept firmly
in mind throughout the analysis.
\end{quotation}

\subsection{Implications for measurement}

\label{subsec:Measurement}

\begin{figure}
\includegraphics[scale=0.6]{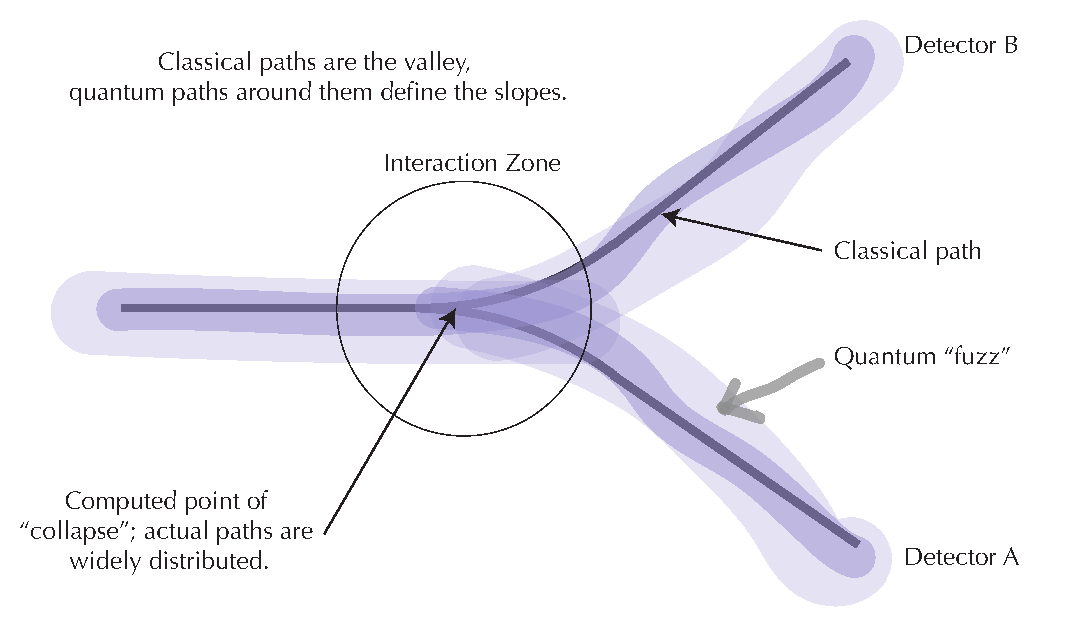}

\caption{\protect\label{fig:Classical-valley;-quantum}Classical valley; quantum
slopes}
\end{figure}

We start with two critical facts (both observations we originally
had from Feynman): 
\begin{enumerate}
\item Everything is made of atoms. At least everything we care about locally. 
\item Atoms do not exist classically. A classical charged particle in orbit
gives off Larmor radiation, thereby loses energy, and thereby spirals
into the nucleus. The time for this is extremely rapid \cite{Feynman:1965ah,Olsen:2017aa}.
\end{enumerate}
\emph{The implication is that there is no classical world, therefore
no transition from quantum to classical, and therefore no collapse.}
On the other hand, the collapse picture has given a good description
of what is going on. We can reconcile these two perspectives by asking
the question when is the classical picture ``good enough''? and
when do we need to resort to a fully quantum description? so the key
question is not: when/where does the collapse happen? but rather\emph{:
when should we use the quantum description and when the classical?}

The Feynman path integral approach has as one of its strengths a simple
and elegant way to address this question. The classical paths are
the river valleys; the quantum fluctuations are the slopes around
the valleys. Let's consider the above picture, which may be treated
as picture of a Stern-Gerlach experiment \cite{Gerlach:1922qv,Gerlach:1922sz,Gerlach:1922zu}.
The picture is of all the paths from start to finish. Most of them
go from start to top or start to bottom, with of course some just
wandering off and hitting a stray bit of laboratory apparatus.

If we ask a yes/no question: top or bottom, we sum over all the paths.
That's the answer. Binary, classical, clear. Because we asked a classical
question. But if we have some kind of continuous detector then we
will get a continuous fractional answer: near the middle of the top,
near the bottom of the bottom. This is a Bohr-like answer. We have
to consider not only the apparatus as a whole but also what question
are we asking?

\paragraph{When does the collapse happen?}

This picture gives us a way to say exactly when the ``collapse''
occurs: we analyze the sum over all paths, look at the valley defined
by the stationary phase method, identify the point where the initial
valley splits into two, declare that the location of the collapse.
It is not a physical event, it is a way to reduce to a number a complex
situation. (And a nice example of catastrophe theory in action \cite{Arnold:1986qd}). 

So it is only retrospectively that we can give a well-defined answer
to the question of when does the collapse happens: it happens when
the two center lines defining the valleys intersect, which will normally
be somewhere in the center of the interaction zone, but is free to
wander a bit, even perhaps to before the interaction zone. 

This is similar to the problem historians have: what was the \emph{critical}
event with respect to the ultimate outcome? This is often obscure
at the time; it is only after we have seen how events played out that
we can go back, draw some conveniently straight lines back to their
point of intersection, and put a pin on the ``critical moment''. 

\paragraph{Schrödinger's cat}

Let's apply this approach to a few of the standard questions in measurement.
First our friend the ``Schrödinger's cat'' \cite{Schrodinger:1935},
the \emph{locus classicus} of the quantum measurement problem. In
reality the cat is like the magician's assistant whose graceful moves
and striking gestures are intended to keep our attention away from
the place where the trick actually occurs. This is the place where
the radioactive atom does and does not emit a particle to trigger
the cascade of events that does and does not result in the cat's death. 

Inside that part of the apparatus, the sets of paths diverge, one
of the two river valleys leading to safety, the other alas not. The
analysis of the Stern-Gerlach experiment covers that. The cat remains,
throughout, a system which can be described in entirely classical
terms.

\paragraph{Wigner's friend}

The same kind of analysis may be applied to the problem of Wigner's
friend. Wigner is observing a quantum mechanical system; he describes
it in terms of a collapse of the wave function. But now Wigner's friend
is observing Wigner.\footnote{Or perhaps Wigner is looking at himself in a mirror.}
Is the collapse postponed until Wigner's friend makes his observation?
(Wigner reflects on the question in \cite{Wigner:1967pp}.)

Our answer here is: the system must ultimately be analyzed as a whole.
But in most practical cases, the presence of Wigner's friend will
not impact in a measurable way the various quantum probabilities being
calculated. We can usually position the collapse -- the point at
which we change languages -- to before Wigner. But if a stray laser
beam reflects off the friend's glasses and back into the apparatus,
then we may have to extend the quantum description to post-Wigner.

\begin{figure}
\includegraphics[scale=0.6]{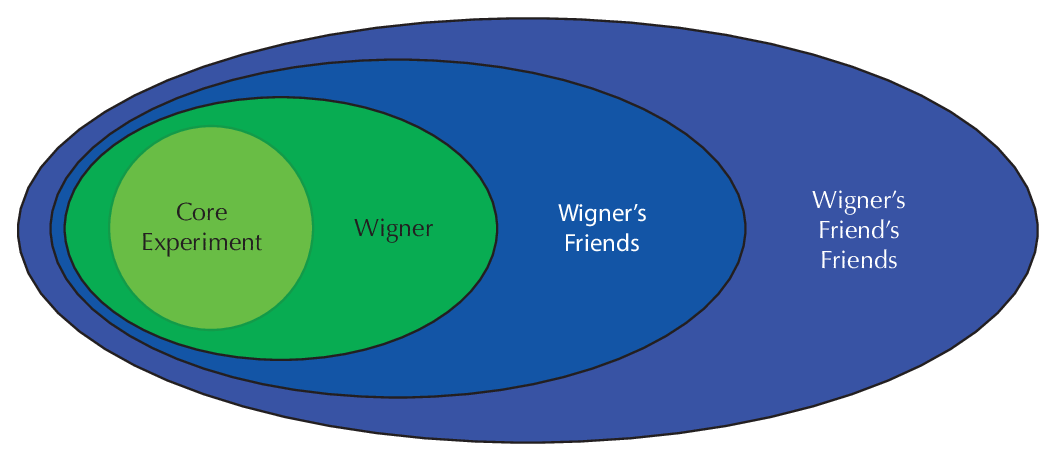}

\caption{\protect\label{fig:Wigner's-friends}Wigner's friends}
\end{figure}

\paragraph{Decoherence}

Of course even if the world is ultimately quantum, classical descriptions
do an excellent job of describing the bits of it we care about. 

The simplest and most general explanation we have seen of this is
Zeh's decoherence approach \cite{Zeh:1993wy,Zeh:1996aa,Zeh:2000sj,Zeh:2005ix,Zeh:2008fr,Zurek:2002ky,Zurek:2003oi,Zurek:2012aa,Zurek:2014aa,Zurek:2021ta}.
Decoherence clearly provides at least part of the answer \cite{Omnes:1994ao,Giulini:1996vp,Heiss:2002pd,Joos:2003gf,Schlosshauer:2007rr}.
But decoherence is not yet as quantitative as one might like (although
see Venugopalan, Qureshi, and Mishra \cite{Venugopalan:2018sf}). 

Decoherence is of course mutual. The system and the environment rub
against each other like two pieces of sand-paper, reciprocally smoothing
each other down, as a stream turns rocks into smooth pebbles, each
rock wearing the jagged bits off all the others.

\paragraph{Bound states as observers in miniature}

In general atoms are not that far from being classical objects: for
instance they tend to be found in specific states. We have seen in
this paper a candidate explanation of this. Small random ``insults''
tend to reinforce on-shell states; but interfere destructively with
off-shell states. Tannor \cite{Tannor:2007qz} argues that this mechanism
can help to explain the Bohr orbitals.

So what we are left with is that bound states will often provide a
useful model of a classical observer. The specific state of an atom
may be often treated as a proxy for a yes/no answer. But in some cases,
not.

\paragraph{One reality two languages}

So what we have is one reality; two languages. Coarse grained questions
will tend to be mostly simply addressed at the classical level; but
very finely grained questions may need a full path-integral or other
analysis. And in the most difficult cases, we may need to include
the wave function of the detector/observer as part of the analysis. 

Further two different observers, with different objectives or measuring
apparatus, might easily define ``close enough to classical'' differently.
To one the ``collapse'' takes place at one time; to the other, another.
Same situation; different definitions of when a classical description
is acceptable.

The problem goes from being a problem with the physics to a problem
of the choice of classical or quantum language, depending on the situation
and objectives. And of course pidgins are also possible, as semi-classical
descriptions.

\paragraph{Entanglement}
\begin{quote}
``Bohr embraced entanglement, seeing in it the roots of complementarity.
Einstein rejected entanglement as incompatible with the principle
of the spatial separability of systems, a principle that he thought
not only a necessary feature of any field theory like general relativity
but also a necessary condition for the very intelligibility of science.''
-- Don Howard \cite{Howard:2007aa}
\end{quote}
Entanglement in time plays a critical role in TQM. We need it to get
the integrals to converge slice-by-slice, to contain the UV divergences,
to define what we mean by a measurement. Entanglement in time is the
most important single expression of time as observable.

But Einstein's point is well-taken. Divide-and-conquer is perhaps
the most important single tool in analysis. How to reconcile with
omnipresent entanglement? We suggest that when we are dividing the
system into observed/observer or microscopic/macroscopic and so on,
we need to keep in mind several principles:
\begin{enumerate}
\item We should start with a view of the whole system, including the bits
before, after, and outside the conceptual box containing the experiment,
\item We may need to proceed by successive approximation,
\item We will need to use appropriate mathematics to map from the complex
realities of the quantum to the simplified, but hopefully not over-simplified
realities of the classical picture. For instance quantum information
analyses typically provide a map from the complex wave function to
the positive definite and bounded numbers of an information metric.
\end{enumerate}

\subsection{Possible experiments}

\label{sec:Experimental-tests}
\begin{quotation}
``In other words, we are trying to prove ourselves wrong as quickly
as possible, because only in that way can we find progress.'' --
Richard P. Feynman (p152 \cite{Feynman:1965jb}).
\end{quotation}

\paragraph{Falsifiability}

The major problem here is that $a_{0}$ is a small number: $53pm$
as a length, $.177as$ as a time. It sets the scale for hydrogen.
The $.177as$ is less than the shortest time measurement we are aware
of, about $1as$.

Hydrogen is of course the smallest atom. We are going to need a bigger
atom. We can try cesium, the biggest. This has a radius of $265pm$,
five times that of hydrogen. With the time dispersion scaling at $r^{3/4}$
this takes us up to $3as$.\footnote{Before accepting the estimate of $3as$ we would have to re-apply
the analysis for the cesium atom, to include the effects of $Z=55$.} And therefore (barely) to falsifiability. However since many of our
numbers are rough, we would like to have a margin, ideally a factor
of ten or better. Our target is therefore $10as$.

We can do this by working with Rydberg atoms. These have the mathematical
structure of hydrogen atoms but principal quantum numbers $n$ of
100 or more. Since the atomic radius scales as $n^{2}$, we have that
the time dispersion of Rydberg atoms should scale as $n^{3/2}$. Multiplying
the Bohr time by $100^{3/2}=1000$ we get $177as$ as the estimated
size. Since our target for falsifiability is $10as$, we have achieved
falsifiability.

\paragraph{General difficulties}

\label{par:General-difficulties}Beyond the size problem there are
a number of general difficulties.

The most important is that averaging will tend to hide the effects.
The archetypal effect is an increase in the dispersion of time-of-arrival
due to the presence of dispersion in time. But we cannot have a meaningful
time-of-arrival unless we have also the corresponding time-of-departure.
If the particles are part of a beam we will not, in general, have
that.

Further the classical paths are the same in TQM as in SQM. Many effects
depend primarily on the classical paths. For instance in the double
slit experiment, the number of wave lengths along each of the two
paths will be the same along the two paths for both TQM and SQM, so
the macroscopic interference pattern will be the same. 

And the customary use of plane waves or delta functions for the initial
wave function is suspect. We need normalizable starting wave functions.

We also need to formulate the criteria for falsified/confirmed appropriately.
We can look at this in roughly same way we do $\epsilon-\delta$ proofs
in calculus: 
\begin{enumerate}
\item Define the target accuracy, as by a $5\sigma$ test,
\item Assess initial dispersion in time needed to meet this,
\item Find experimental setup that should provide required dispersion,
\item Verify confirmation/disconfirmation by running experiment.
\end{enumerate}
Non-trivial. But at least there are a large number of possible experimental
setups.

\paragraph{Heisenberg uncertainty principle in time/energy}

\label{par:Heisenberg-uncertainty-principle}

\begin{figure}
\includegraphics[scale=0.5]{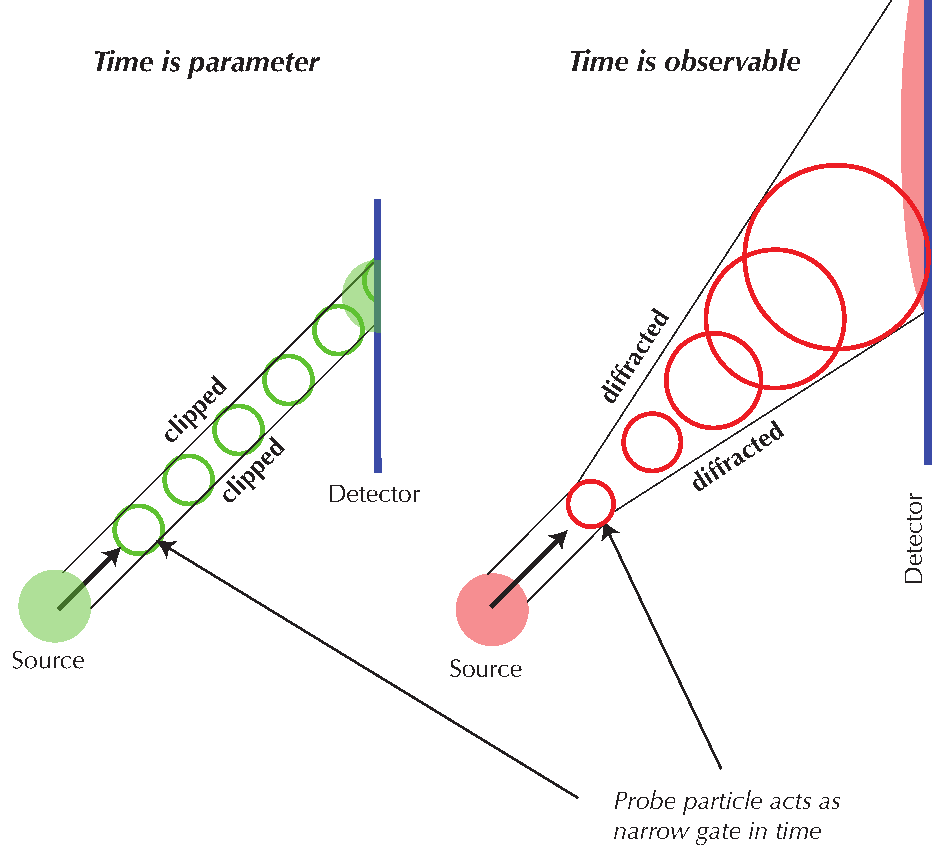}

\caption{\protect\label{fig:Heisenberg-uncertainty-principle}Heisenberg uncertainty
principle in energy and time}
\end{figure}

Currently in SQM, as Busch \cite{Busch-2001} puts it ``\ldots{} different
types of time energy uncertainty can indeed be deduced in specific
contexts, but \ldots{} there is no unique universal relation that could
stand on equal footing with the position-momentum uncertainty relation.''
See also \cite{Pauli:1980wd,Butterfield:2014aa,Olkhovsky:2008aa,Dirac:1958ty,Muga:2002ft,Muga:2008vv}. 

A key point in SQM analyses of the Heisenberg uncertainty principle
(HUP) in time/energy is the assumption of a lower bound to the energy.
In TQM this assumption is dropped. A full equivalence of the HUP in
time/energy to the HUP in space/momentum is assumed.\footnote{In TQM particles with negative frequency components are no more problematic
than particles with negative x-momentum (going left say) in SQM. The
overall direction in time is determined by the sign of the expectation
of the coordinate energy just as the overall direction along the x-axis
is determined by the sign of the expectation of the $p_{x}$ operator.}

The single slit experiment provides a direct test. For instance suppose
we have a test particle of calculated width in time being scattered
by a probe particle also of calculated width in time. If the probe
particle is built to have a narrow width in time it will act as a
single slit in time. If TQM is correct we will see diffraction in
time: the resulting time-of-arrival of the test particle can be made
arbitrarily wide by making the probe particle arbitrarily narrow.
This is of course the Heisenberg uncertainty principle in time/energy
in action. But if TQM is false, then the probe particle will act as
a gate and the resulting width in time-of-arrival will be arbitrarily
small. The effect will go in opposite directions depending on whether
TQM or SQM is correct. This is an \emph{experimentum crucis}: a decisive
or crucial experiment.

\paragraph{Direct effects of dispersion in time\protect\label{par:Many-possible-experiments}. }

In addition to the HUP in time/energy, there are a great number of
other candidate experiments. As noted, with TQM all quantum effects
in space are expected to have the corresponding effect ``in time''.
If we go to any compendium of foundational experiments in quantum
mechanics \cite{Auletta:2000vj,Ghose:1999zy,Lamoreaux:1992nh} we
can flip the time and space axis and see if we get something interesting.
This is how we found the single slit in time, for instance.

There are also some novel effects. Dispersion in time implies hysteresis
in time, as in the photon scattering experiments discussed earlier.

And some effects can be picked out statistically. For instance, if
there is dispersion in time particles will start to interact a bit
sooner than might otherwise have been the case. These imply forces
of anticipation. And leave off interacting later than they might have.
These imply forces of regret. In a statistical sample these effects
may be detectable.

In general high frequency/short duration phenomena -- ``chirps''
-- are most likely to be useful. The results of ``chirp-chirp''
scattering may be particularly helpful. (The proposed test of the
HUP is a chirp-chirp experiment.)

\paragraph{Shift from focus on time to focus on energy}

Normally we don't think of energy or velocity measurements as direct.
For instance, we usually measure velocity by measuring two successive
positions and dividing the distance by the time interval. But we can
measure the velocity more directly by using diffraction measurements.
The complete equivalence of time and space implies a similar complete
equivalence of energy and momentum. We can see ``diffraction in time''
as a direct measurement of energy.

\paragraph{Effects of shift from virtual to real particles}

Another class of experiments may be developed by taking seriously
the point that in TQM the virtual particles are promoted to real.
This means experimental questions may be asked of what had previously
been virtual particles. For instance, a probe particle may be aimed
at the ``inside'' of a Feynman diagram, looking to nose around and
see what is actually going on there. In some cases, the SQM calculation
may capture the associated effects. But in the most interesting cases,
a TQM analysis will be required.

Potentially interesting cases include, but are not limited to:
\begin{enumerate}
\item The photons that hold a bound state together must be treated as real.
In fact, a bound state should properly be seen as not a nucleus and
a particle exchanging photons, but rather as each particle absorbing
from and emitting to a ``bound photon cloud'': with photons and
particles in an elaborate dance to create the bound state.
\item In a circuit diagram, the fields that the charges and currents interact
with should be seen as ensembles of photons. One could do this in
any case, but TQM becomes particularly relevant when looking at time
dependent elements as memristors \cite{Chua:1971ty,Omid-Kavehci:2009aa,Pfeiffer:2015fk,Ventra:2022aa}
or at hysteresis effects.
\end{enumerate}

\paragraph{Already seen but not observed}

Because small and not looked for the effects of TQM may have already
been seen but not, as Mr. Holmes would put it, observed.\footnote{``You see but you do not observe. The distinction is clear.'' --
Mr. Sherlock Holmes in \emph{A Study in Scarlet} \cite{Doyle:1888qj}} Reexamination of existing data sets may be useful.

\paragraph{Higher order, subtle, and esoteric effects}

There are higher order effects which may be of interest as well. For
instance, the calculation of loop diagrams may need to include the
effective cutoffs provided by the dispersion in time of the initial
wave functions \eqref{par:Ultraviolet-(UV)-divergences}. The value
of $c$, the speed-of-light. becomes subject to small quantum fluctuations.
And while we have been focused on the time dispersion of the outer
orbitals, really it is the entire wave function of the atom - including
all electrons and their attendant photon cloud -- which is expected
to have dispersion in time.

\paragraph{Need for AI and automated analysis }

One of the problems in devising appropriate experiments is that we,
as humans, have the ``clock time/parameter'' picture of time deeply
wired in our conceptual frameworks.\footnote{To the extent that disorganization of the time sense can be associated
with mental illness, as schizophrenia. \cite{Amadeo:2022ab,Friedman:1990lq,Block:1990es,Block:1990ii}} It might be useful to program AIs to do systematic searches of possible
experiments, both because of the large number of possible experiments
and because our own hard-wired instincts may leave us blind to some
interesting possibilities.

\paragraph{Maximizing the size of the effects}

Scattering from resonance cavities or time crystals \cite{Hannaford:2022aa,Autti:2022aa,Khemani:2019aa,Sacha:2018aa,Sacha:2018tm,Shapere:2012wz,Wilczek:2012uq}
may provide a way to amplify otherwise small effects. We note that
Rydberg atoms are frequently used in the study of time crystals (\cite{Fan:2020aa,Yu:2018eu,Gambetta:2018uq,Nieto:1996vj}).
For instance, we could send a beam of Rydberg atoms through a time
crystal, also made of Rydberg atoms, looking for diffraction patterns.

\emph{The combination of the much larger size of the basic effect
in Rydberg atoms and the resonance amplification from time crystals
may make such setups ideal for the study of dispersion in time.}

\subsection{Practical uses}

\label{sec:Uses}

It is of course only one step from an interesting experiment to a
useful application. 

\paragraph{Uses assuming the hypothesis is still tentative or even disconfirmed}

In some cases the simple addition of the phrase ``in time'' to existing
phenomena can suggest new possibilities. This is how Wilczek came
up with the idea for time crystals. He asked: we have crystals in
space, what would it mean to have 'crystals in time'? \cite{Wilczek:2019aa,Wilczek:2012uq}. 

The idea of ``virtual work'' has been useful in mechanics. Treating
``virtual particles'' as real (even if they are not) may also suggest
useful insights. Atoms themselves started life as a ``useful hypothesis''.

\paragraph{Uses on confirmation}

There are many areas with potentially interesting applications:
\begin{enumerate}
\item \emph{Circuit element design:} Dispersion in time may offer opportunities
to help with design of time dependent circuits, as memristors.
\item \emph{Communication/Quantum networks:} An additional dimension to
play with means greater bandwidth. We could use the ``time channel''
for fingerprinting, covert communications, and so on.
\item \emph{Direct measurement of energy/velocity:} As discussed above.
\item \emph{High speed sensing: }New ways to make high speed sensors.
\item \emph{Neural networks}: Biologic neurons can sense change in time:
they can be alerted by one set of inputs, but then delay firing until
the inputs are repeated. Using time dispersion to help add this facility
to software neurons may help improve their performance/energy use
ratio.
\item \emph{Quantum computing}: Time dispersion gives an additional channel
to help with stabilization, and additional possibilities for parallelizing
computations.
\item \emph{Fusion research}: Rydberg atoms have found applications in commercial
fusion research \cite{EFS:2015aa}. The expected large dispersion
in time of Rydberg atoms may have implications for the control of
the associated fusion reactions. Given that control is the essential
problem in developing fusion, this may be of interest.
\item \emph{Security applications}: The extension of wave functions to time
gives a whole new dimension to work with, to store, hide, or probe
information.
\end{enumerate}
There are some potential cross applications among these cases of course,
for instance improved memristors have possible applications to neural
nets, photonics, and other areas\cite{Sanz:2018aa,Sun:2021aa,Domaradzki:2020aa}. 

\paragraph{Uses in either case}

We expect some interesting mathematical questions will be raised by
TQM. 

As an example, Kåhre \cite{Kahre:2002xj}'s Law of Diminishing Information
acts as a de facto measure of disentanglement. The law states that
to be considered as a measure of entropy a measure must show the relative
entropy between two systems as zero when they are uncorrelated (disentangled).
Conformant entropy measures include both the Shannon and von Neumann
measures, as well as many others. Conformant measures provide a natural
way to classify systems as entangled/disentangled. Since the definition
of entropy is at the core of modern communications and computation,
Kåhre's approach should provide an interesting line of attack on problems
in these areas.

\section{Discussion}

\label{sec:Discussion}
\begin{quotation}
\textit{\emph{``Look, I don't care what your theory of time is. Just
give me something I can prove wrong.}}'' -- Nathan Gisin \cite{Gisin:2009aa}
\end{quotation}
Our objective here has been to develop the hypothesis that time is
properly treated as an observable in a way that is:
\begin{enumerate}
\item Consistent with known results,
\item Manifestly covariant,
\item And falsifiable with current technology. 
\end{enumerate}
The basic strategy has been to start with the Feynman path integral
approach, promote the paths from 3D to 4D, but leave everything else
unchanged. The key turning point was recovering the Feynman-Stückelberg
equation but with the time parameter identified as the clock time.
This immediately gave us both a way to describe experiments in the
non-relativistic domain and a starting point to build the Feynman
diagrams in the QED domain. 

The extension of these two cases to the bound case was not entirely
straightforward. We worked by using a hybrid method, working up from
the FS/T and down from QED, meeting in the middle. In the case of
hydrogen atom, the resulting estimate is only .$177as$, the time
a photon takes to get from nucleus to atomic electron. This may be
too small to see with current technology. 

However since the time dispersion should scale as the principal quantum
number $n$ to the $3/2$, the use of Rydberg atoms, with $n\sim100$,
should let us increase the time dispersion by a factor of a thousand,
from $.177as\to177as$. 

Since the minimum size for detection is $1as$ and since we want to
allow a factor of $10$ margin for errors in the estimates and the
like we have a target of $10as$. $177as$ is well past this; we have
therefore achieved falsifiability.

What are the odds of a positive result? We can put this question on
a more philosophically sound basis by applying a set of criteria developed
by Schick and Vaughan \cite{Schick-Jr.:1995je}. They propose five
questions: \emph{is the hypothesis in question}:
\begin{enumerate}
\item \emph{Consistent with established results?} We have argued that outside
the ``chirp'' zone (high frequency/short lived phenomena), SQM and
TQM are roughly equivalent.
\item \emph{Reasonably simple?} TQM is definitely more complex from a calculational
point of view; witness this paper. But it is conceptually simpler:
time and space are handled in a fully covariant way, the UV divergences
do not appear, there is no need for quantum jumps, and the transition
from quantum to classical becomes a problem in choice of language,
not requiring new forces or infinite numbers of universes.
\item \emph{Broad in scope?} TQM approximates SQM as long times and low
energies so includes SQM as a subset. TQM itself is at its best at
short times and high energies, therefore well-suited to discussions
of the strong force, quantum gravity, the Big Bang and times before
the Big Bang. 
\item \emph{Testable?} until the last decade or so, the response to this
question would have to have been a regretful no. But with the extraordinary
progress in attosecond physics, it should be only a matter of time
before TQM is put to a direct test, ideally several of them.
\item \emph{Fruitful?} As noted, there are many possible experimental tests.
And the admitted difficulties may themselves be productive of new
technology.
\end{enumerate}
We make two further arguments:

From Gell-Mann's totalitarian principle \cite{Gell-Mann:1956uw}:
`\emph{Anything not forbidden is compulsory}.' From the arguments
above, the extension of quantum mechanics to include the time dimension
is not forbidden.

And finally from beauty: in special and general relativity, the simultaneous
treatment of time and space has, as Minkowski famously noted, a deep
beauty.
\begin{quote}
``Henceforth space by itself, and time by itself, are doomed to fade
away into mere shadows, and only a kind of union of the two will preserve
an independent reality.'' -- Minkowski \cite{Minkowski:1908yi}
\end{quote}
But in standard quantum mechanics the merger is incomplete: time is
left to trudge forward at a steady pace, the wave function fluctuating
all around it in space but never in time. How can this discordant
behavior be fully consistent with relativity?

We propose that in quantum mechanics time and space should be treated
as complementary, opposite parts of the same dance. The allowance
of some off-shell extension in time allows for a bit of syncopation:
at one point the space side taking the lead, at another point time.
But they are complementary, both part of the same dance, partners
keeping time together.

\section*{Acknowledgments}

\label{sec:Acknowledgments}

I thank my long time friend Jonathan Smith for invaluable encouragement,
guidance, and practical assistance.

I thank Ferne Cohen Welch for extraordinary moral and practical support.

I thank Martin Land, L. P. Horwitz, James O'Brien, Tepper Gill, Petr
Jizba, Andras Kovacs, Matthew Trump, and the other organizers of the
International Association for Relativistic Dynamics (IARD) 2018, 2020,
2022, and 2024 Conferences for encouragement, useful discussions,
and for hosting talks on the papers in this series in the IARD conference
series. I also thank my fellow participants in the 2024 conference
-- especially Mayeul Arminjon, Bei-Loc Hu, Jussi Lindgren, Bruce
Mainland, Theophilos Maltezopoulos, Anatolij Prykarpatski, and others
-- for many excellent questions and discussions.

I thank the organizers of several QUIST, DARPA, Perimeter Institute
conferences I've attended and the very much on topic conferences \textit{Quantum
Time} in Pittsburgh in 2014 and \textsl{Time and Quantum Gravity}
in San Diego in 2015.

I thank Robert Broberg for some very energetic and suggestive discussions.
I thank Grigory Volovik for a very interesting discussion of Time
Crystals and a tour of his Low Temperature Laboratory at Aalto University.
I thank Steven Libby for several useful conversations and in particular
for insisting on the extension of the original ideas to the high energy
limit and therefore to QED. I thank Larry Sorensen for many helpful
references. I thank Ashley Fidler for helpful references to the attosecond
physics literature. I thank Avi Marchewka for an interesting and instructive
conversation about various approaches to time-of-arrival measurements.
I thank Hou Yau for an interesting discussion of variations on these
themes and for directing my attention to his paper \cite{Yau:2021ab}.
I thank Asher Yahalom for insisting that a better explanation of the
clock time than simply what clocks measure was required. I thank Thomas
Cember for useful clarifications of several points in the argument.
I thank the reviewer who drew my attention to Horwitz's work on gravity
\cite{Horwitz:2018aa}. I think Danko Georgiev of the journal Quanta
for very practical suggestions and advice. I thank Y. S. Kim for organizing
the invaluable Feynman Festivals, for several conversations, and for
general encouragement and good advice.

I thank Catherine Asaro, Julian Barbour, Gary Bowson, Howard Brandt,
Daniel Brown, Ron Bushyager, John G. Cramer, J. Ferret, Robert Forward,
Fred Herz, J. Peřina, Linda Kalb, A. Khrennikov, David Kratz, Andy
Love, Walt Mankowski, O. Maroney, John Myers, Paul Nahin, Marilyn
Noz, R. Penrose, Stewart Personick, V. Petkov, H. Price, Matt Riesen,
Terry Roberts, J. H. Samson, Lee Smolin, L. Skála, Arthur Tansky,
R. Tumulka, Joan Vaccaro, L. Vaidman, A. Vourdas, H. Yadsan-Appleby,
S. Weinstein, and Anton Zeilinger for helpful conversations over the
years.

And none of the above are in any way responsible for any errors of
commission or omission in this work.

\appendix

\section{Gaussian test functions}

\label{sec:=000020Gaussian=000020test=000020functions}

By Gaussian test functions (GTFs) we mean functions of the general
form:

\begin{equation}
{\varphi_{0}}\left(x\right)=\sqrt[4]{{\frac{1}{{\pi{\sigma^{2}}}}}}{e^{\imath{p_{0}}x-\frac{{{\left({x-{x_{0}}}\right)}^{2}}}{{2{\sigma^{2}}}}}}\label{eq:gtf-time}
\end{equation}

We generally take them as normalized to one. We refer to $\sigma$
as the dispersion. The uncertainty in the associated dimension is
given by the dispersion divided by $\sqrt{2}$:

\begin{equation}
\Delta x\equiv\sqrt{\left\langle {{\left({x-\left\langle x\right\rangle }\right)}^{2}}\right\rangle }=\frac{{\sigma}}{{\sqrt{2}}}
\end{equation}
We can get a rough approximation of a typical normalizable wave function
by using the GTF with the same uncertainty:

\begin{equation}
\varphi_{0}\left(x\right)\equiv\sqrt[4]{{\frac{1}{{2\pi{{\left({\Delta x}\right)}^{2}}}}}}{e^{\imath\left\langle {p_{0}}\right\rangle x-\frac{{{\left({x-\left\langle {x_{0}}\right\rangle }\right)}^{2}}}{4{{{\left({\Delta x}\right)}^{2}}}}}}
\end{equation}

Using Morlet Wavelet Analysis (MWA) (see \cite{Ashmead:2012kx}) we
can write an arbitrary normalizable wave function as a sum over Morlet
wavelets. Since each Morlet wavelet is itself comprised of a pair
of GTFs, any normalizable wave function may be written as a sum over
GTFs. GTFs are therefore both typical and complete. We can therefore
achieve both generality and physical reasonableness by using a basis
of Morlet wavelets \cite{Morlet:1982mw,Chui:1992be,Meyer:1992hl,Kaiser:1994ph,Berg:1999mo,Bratteli:2002ar,Addison:2002jk,Visser:2003ga,Antoine:2004rm,Ashmead:2012kx}.

\label{subsec:gtf-In-time-energy}

Here we are most concerned with GTFs in time and energy. In time:

\begin{equation}
{{\varphi}_{0}}\left(t\right)\equiv\sqrt[4]{{\frac{1}{{\pi\sigma_{t}^{2}}}}}{e^{-\imath{E_{0}}t-\frac{{{\left({t-{t_{0}}}\right)}^{2}}}{{2\sigma_{t}^{2}}}}}\label{eq:time-WF}
\end{equation}

with Fourier transform into energy:
\begin{equation}
{{\varphi}_{0}}\left(E\right)\equiv\sqrt[4]{{\frac{1}{{\pi\sigma_{E}^{2}}}}}{e^{\imath\left({E-{E_{0}}}\right){t_{0}}-\frac{{{\left({E-{E_{0}}}\right)}^{2}}}{{2\sigma_{E}^{2}}}}}\label{eq:gtf-energy}
\end{equation}
where ${\sigma_{E}}=\frac{1}{{\sigma_{t}}}$. 

We are using standard conventions for the Fourier transform:

\begin{equation}
\begin{array}{c}
{{f}\left({t}\right){=}\frac{1}{\sqrt{{2}\mathit{\pi}}}\int{{dw}\exp\left({{-}\imath{wt}}\right)}\hat{f}\left({w}\right)}\\
{\hat{f}\left({w}\right){=}\frac{1}{\sqrt{{2}\mathit{\pi}}}\int{{d}{t}\exp\left({\imath{wt}}\right)}{f}\left({t}\right)}
\end{array}
\end{equation}

Similarly we have for the space and momentum forms:

\begin{equation}
{\varphi}_{0}\mbox{ \ensuremath{\left({x}\right)}}\equiv\sqrt[4]{\frac{1}{{\pi\sigma}_{x}^{2}}}{e}^{\imath{p}_{0}{x}{-}\frac{{\mbox{ \ensuremath{\left({{x}{-}{x}_{0}}\right)}}}^{2}}{2{\sigma}_{x}^{2}}}
\end{equation}

with Fourier transform into momentum:

\begin{equation}
{\varphi}_{0}\mbox{ \ensuremath{\left({{p}_{x}}\right)}}\equiv\sqrt[4]{\frac{1}{\pi{\sigma}_{{p}_{x}}^{2}}}{e}^{{-}\imath\mbox{ \ensuremath{\left({{p}{-}{p}_{0}}\right)}}{x}_{0}{-}\frac{{\mbox{ \ensuremath{\left({{p}_{x}{-}{p}_{{x}_{0}}}\right)}}}^{2}}{2{\sigma}_{{p}_{x}}^{2}}}
\end{equation}

For the space/momentum Fourier transforms we use the opposite sign
for the argument of the exponential. This simplifies the Fourier transforms
of four-vectors since we can then write:

\begin{equation}
\begin{array}{c}
{{f}\left({x}\right){=}\frac{1}{{\sqrt{{2}\mathit{\pi}}}^{4}}\int{{d}^{4}{p}\exp\left({{-}\imath{px}}\right)}\hat{f}\left({p}\right)}\\
{\hat{f}\left({p}\right){=}\frac{1}{{\sqrt{{2}\mathit{\pi}}}^{4}}\int{{d}^{4}{x}\exp\left({\imath{px}}\right)}{f}\left({x}\right)}
\end{array}
\end{equation}

rather than showing the time and space parts separately.

\section{Solution of free equation for massive particles}

\label{Solution=000020of=000020free=000020equation=000020for=000020massive=000020particles}

We start with the non-relativistic free FS/T:

\begin{equation}
{2}{m}\imath\frac{\partial{\psi}_{\tau}}{\partial\tau}{=}{-}\left({{E}^{2}{-}{\vec{p}}^{2}}\right){\psi}_{\tau}
\end{equation}

The kernel is defined by:

\begin{equation}
\left({\imath\frac{\partial}{\partial\mathit{\tau}}{+}\frac{{E}^{2}{-}{\vec{p}}^{2}}{2m}}\right){K}_{\mathit{\tau}}\left({x;x'}\right){=}{i}{\mathit{\delta}}^{4}\left({{x}{-}{x}{'}}\right)\delta\left(\tau-\tau'\right)
\end{equation}

We can solve by inspection in momentum space:

\begin{equation}
{K}_{\tau}\mbox{\ensuremath{\left({p;p'}\right)}}{=}\exp\mbox{\ensuremath{\left({{-}\imath{\varpi}_{p}\tau}\right)}}{\delta}^{4}\mbox{\ensuremath{\left({{p}{-}{p}{'}}\right)\mathit{\theta}\left({\mathit{\tau}}\right)}}\label{eq:gtf-kernel-tqm-momentum-space}
\end{equation}

with:
\[
{\varpi}_{p}\equiv{-}\frac{{E}^{2}{-}{\vec{p}}^{2}{-}{m}^{2}}{2m}
\]

The Fourier transform is:

\begin{equation}
{K_{\tau}}\left({x;x'}\right)=-\imath\frac{{m^{2}}}{{4{\pi^{2}}{\tau^{2}}}}{e^{-\frac{{\imath m}}{{2\tau}}{{\left({x-x'}\right)}^{2}}-\imath\frac{m}{2}\tau}}\mathit{\theta}\left({\mathit{\tau}}\right)\label{eq:4D-kernel-in-coordinate-space}
\end{equation}

This matches the non-relativistic kernel found in introductory quantum
mechanics textbooks except that it now includes paths in time. This
is the product of a time kernel by the usual space kernel (which is
developed in Merzbacher and many other references \cite{Feynman:1948,Feynman:1949sp,Feynman:1949uy,Feynman:2005ng,Feynman:2010bt,Grosche:1998uj,Huang:1998bd,Kleinert:1990qa,Kleinert:2009hw,Marchewka:2000ys,Rivers:1987ma,Schulman:1981um,Swanson:1992ju,Zinn-Justin:2005nx}):

The time part of this is the center of interest here:

\begin{equation}
\text{\ensuremath{{K}_{\tau}}\mbox{ \ensuremath{\left({{t}'';{t}\prime}\right)}}{=}\ensuremath{\sqrt{{-}\imath\frac{m}{{2}\pi\tau}}{e}^{{-}\imath{m}\frac{{\mbox{ \ensuremath{\left({{t}''{-}{t}'}\right)}}}^{2}}{{2}\tau}}}}\label{eq:kernel-in-time}
\end{equation}

We apply this to $\varphi_{0}$ to get:

\begin{equation}
\varphi_{\tau}\left(t\right)=\sqrt[4]{\frac{1}{\pi\sigma_{t}^{2}}}\sqrt{\frac{1}{f_{\tau}^{\left(t\right)}}}e^{-\imath E_{0}t+\imath\frac{E_{0}^{2}}{2m}\tau-\frac{1}{2\sigma_{t}^{2}f_{\tau}^{\left(t\right)}}\left(t-t_{0}-\frac{E_{0}}{m}\tau\right)^{2}}\label{eq:wf-time-of-tau}
\end{equation}

with the side definitions: 
\[
{\sigma_{E}}=\frac{1}{{\sigma_{t}}}
\]
\[
f_{\tau}^{\left(t\right)}\equiv1-\imath\frac{\tau}{m\sigma_{t}^{2}}
\]

The corresponding probability density is:

\label{probability-density-in-time}

\begin{equation}
{\rho_{\tau}^{T}}\left(t\right)=\sqrt{\frac{1}{{\pi\sigma_{t}^{2}\left({1+\frac{{\tau^{2}}}{{{E_{0}^{2}}\sigma_{t}^{4}}}}\right)}}}\exp\left({-\frac{{{\left({t-\left\langle t_{\tau}\right\rangle }\right)}^{2}}}{{\sigma_{t}^{2}\left({1+\frac{{\tau^{2}}}{{{E_{0}^{2}}\sigma_{t}^{4}}}}\right)}}}\right)
\end{equation}

With $t_{\tau},$ the relative time, is defined as:

\[
\left\langle t_{\tau}\right\rangle \equiv t_{0}-\frac{E_{0}}{m}\tau
\]

At non-relativistic energies $E_{0}\approx m$ so that $t_{\tau}\approx t_{0}-\tau$,
which is the form we use in the text.

We can extend the non-relativistic approach to high energy beams if
the beams are reasonably well-collimated in velocity, i.e. $\frac{\Delta{v}}{v}\ll{1}$.

With this assumption it is legitimate to replace the mass in the denominator
within the exponential with the average energy plus a correction:

\begin{equation}
{m}\rightarrow E,E\to\left|{E}\right|+{\delta E}
\end{equation}

Expanded, the denominator gives a series:

\[
\frac{1}{E}\rightarrow\frac{1}{\bar{E}}{-}\frac{\delta{E}}{{\bar{E}}^{2}}{+}\frac{{\mbox{ \ensuremath{\left({\delta{E}}\right)}}}^{2}}{{\bar{E}}^{3}}{-}
\]

so that as long as the beam is well-collimated, we may keep just the
first two powers in $\delta E$ and thereby approximate the wave function
by a Gaussian: 

\begin{equation}
{K}_{\tau}\mbox{ \ensuremath{\left({p;p'}\right)}}\approx\exp\mbox{ \ensuremath{\left({\imath\left({\frac{{\mbox{ \ensuremath{\bar{E}}}}^{2}{-}{m}^{2}}{2\mbox{ \ensuremath{\bar{E}}}}{+}\mathit{\delta}{E}\left({\frac{{\mbox{ \ensuremath{\bar{E}}}}^{2}{+}{m}^{2}}{2\mbox{ \ensuremath{\bar{E}}}}}\right){-}{\left({\mathit{\delta}{E}}\right)}^{2}\frac{{m}^{2}}{2{\mbox{ \ensuremath{\bar{E}}}}^{3}}}\right)\tau}\right)}}{\delta}^{4}\mbox{ \ensuremath{\left({{p}{-}{p}{'}}\right)}}\theta\mbox{ \ensuremath{\left({\tau}\right)}}
\end{equation}

\section{Machian approach to clock time}

\label{Machian=000020approach=000020to=000020clock=000020time}

Whether the clock time term on the left of the Feynman/Stueckelberg
equation should go as $m\imath\frac{\partial}{\partial\mathit{\tau}}$
or $E\imath\frac{\partial}{\partial\mathit{\tau}}$ is key in the
extension of the single particle results to general Feynman diagrams.
We use a Machian approach to resolve this (we are reprising an argument
from \cite{Ashmead:2023aa}).

In the spirit of Mach we assume that the Klein-Gordon (KG) equation
for a specific particle is given not absolutely but relative to the
rest of the universe. So we make the replacement $p\to p-\mathcal{P}$
(where $\mathcal{P}$is the four vector for the rest of the universe).
We define $\mathcal{Q}\equiv p-\mathcal{P}$.

Then the Klein-Gordon equation becomes:

\begin{equation}
{\mathcal{Q}^{2}}\psi={m^{2}}\psi\label{eq:vacuum=000020KG=000020equation}
\end{equation}

Consider the wave function $\Psi$ of ``the rest of the universe''
(or perhaps just the local laboratory). We assume the particles in
$\Psi$ obey the KG equation. Then we have:

\begin{equation}
\left({{\mathcal{P}^{2}}-{\mathcal{M}^{2}}}\right)\left|{\Psi}\right\rangle =0
\end{equation}

where $\mathcal{P}$ and $\mathcal{M}$ are to be understood as averages
over large volumes of particles.

The total wave function is given by the direct product of laboratory
and particle: $\left.{\left|{\Psi}\right.}\right\rangle \left.{\left|{\mathit{\psi}}\right.}\right\rangle $.
It obeys the equation:

\begin{equation}
\left({2{\mathcal{P}^{2}}-2p\mathcal{P}+{p^{2}}-{m^{2}}-{\mathcal{M}^{2}}}\right)\left|\left.{\left|{\Psi}\right.}\right\rangle \psi\right\rangle =0
\end{equation}

We can replace the $\mathcal{P}^{2}\to{\mathcal{M}^{2}}$, leaving
a constant factor, which is not physically significant.

Applying $\left.{\left\langle {\Psi}\right.}\right|$ on the left
we get:

\begin{equation}
{-}{2}{p}{\mathcal{P}}\mathit{\psi}{=}{-}\left({{p}^{2}{-}{m}^{2}}\right)\mathit{\psi}
\end{equation}

We take $\mathcal{X}$ as the coordinates corresponding to $\mathcal{P}$.
We have:

\begin{equation}
{\mathcal{P}}_{0}\equiv\imath\frac{\partial}{\partial{\mathcal{X}}_{0}}
\end{equation}

The sense of the time coordinate of the particle is the same of the
sense of the laboratory clock, both generally positive. But the sense
of the time coordinate of the particle should be opposite that of
the time coordinate of $\Psi$ (since $\mathcal{Q}_{0}=p_{0}-\mathcal{P}_{0}$)
so we take:

\begin{equation}
\frac{\partial}{{\partial{\mathcal{{X}}_{0}}}}=-\frac{\partial}{{\partial\tau}}\label{eq:vacuum=000020kicks=000020like=000020frog}
\end{equation}
And we thereby recover the FS/T (in the rest frame of $\Psi$):

\begin{equation}
-2E\imath\frac{\partial}{{\partial\tau}}={p^{2}}-{m^{2}}\label{eq:seqn=000020for=000020particle=000020in=000020a=000020vacuum}
\end{equation}

\section{Entropic estimates of initial wave functions}

\label{sec:Entropic-estimate}

\paragraph{Estimate from maximum entropy with constraints}

\label{subsec:Estimate-from-maximum-entropy}

Calculations in TQM typically require we specify the initial wave
function. If we do not have more specific information, then we can
use available constraints and the Lagrange multiplier method to construct
an estimate with the maximum entropy consistent with the constraints.
Such maximum entropy estimates are likely to give a reasonable estimate
of the initial wave function.

To do this we estimate a ``most typical'' GTF as the one having
the maximum entropy consistent with the constraints. In \cite{Ashmead:2019aa}
we did this using the Lagrange multiplier approach. Here we summarize
the results of that analysis.

We start with the constraints:
\begin{equation}
\begin{array}{l}
\left\langle 1\right\rangle =1\\
\left\langle E\right\rangle =\sqrt{m^{2}+\left\langle \vec{p}\right\rangle ^{2}}\\
\left\langle E^{2}\right\rangle =\left\langle m^{2}+\vec{p}^{2}\right\rangle =m^{2}+\left\langle \vec{p}^{2}\right\rangle 
\end{array}\label{eq:constraints-1-1}
\end{equation}

From the third constraint we have:

\begin{equation}
{\left({\Delta{E}}\right)}^{2}={\left({\Delta{p}_{x}}\right)}^{2}{+}{\left({\Delta{p}_{y}}\right)}^{2}{+}{\left({\Delta{p}_{z}}\right)}^{2}
\end{equation}

Or in a more compact notation:

\[
{\mbox{ \ensuremath{\left({\Delta{E}}\right)}}}^{2}{=}{\mbox{ \ensuremath{\left({\Delta\mbox{ \ensuremath{\vec{p}}}}\right)}}}^{2}
\]

With the uncertainty in energy we have the ``most typical'' probability
density as a GTF:

\begin{equation}
{\rho}_{0}\mbox{ \ensuremath{\left({E}\right)}}{=}\sqrt{\frac{1}{{2}\pi{\mbox{ \ensuremath{\left({\Delta{E}}\right)}}}^{2}}}{e}^{{-}\frac{{\mbox{ \ensuremath{\left({{E}{-}\mbox{ \ensuremath{\bar{E}}}}\right)}}}^{2}}{\mbox{ \ensuremath{4\left({\Delta{E}}\right)}}}}
\end{equation}

And we have the wave function in energy as the square root of this:

\begin{equation}
{\varphi}_{0}\mbox{ \ensuremath{\left({E}\right)}}{=}\sqrt[4]{\frac{1}{{\pi\sigma}_{E}^{2}}}{e}^{\imath\mbox{ \ensuremath{\left({{E}{-}\mbox{ \ensuremath{\bar{E}}}}\right)}}\tau{-}\frac{{\mbox{ \ensuremath{\left({{E}{-}\mbox{ \ensuremath{\bar{E}}}}\right)}}}^{2}}{2{\sigma}_{E}^{2}}}
\end{equation}

with:

\[
{\mathit{\sigma}}_{E}^{2}\equiv2\left(\Delta{E}\right)^{2}
\]

In coordinate time:

\begin{equation}
{\varphi}_{0}\mbox{\ensuremath{\left({t}\right)}}{=}\sqrt[4]{\frac{1}{{\pi\sigma}_{t}^{2}}}{e}^{{-}\imath{\bar{E}}_{}\mbox{\ensuremath{\left({{t}{-}\tau}\right)}}{-}\frac{{t}^{2}}{2{\sigma}_{t}^{2}}}
\end{equation}

with:

\[
{\sigma}_{t}^{2}{=}\frac{1}{{\sigma}_{E}^{2}}
\]

\paragraph{Entropic estimate of bound state wave functions}

\label{subsec:Entropic-estimate-bound-states}

We are looking for a direct-product/disentangled approximation to
the bound state wave function:

\begin{equation}
{\psi}_{\tau}\mbox{\ensuremath{\left({E,\mbox{\ensuremath{\vec{p}}}}\right)}}\approx{\psi}_{\tau}^{\mbox{T}}\mbox{\ensuremath{\left({E}\right)}}{\psi}_{\tau}^{\mbox{\ensuremath{S}}}\mbox{\ensuremath{\left({\mbox{\ensuremath{\vec{p}}}}\right)}}
\end{equation}

In SQM the energy is forced by the momentum:

\begin{equation}
{{\mathit{\psi}}_{\mathit{\tau}}^{T}\left({E}\right){=}\sqrt{\mathit{\delta}\left({{E}{-}\sqrt{{m}^{2}{+}{\vec{p}}^{2}}}\right)}}
\end{equation}

For each ray in momentum space, we get $E$ as a corresponding delta
function:

\begin{equation}
\sqrt{\delta\mbox{ \ensuremath{\left({x}\right)}}}\rightarrow\mathop{\lim}\limits_{\varepsilon\rightarrow{0}}\frac{1}{\sqrt[4]{2{\mathit{\pi}\varepsilon}^{2}}}\exp\mbox{ \ensuremath{\left({-\mbox{{\large \ensuremath{\frac{{x}^{2}}{4{\varepsilon}^{2}}}}}}\right)}}
\end{equation}

We have the initial wave function ground state in momentum space from
Podolsky and Pauling \cite{Podolsky:1929aa} and more recently from
Hage-Hassan \cite{Hage-Hassan:2008aa}. We can also compute them by
Fourier transform from the familiar coordinate forms \cite{Wang:2008aa}.
The ground state in momentum space is given by: 

\begin{equation}
{\psi}_{100}\left({p}\right){=}\sqrt{\frac{32}{\pi}}{\sqrt{{a}_{0}}}^{3}\frac{1}{{\mbox{ \ensuremath{\left({{p}^{2}{a}_{0}^{2}{+}{1}}\right)}}}^{2}}\label{eq:ground-state-sqm-momentum}
\end{equation}

The average energy is:

\begin{equation}
{\left\langle {E}\right\rangle \equiv\int{{dEd}\vec{p}\mathit{\delta}\left({{E}{-}\sqrt{{m}^{2}{+}{\vec{p}}^{2}}}\right){E}{\mathit{\rho}}_{\mathit{\tau}}^{S}\left({\vec{p}}\right)}}
\end{equation}

We can get the uncertainty in energy by direct calculation from the
momentum space wave function:

\begin{equation}
\Delta{E}\equiv\sqrt{\left\langle {{E}^{2}}\right\rangle {-}{\left\langle {E}\right\rangle }^{2}}
\end{equation}

Taking advantage of the fact that we are working non-relativistically: 

\begin{equation}
\Delta{E}\equiv\sqrt{\left\langle {\left({m}^{2}{+}{\left|{\vec{p}}\right|}^{2}\right)^{2}}\right\rangle {-}{\left\langle {\sqrt{{m}^{2}{+}{\left|{\vec{p}}\right|}^{2}}}\right\rangle }^{2}}\approx\frac{1}{2m}\sqrt{\left\langle {{\left|{\vec{p}}\right|}^{4}}\right\rangle {-}{\left\langle {{\left|{\vec{p}}\right|}^{2}}\right\rangle }^{2}}
\end{equation}

Specializing to the ground state:

\begin{equation}
\Delta{E}_{100}\approx\frac{1}{2m}\sqrt{\frac{5}{{a}_{0}^{4}}{-}\frac{1}{{a}_{0}^{4}}}{=}\frac{1}{2m}\frac{2}{{a}_{0}^{2}}=\alpha^{2}m
\end{equation}

using $a_{0}=\frac{1}{\alpha m}$.

The ground state energy is given by:

\begin{equation}
{13}{.}{6}{\mathrm{eV}}=-\frac{{\mathit{\alpha}}^{2}m}{2}
\end{equation}

so we have $\Delta E$ as twice that:

\begin{equation}
\Delta{E}={27}{.}{2}{eV}\label{eq:ground-state-energy-uncertainty}
\end{equation}

\section{Estimate of the radial momentum}

\label{sec:bar-kappa}

The radial momentum $\kappa$ is needed to define the time/energy
part of the propagator. However it also entangles the time/energy
part with the space/momentum part. By replacing the radial momentum
$\kappa$ by its average we can disentangle these two parts. This
averaging is a formal device, since ultimately both parts have to
be thoroughly entangled, but it is a useful way to give a rough overall
scale of the dispersion in time, start off a perturbation analysis,
and so on.

We can parameterize this estimate by writing:

\begin{equation}
\bar{\kappa}=\frac{\mu\left(r\right)}{r}
\end{equation}

since $\bar{\kappa}$, from symmetry, cannot depend on anything but
the radius. 

We can compute $\bar{\kappa}$ by computing the associated integral
over the Green's function in time/energy:

\begin{equation}
\exp\mbox{ \ensuremath{\left(\imath\bar{\kappa}\tau+{\imath\delta{w}\tau{-}\imath\mbox{ \ensuremath{\frac{{\mbox{ \ensuremath{\left({\delta{w}}\right)}}}^{2}}{2\bar{\kappa}}}}\tau}\right)}}{=}\left\langle {{G}_{\mathit{\tau}}^{T}\left({{w}{,}\mathit{\kappa}}\right)}\right\rangle 
\end{equation}

where the average is taken over all values of $\kappa$, weighted
by the SQM part of the propagator:

\begin{equation}
\left\langle {{G}_{\tau}^{T}\mbox{ \ensuremath{\left({{w}{,}\kappa}\right)}}}\right\rangle {=}\mathop{\int}\limits_{0}^{\infty}{{d}{\mathit{\kappa}\mathit{\kappa}}^{2}{G}_{\tau}^{T}\mbox{ \ensuremath{\left({{w}{,}\kappa}\right)}}{G}_{\mathit{\tau}}^{S}\left({\mathit{\kappa}}\right)}
\end{equation}

We have:

\begin{equation}
{G}_{\mathit{\tau}}^{S}\left({\mathit{\kappa}}\right){=}\frac{1}{{\mathit{\kappa}}^{2}}
\end{equation}

So the average over $G^{T}$ is:

\begin{equation}
\left\langle {{G}_{\tau}^{T}\mbox{ \ensuremath{\left({{w}{,}\kappa}\right)}}}\right\rangle {=}\exp\mbox{ \ensuremath{\left({\imath\delta{w}\tau}\right)}}\mathop{\int}\limits_{0}^{\infty}{{d}\kappa\exp\mbox{ \ensuremath{\left({\imath\kappa\tau{-}\imath\mbox{{\large \ensuremath{\frac{{\mbox{ \ensuremath{\left({\delta{w}}\right)}}}^{2}}{2\kappa}}}}\tau}\right)}}}\label{eq:bar-kappa-integral}
\end{equation}

The integral is not convergent as it stands. We can try applying it
to the original photon wave function, using the original wave function
as an organic regularizing factor. However this is not physically
reasonable. The initial photon wave function has a dispersion of order
100 MeV. We expect that an appropriate regularizing factor would have
a dispersion of order KeV. 

We can find such a regularizing factor in what we might call the ``final
photon wave function''. Just as in creating the initial photon wave
function we took advantage of the fact that the initial photon momentum
is given by the difference of the initial and final proton momenta,
we can take advantage of the fact that final photon momentum must
be given by the difference of the final and initial atomic electron
momenta (${\mathit{\delta}}^{3}\left({{\vec{p}}_{2}{-}\vec{k}{-}{\vec{p}}_{1}}\right)$).
We could, in principle, use this to compute the average value of $\kappa$
directly. Unfortunately the equivalent ``final integral'' does does
not converge for the ground state: after all powers of the momentum
are accounted for, we are left with a logarithmic divergence.

Our suspicion is that these two failures may be the mathematics's
way of telling us that we are attacking the problem from the wrong
direction; if we are investigating the TQM solution we should do so
starting from TQM itself rather than treating it as a bolt-on to SQM.

For the time being we will therefore resign ourselves to estimating
the value of $\mu$ by comparing this to the GHO estimate. Per the
discussion above (eqn \ref{eq:harmonic-oscillator-comparison}) $\mu=1$.

\bibliographystyle{hplain}

\bibliography{/Volumes/Ishtar/Projects/bibliography/taqm, /Volumes/Ishtar/Projects/bibliography/general}

\end{document}